\documentclass{elsart}
\usepackage{amssymb}
\usepackage{amsmath}
\usepackage{graphicx}
\usepackage{dcolumn}
\usepackage{bm}
\newlength{\defbaselineskip}
\newcommand{\hs}{\hspace*{0.5cm}}
\newcommand{\vs}{\vspace*{0.5cm}}
\newcommand{\be}{\begin{equation}}
\newcommand{\ee}{\end{equation}}
\newcommand{\bea}{\begin{eqnarray}}
\newcommand{\eea}{\end{eqnarray}}
\newcommand{\nn}{\nonumber}
\newcommand{\crn}{\nonumber \\}

\newcommand{\+}{\dagger}
\newcommand{\al}{\alpha}
\newcommand{\la}{\lambda}
\newcommand{\bet}{\beta}
\newcommand{\ga}{\gamma}
\newcommand{\va}{\varphi}
\newcommand{\om}{\omega}
\newcommand{\pa}{\partial}
\newcommand{\fr}{\frac}
\newcommand{\bc}{\begin{center}}
\newcommand{\ec}{\end{center}}
\newcommand{\Ga}{\Gamma}
\newcommand{\de}{\delta}
\newcommand{\De}{\Delta}
\newcommand{\ep}{\epsilon}

\newcommand{\ka}{\kappa}

\newcommand {\ba}{\begin{array}}
\newcommand {\ea}{\end{array}}
\newcommand{\ben}{\begin{enumerate}}
\newcommand{\een}{\end{enumerate}}

\begin{document}
\flushright{KEK-TH-1245}

\vs

\runauthor{Dong, and Long}
\begin{frontmatter}
\title{The economical $\mathrm{SU}(3)_C\otimes \mathrm{SU}(3)_L \otimes
\mathrm{U}(1)_X$ model}
\author{P. V. Dong\thanksref{D}}
\address{Theory Group, KEK, 1-1 Oho, Tsukuba, 305-0801, Japan}
\author{H. N. Long\thanksref{L}}
\address{Institute of Physics, VAST, P.O. Box 429, Bo Ho, Hanoi
10000, Vietnam}
\thanks[D]{Email: pvdong@iop.vast.ac.vn; on leave from Institute
of Physics, VAST, Vietnam}
\thanks[L]{Email: hnlong@iop.vast.ac.vn}

\begin{abstract}
 In this report the $\mathrm {SU}(3)_C\otimes \mathrm
{SU}(3)_L \otimes {\mathrm U}(1)_X$ gauge model with minimal scalar
sector, two Higgs triplets, is presented in detail. One of the
vacuum expectation values $u$ is a source of lepton-number
violations and a reason for mixing among charged gauge bosons---the
standard model $W^\pm$ and the bilepton gauge bosons $Y^\pm$ as well
as among the neutral non-Hermitian bilepton $X^0$ and neutral gauge
bosons---the $Z$ and the new $Z'$. An exact diagonalization of the
neutral gauge boson sector is derived and bilepton mass splitting is
also given. Because of these mixings, the lepton-number violating
interactions exist in both charged and neutral gauge boson sectors.
Constraints on vacuum expectation values of the model are estimated
and $u \simeq \mathcal{O}(1)\ \textrm{GeV}, v \simeq
v_{\mathrm{weak}} = 246 \ \textrm{GeV}$, and $\om  \simeq
\mathcal{O}(1)\ \textrm{TeV}$. In this model there are three
physical scalars, two neutral and one charged, and eight Goldstone
bosons---the needed number for massive gauge bosons. The minimal
scalar sector can provide all fermions including quarks and
neutrinos consistent masses in which some of them require one-loop
radiative corrections.

\end{abstract}

\journal{Advances in High Energy Physics}

\end{frontmatter}

\newpage

\tableofcontents
\newpage

\section{\label{Ch:Intro}Introduction}

In spite of all the successes of the standard model it is unlikely
to be the final theory. It leaves many striking features of the
physics of our world unexplained. In the following we list some of
them which leads to the model's extensions. In particular the models
with $\mathrm {SU}(3)_C\otimes \mathrm {SU}(3)_L \otimes {\mathrm
U}(1)_X$ (3-3-1) gauge group are presented.

\subsection{Generation Problem and 3-3-1 Models}

In the standard model the fundamental fermions come in generations.
In writing down the theory one may start by first introducing just
one generation, then one may repeat the same procedure by
introducing copies of the first generation. Why do quarks and
leptons come in repetitive structures (generations)? How many
generations are there? How to understand the inter-relation between
generations? These are the central issues of the weak interaction
physics known as the generation problem or the flavor question.
Nowhere in physics this question is replied~\cite{glashow}. One of
the most important experimental results in the past few years has
been the determination of the number of these generations within the
framework of the standard model. In the minimal electroweak model
the number of generations is given by the number of the neutrino
species which are all massless, by definition. The number of
generations is then computed from the invisible width of the $Z^0$,
\begin{equation}
\Gamma_{\rm inv}\equiv \Gamma_{Z^0} - (\Gamma_h + \sum_l
\Gamma_l),\nn
\end{equation}
where $\Gamma_{Z^0}$ denotes the total width, the subscript $h$
refers to hadrons and $\Gamma_l$ $(l = e,\ \mu,\ \tau)$ is the
width of the $Z^0$ decay into an $l\bar l$ pair. If $\Gamma_\nu$
is the theoretical width for just one massless neutrino, the
number of generations is $ N_{\rm \mathrm{gen}} = N_\nu =
\Gamma_{\rm inv}/\Gamma_\nu$ and recent results give a value very
close to three $N_{\rm \mathrm{gen}} = 2.99\pm 0.03$
\cite{dois,pdg} but we do not understand why the number of
standard model generations is three.

The answer to the generation problem may require a radical change
in our approaches. It could be that the underlying objects are
strings and all the low energy phenomena will be determined by
physics at the Planck scale. Grand Unified Theories (GUTs) have
had a major impact on both cosmology and astrophysics; for
cosmology they led to the inflationary scenario, while for
astrophysics supernova, neutrinos were first observed in
proton-decay detectors. It remains for GUTs to have impact
directly on particle physics itself~\cite{doisA}. GUTs cannot
explain the presence of fermion generations. On the other side,
supersymmetry (SUSY) for the time being is an answer in search of
question to be replied. It does not explain the existence of any
known particle or symmetry. Some traditional approaches to the
problem such as GUTs, monopoles and higher dimensions introduce
quite speculative pieces of new physics at high and experimentally
inaccessible energies. Some years ago there were hopes that the
number of generations could be computed from first principles such
as geometry of compactified manifolds but these hopes did not
materialize.

A very interesting alternative to explain the origin of
generations comes from the cancelation of chiral anomalies of a
gauge theory in all orders of perturbative expansion, which
derives from the renormalizability condition. This constrains the
fermion representation content. Three perturbative anomalies have
been identified~\cite{GeMa}
 for chiral gauge theories in four dimensional space-time:
(i) The triangle chiral gauge anomaly~\cite{ABJ} must be cancelled
to avoid violations of gauge invariance and the renormalizability
of the theory; (ii) The global non-perturbative SU(2) chiral gauge
anomaly,~\cite{Witten} which must be absent in order for the
fermion integral to be defined in a gauge invariant way; (iii) The
mixed perturbative chiral gauge gravitational
anomaly~\cite{DelSalam,LAGW} which must be cancelled in order to
ensure general covariance. The general anomaly-free condition is
\begin{equation}
A^{ijk}\equiv\mathrm{Tr}[\{T^i,T^j\}T^k]=\sum_{\rm
representations} {\rm Tr}[\{T^i_L, T^j_L\}T^k_L - \{T^i_R, T^j_R\}
T^k_R] = 0,
\end{equation}
where $T^i$ is the representation of the gauge algebra on the set
of all left-handed fermion and anti-fermion fields put in a single
column $\psi$, and ``Tr'' denotes a sum over these fermion and
anti-fermion species; $T^i_{L,R}$ are the coupling matrices of
fermions $\psi_{L,R}$ to the current $J^i_\mu=\bar{\psi}_L\ga_\mu
T^i_L\psi_L+\bar{\psi}_R\ga_\mu T^i_R\psi_R$, respectively. The
$i$ index runs over the dimension of a simple SU($n$) group, $i=
1,2,...,n^2-1$, with a rank $n-1$, and $i=0$ for the Abelian
factor.

First let us consider the relationship between anomaly cancelation
and flavor problem in the standard model. The individual generations
have the following structure under the $\mathrm{SU}(3)_C\otimes
\mathrm{SU}(2)_L\otimes \mathrm{U}(1)_Y$ (3-2-1) gauge group, \bea
(\nu_{aL},l_{aL})&\sim& (1,2,-1),\hs l_{aR}\sim (1,1,-2),\crn
(u_{aL},d_{aL})&\sim& (3,2,1/3),\hs u_{aR}\sim (3,1,4/3),\hs
d_{aR}\sim (3,1,-2/3). \eea The values in the parentheses denote
quantum numbers based on the $(\mathrm{SU}(3)_C$,\
$\mathrm{SU}(2)_L$,\ $\mathrm{U}(1)_Y)$ symmetry, where the
subscripts $C$, $L$ and $Y$, respectively, indicate to the color,
left-handed, and hypercharge. The electric charge operator is
defined as $Q=T^3+\fr{1}{2}Y$ where $T^i=\fr{1}{2}\sigma^i$
$(i=1,2,3)$ with $\sigma^i$ are Pauli matrices. The weak isospin
group SU(2)$_L$ is a safe group due to the fact that
\begin{equation}
{\rm Tr}[\{\sigma^i, \sigma^j\}\sigma^k] = 2\delta^{ij}{\rm
Tr}[\sigma^k] = 0.
\end{equation}
However, in the case where at least one of the generators is
hypercharge we have:
\begin{equation}
{\rm Tr}[\sigma^i Y Y]\propto {\rm Tr}[\sigma^i]=0,\hs {\rm
Tr}[\{\sigma^i, \sigma^j\}Y] = 2\delta^{ij}\,{\rm Tr}[Y].
\end{equation}
The anomaly contribution in the last condition is proportional to
the sum of all fermionic discrete hypercharge values on the color,
flavor, and weak-hypercharge degrees of freedom
\begin{equation}
{\rm Tr}[Y]=\sum_{\rm lepton} (Y_L + Y_R) + \sum_{\rm quark} (Y_L
+ Y_R). \nn
\end{equation}
The Tr$[Y]$ vanishes for the fermion content in the
$a^{th}$-generation because \bea \sum_{\rm lepton}(Y_L + Y_R) &=&
Y(\nu_{aL}) + Y(l_{aL}) + Y(l_{aR}) = -4 ,\crn \sum_{\rm quark}
(Y_L + Y_R) &=& 3[Y(u_{aL}) + Y(d_{aL})+Y(u_{aR}) + Y(d_{aR})] =
+4,\nn\eea where the 3 factor takes into account the number of
quark colors. In the last case all the generators are hypercharge:
\begin{equation}
{\rm Tr}[Y^3]\propto {\rm Tr}[Q^2T_3 - QT_3^2],
\end{equation}
where we used the fact that the electromagnetic vector neutral
current vertices do not have anomalies. For the $a^{th}$-generation,
we have \bea \sum_{\rm lepton} (Q^2T_3 - QT^2_3) &=& [(0)^2(1/2) -
(0)(1/2)^2] \crn &&+ [(-1)^2(-1/2) - (-1)(-1/2)^2] =
-\frac{1}{4},\eea \bea \sum_{\rm quark} (Q^2T_3 - QT_3^2) & = &
3[(2/3)^2(1/2)-(2/3)(1/2)^2] \crn &&+ 3[(-1/3)^2(-1/2) -
(-1/3)(-1/2)^2] = + \frac{1}{4}.\eea It yields that the anomaly in
standard model cancels within each individual generation, but not by
generations. Flavor question and anomaly-free conditions do not seem
to have any connection in the standard model. This leads us to
questions when going beyond this model: Are the anomalies always
canceled automatically within each generation of quarks or leptons?
Do the anomaly cancelation conditions have any connection with
flavor puzzle?

We wish to show that some very fundamental aspects of the standard
model, in particular the flavor problem, might be understood by
embedding the three-generation version in a Yang-Mills theory with
the $\mathrm{SU(3)}_C\otimes\mathrm{SU(3)}_L\otimes\mathrm{U(1)}_X$
semisimple gauge group with a corresponding enlargement of the
lepton and quark representations \cite{ppf,flt,flaque}. In
particular, the number of generations will be related by anomaly
cancelation to the number of quark colors, and one generation of
quarks will be treated differently from the two others; in the 3-2-1
low-energy limit all three generations appear similarly and cancel
anomalies separately. Let us consider the following 3-3-1 fermion
representation content
\bea \psi_{aL}&=&\left(%
\begin{array}{c}
  \nu_{aL} \\
  l_{aL} \\
  \nu^c_{aR} \\
\end{array}%
\right)\sim \left(1,3,-\fr 1 3\right),\hs l_{aR}\sim (1,1,-1),\hs
a=1,2,3,\crn Q_{1L}&=&\left(%
\begin{array}{c}
  u_{1L} \\
  d_{1L} \\
  U_L \\
\end{array}%
\right)\sim \left(3,3,\fr 1 3\right),\ Q_{\al L}=\left(%
\begin{array}{c}
  d_{\al L}\\
  -u_{\al L}\\
  D_{\al L}\\
\end{array}%
\right)_L\sim (3,3^*,0),\ \al=2,3,\label{parcontrhn}\\ u_{a
R}&\sim&\left(3,1,\fr 2 3\right),\ d_{a R} \sim \left(3,1,-\fr 1
3\right),\ U_{R}\sim \left(3,1,\fr 2 3\right),\ D_{\al R} \sim
\left(3,1,-\fr 1 3\right).\nn\eea The quantum numbers in the
parentheses are based on the $(\mathrm{SU}(3)_C$,
$\mathrm{SU}(3)_L$, $\mathrm{U}(1)_X)$ symmetry. The right-handed
neutrinos $\nu_R$ and the exotic quarks $U$ and $D_\al$ are composed
along with that of the standard model. We call 3-3-1 model with
right-handed neutrinos. The electric charge operator in this case
takes a form $Q=T^3-\fr{1}{\sqrt{3}}T^8+X$ with $T^i=\fr{1}{2}\la^i$
$(i=1,2,...,8)$ and $X$ standing for $\mathrm{SU}(3)_L$ and
$\mathrm{U}(1)_X$ charges, respectively. Electric charges of the
exotic quarks are the same as of the usual quarks, i.e., $q_{U}=\fr
2 3$ and $q_{D_\al}=-\fr 1 3$.

The SU(3)$_L$ group is not safe in the sense of the standard model
SU(2)$_L$ with the vanishing
$\mathrm{Tr}[\{\sigma^i,\sigma^j\},\sigma^k]=0$. The SU(3)$_L$
generators proportional to the Gell-Mann matrices close among them
the Lie algebra structure, \be [\lambda^i,\lambda^j]=
2if^{ijk}\lambda^k,\hs \{\lambda^i,\lambda^j\} =
\frac{4}{3}\delta^{ij} + 2d^{ijk}\lambda^k,\ee where the structure
constant $f^{ijk}$ is totally antisymmetric and $d^{ijk}$ is totally
symmetric under exchange of the indices. We can normalize the
$\lambda$-matrices such that $\mathrm{Tr}[\la^i\la^j]=2\de^{ij}$.
Therefore, $f^{ijk}$ and $d^{ijk}$ are calculated by
\begin{eqnarray}
f^{ijk} =\frac{1}{4i}\,{\rm Tr}\left[[\lambda^i,
\lambda^j]\lambda^k\right],\hs d^{ijk} =\frac{1}{4}\,{\rm
Tr}\left[\{\lambda^i,
\lambda^j\}\lambda^k\right].\nn\end{eqnarray} The anomaly is
proportional to $d^{ijk}$ in general, and of course such
coefficients vanish in the case of the SU(2)$_L$ generators.

In the 3-3-1 model there are six triangle anomalies which are
potentially troublesome; in a self-explanatory notation these are
$(3_C)^3$, $(3_C)^2X$, $(3_L)^3$, $(3_L)^2X$, $X^3$, and
$(\mathrm{graviton})^2X$. The quantum chromodynamics anomaly
$(3_C)^3$ is absent because the theory mentioned is vectorlike
(i.e., $T^i_L=U^{-1}T^i_RU$ with some unitary matrix $U$), and hence
the conditions $A^{ijk}=0$ are automatically satisfied. For any $D$
fermion representation, it satisfies the condition $A(D) = -A(D^*)$
where $A(D^*)$ is the anomaly of the conjugate representation of $D$
\cite{G84}. The pure SU(3)$_L$ anomaly $(3_L)^3$ therefore vanishes
because there is an equal number of triplets $3_L$ and antitriplets
$3^*_L$ in the given fermion content. The remaining anomaly-free
conditions are explicitly written as follows \ben \item \label{it1}
$\mathrm{Tr}[\mathrm{SU}(3)_C]^2[\mathrm{U}(1)_X]=0:$ \be
3\sum_{\mathrm{generation}}X^L_q-\sum_{\mathrm{generation}}
{\sum_{\mathrm{singlet}}X^R_q}=0,\nn\ee
\item \label{it2} $\mathrm{Tr}[\mathrm{SU}(3)_L]^2[\mathrm{U}(1)_X]=0:$
\be \sum_{\mathrm{generation}}X^L_l
+3\sum_{\mathrm{generation}}X^L_q =0,\nn\ee \item \label{it3}
$\mathrm{Tr}[\mathrm{U}(1)_X]^3=0:$\bea
&&3\sum_{\mathrm{generation}}(X^L_l)^3+
9\sum_{\mathrm{generation}}(X^L_q)^3 -
3\sum_{\mathrm{generation}}\sum_{\mathrm{singlet}}(X^R_q)^3\crn
&&-\sum_{\mathrm{generation}}\sum_{\mathrm{singlet}}(X^R_l)^3=0,\nn\eea
\item \label{it4} $\mathrm{Tr}[\mathrm{graviton}]^2[\mathrm{U}(1)_X]=0$
\bea
&&3\sum_{\mathrm{generation}}X^L_l+9\sum_{\mathrm{generation}}X^L_q-3
\sum_{\mathrm{generation}}\sum_{\mathrm{singlet}}X^R_q \crn
&&-\sum_{\mathrm{generation}}\sum_{\mathrm{singlet}}X^R_l=0,\nn\eea\een
where $X^L_l$, $X^L_q$ and $X^R_l$, $X^R_q$ indicate to the
$\mathrm{U}(1)_X$ charges of the left-handed lepton, quark triplets
or antitriplets and the right-handed lepton, quark singlets,
respectively. It is worth noting that some 3 factors in the
conditions (\ref{it2}), (\ref{it3}) and (\ref{it4}) take into
account the number of quark colors. With the fermion content as
given, it is easily checked that all the above anomaly-free
conditions are satisfied. For example, let us take the condition
(\ref{it2}). We first calculate the $3^2_LX$ anomaly for the first
generation: $-1/3+3\times(1/3)=2/3$. The anomaly of the second or
the third generation is $-1/3+3\times 0=-1/3$. It is especially
interesting that this anomaly cancelation takes place between
generations, unlike those of the standard model. Each individual
generation possesses non-vanishing $(3_L)^3,\ (3_L)^2X,\ X^3$, and
$(\mathrm{gravion})^2X$ anomalies. Only with a matching of the
number of generations with the number of quark colors does the
overall anomaly vanish.

Next let us introduce an alternative fermion content where the three
known left-handed lepton components for each generation are
associated to three $\mathrm{SU}(3)_L$ triplets such that
$(\nu_{aL},l_{aL},l^c_{aR})^T\sim(1,3,0)$ (called minimal 3-3-1
model). Canceling the pure SU(3)$_L$ anomaly requires that there are
the same number of triplets and antitriplets, thus $Q_{1L} =
(u_{1L},d_{1L},J_L)^T \sim (3,3,2/3)$, $Q_{\al L}=(d_{\al L},-u_{\al
L},J_{\al L})^T\sim (3,3^*,-1/3)$. The respective right-handed
fields are singlets: $u_{aR}\sim (3,1,2/3)$ and
$d_{aR}\sim(3,1,-1/3)$ for the ordinary quarks; $J_R\sim(3,1,5/3)$
and $J_{\al R}\sim (3,1,-4/3)$ for the exotic quarks. Similarly to
the previous 3-3-1 model, the $(3_L)^3,\ (3_L)^2X,\ X^3$ anomalies
vanish only if three generations of quarks and leptons take into
account.

In a general case, we can verify that the number of generations must
be multiple of the quark-color number in order to cancel the
anomalies. On the other hand, if we suppose that the exotic quarks
also contribute to the running of the coupling constants, the
asymptotic-freedom principle requires that the number of quark
generations is no more than five. It follows that the number of
generations is {\it just} three. This provides a first step towards
answering the flavor question. The asymmetric treatment of one
generation of quarks breaks generation universality. This might
provide an explanation of why the top quark is uncharacteristically
heavy \cite{frampton1,longvan}. An interesting alternative feature
is that the electric charge quantization in nature might also be
explained in this framework \cite{ecquantize}. Just enlarging
SU(2)$_L$ to SU(3)$_L$, we have thus presented the simplest gauge
extension of the standard model for the flavor question. The new
models get five additional gauge bosons contained in a gauge adjoint
octet: $8 = 3 + (2 +2) + 1$ under SU(2)$_L$. The $1$ is a neutral
$Z^\prime$ and the two doublets are readily identifiable from the
leptonic contents as non-Hermitian bilepton gauge bosons $(X,Y)^T$
and $(X^{*},Y^{*})$. From the renormalization group analysis of the
coupling constants \cite{prampton}, the SU(3)$_L$ breaking scale is
estimated to be lower than some TeV in the minimal 3-3-1 model. This
is due to the fact that the squared sine of the Weinberg angle
$\theta_W$ gets an upper bound, $\sin^2\theta_W<1/4$. There is no
``grand desert'' in this model in comparison to GUTs. In contrast,
the energy scale in the 3-3-1 model with right-handed neutrinos is
very high, even larger than the Planck scale, because of
$\sin^2\theta_W<3/4$. This version might allow the existence of a
``desert''. Anyway, the new physics in these models expected arise
at not too high energies. The new particles such as the bilepton
gauge bosons, $Z'$ and exotic quarks would be determinable in the
next generation of collides.

\subsection{Proposal of Minimal Higgs Sector}

As mentioned above, there are two main versions of 3-3-1
models---the minimal model and the model with right-handed
neutrinos, which have been subjects studied extensively over the
last decade. In the minimal 3-3-1 model \cite{ppf}, the scalar
sector is quite complicated and contains three scalar triplets and
one scalar sextet. In the 3-3-1 model with right-handed neutrinos
\cite{flt,long}, the scalar sector requires three Higgs triplets. It
is interesting to note that two Higgs triplets of this model have
the same $\mathrm{U}(1)_X$ charges with two neutral components at
their top and bottom. Allowing these neutral components vacuum
expectation values (VEVs) we can reduce number of Higgs triplets to
be two. Note that the mentioned model contains very important
advantage, namely, there is no new parameter, but it contains very
simple Higgs sector, therefore the significant number of free
parameters is reduced. To mark the minimal content of the Higgs
sector, this version that includes right-handed neutrinos is going
to be called the {\it economical 3-3-1 model}
\cite{ponce,ponc1,dlns,dls,dln,dhhl,dls1}. The interested reader can
find the suppersymmetric version in Ref. \cite{supeco}.

This kind of model was proposed in Ref. \cite{ponce}, but has not
got enough attention. In Ref. \cite{ponc1}, phenomenology of this
model was presented without mixing between charged gauge bosons as
well as neutral ones. The mass spectrum  of the mentioned scalar
sector has also been presented in~\cite{ponce}, and some couplings
of the two neutral scalar fields with the charged $W$ and the
neutral $Z$ gauge bosons in the standard model were presented. From
explicit expression for the $ZZH$ vertex, the authors concluded that
two VEVs responsible for the second step of spontaneous symmetry
breaking have to be in the same range:  $u \sim v$, or the theory
needs an additional scalar triplet. As we will show in the
following, this conclusion is incorrect.

It is well known that the electroweak symmetry breaking  in the
standard model is achieved via the Higgs mechanism. In the
Weinberg-Salam model there is a single complex scalar doublet,
where the Higgs boson $H$ is the physical neutral Higgs scalar
which is the only remaining part of this doublet after spontaneous
symmetry breaking. In the extended  models there are additional
charged and neutral scalar Higgs particles. The prospects for
Higgs coupling measurements at the CERN Large Hadron Collider
(LHC) have recently been analyzed in detail in Ref.~\cite{logan}.
The experimental detection of the $H$ will be great triumph of the
standard model of electroweak interactions and will mark new stage
in high energy physics.

In extended Higgs models, which would be deduced in the low energy
effective theory of new physics models, additional Higgs bosons
like charged and CP-odd scalar bosons are predicted. Phenomenology
of these extra scalar bosons strongly depends on the
characteristics of each new physics model. By measuring their
properties like masses, widths, production rates and decay
branching ratios, the outline of physics beyond the electroweak
scale can be experimentally determined.

The interesting feature compared with other 3-3-1 models is the
Higgs physics. In the 3-3-1 models, the general Higgs sector is
very complicated~\cite{long98,changlong,tujo,ochoa2} and this
prevents the models' predicability. The scalar sector of the
considering  model is one of  subjects in the present work. As
shown, by couplings of the scalar fields with the ordinary gauge
bosons such as the photon, the $W$ and the neutral $Z$ gauge
bosons, we are able to identify full content of the Higgs sector
in the standard model including the neutral $H$ and the Goldstone
bosons eaten by their associated massive gauge ones. All
interactions among Higgs-gauge bosons in the standard model are
recovered.

Production of the Higgs boson in the 3-3-1 model with right-handed
neutrinos at LHC has been considered in \cite{ninhlong}. In scalar
sector of the considered model, there exists the singly-charged
boson $H_2^\pm$, which is a subject of intensive current studies
\cite{kame,roy}. The trilinear coupling $ZW^\pm H^\mp$ which
differs, at the tree level, from zero  only in the models with Higgs
triplets, plays a special role on study phenomenology of these
exotic representations.  We shall pay particular interest on this
boson.

At the tree level, the mass matrix for the up-quarks has one
massless state and in the down-quark sector, there are two massless
ones. This calls for radiative corrections. To solve this problem,
the authors in Ref. \cite{ponc1} have introduced the third Higgs
triplet. In this sense the economical 3-3-1 model is not realistic.
In the present work we will show that this is a mistake! Without the
third one, at the one loop level, the fermions in this model, with
the given  set of parameters, gain a consistent mass spectrum. A
numerical evaluation leads us to conclusion that in the model under
consideration, there are two scales for masses of the exotic quarks.

At the tree level, the neutrino spectrum is Dirac particles with one
massless and two degenerate in mass $\sim h^\nu v$. This spectrum is
not realistic under the data because there is only one squared-mass
splitting. Since the observed neutrino masses are so small, the
Dirac mass is unnatural. One must understand what physics gives
$h^\nu v \ll h^l v$---the mass of charged leptons. In contrast to
the seesaw cases \cite{seesaw} in which the problem can be solved,
in this model the neutrinos including the right-handed ones get only
small masses through radiative corrections
\cite{radiative,radiative331,dls1,changlong}. We will obtain these
radiative corrections and provides a possible explanation of natural
smallness of the neutrino masses. This is not the result of a
seesaw, but it is due to a finite mass renormalization arising from
a very different radiative mechanism. We will show that the
neutrinos can get mass not only from the standard symmetry
breakdown, but also from the electroweak $\mathrm{SU}(3)_L \otimes
\mathrm{U}(1)_X$ breaking associated with spontaneous lepton-number
breaking (SLB), and even through the explicit lepton-number
violating processes due to a new physics. The total neutrino mass
spectrum at the one-loop level is neat and can fit the data.

This report is organized as follows. In Section \ref{ecn331model} we
give a review of the model with stressing on the gauge bosons,
currents, and constraints on the new physics. The Higgs--gauge
interactions and scalar content are considered in Section
\ref{hginteractions}. Section \ref{fermionmasses} is devoted to
fermion masses. We summarize our results and make conclusions in the
last section---Section \ref{conclutons}.

\section{\label{ecn331model}The Economical 3-3-1 Model}

We first recall the idea of constructing the model. An exact
diagonalization of charged and neutral gauge boson sectors and their
masses and mixings are presented. Because of the mixings, currents
in this model have unusual features which are obtained then.
Constraints on the parameters and some phenomena are sketched.

\subsection{\label{pcontent}Particle Content}

The fermion content which is anomaly free is given by Eq.
(\ref{parcontrhn}) like that of the 3-3-1 model with right-handed
neutrinos. However, contrasting with the ordinary model in which the
third generation of quarks should be discriminating \cite{longvan},
in the model under consideration the first generation has to be
different from the two others. This results from the mass patterns
for the quarks which will be derived in Section~\ref{fermionmasses}.

The 3-3-1 gauge group is broken spontaneously via two stages. In the
first stage, it is embedded in
that of the standard model via a Higgs scalar triplet \be \chi=\left(%
\begin{array}{c}
  \chi^0_1 \\
  \chi^-_2 \\
  \chi^0_3 \\
\end{array}%
\right) \sim \left(1,3,-\fr 1 3\right)\ee with the VEV given by \be
\langle\chi\rangle=\fr{1}{\sqrt{2}}\left(%
\begin{array}{c}
  u \\
  0 \\
  \om \\
\end{array}%
\right).\label{vevc}\ee In the last stage, to embed the standard
model gauge symmetry in $\mathrm{SU}(3)_C\otimes\mathrm{U}(1)_Q$,
another Higgs scalar triplet is needed \be
\phi=\left(%
\begin{array}{c}
  \phi^+_1 \\
  \phi^0_2 \\
  \phi^+_3 \\
\end{array}%
\right)\sim \left(1,3,\fr 2 3\right)\ee with the VEV as follows
\be \langle\phi\rangle=\fr{1}{\sqrt{2}}\left(%
\begin{array}{c}
  0 \\
  v \\
  0 \\
\end{array}%
\right).\label{vevp}\ee

The Yukawa interactions which induce masses for the fermions can be
written in the most general form as follows \be {\mathcal
L}_{\mathrm{Y}}={\mathcal L}_{\mathrm{LNC}} +{\mathcal
L}_{\mathrm{LNV}},\ee  where LNC and LNV respectively indicate to
the lepton number conserving and violating ones as shown below.
Here, each part is defined by \bea {\mathcal
L}_{\mathrm{LNC}}&=&h^U\bar{Q}_{1L}\chi
U_{R}+h^D_{\al\beta}\bar{Q}_{\al L}\chi^* D_{\beta
R}+h^d_{a}\bar{Q}_{1 L}\phi d_{a R}+h^u_{\al a}\bar{Q}_{\al L}\phi^*
u_{aR}\crn &&+h^l_{ab}\bar{\psi}_{aL}\phi
l_{bR}+h^\nu_{ab}\ep_{pmn}(\bar{\psi}^c_{aL})_p(\psi_{bL})_m(\phi)_n
+ \mathrm{H.c.},\label{y1}\\ {\mathcal
L}_{\mathrm{LNV}}&=&s^u_{a}\bar{Q}_{1L}\chi u_{aR}+s^d_{\al
a}\bar{Q}_{\al L}\chi^* d_{a R} +s^D_{ \al}\bar{Q}_{1L}\phi D_{\al
R}+s^U_{\al }\bar{Q}_{\al L}\phi^* U_{R}\crn &&+
\mathrm{H.c.},\label{y2}\eea where $p$, $m$ and $n$ stand for
$\mathrm{SU}(3)_L$ indices.

The VEV $\om$ gives mass for the exotic quarks $U$ and $D_\al$, $u$
gives mass for $u_1, d_{\al}$, while $v$ gives mass for $u_\al,
d_{1}$ and all ordinary leptons. In Section \ref{fermionmasses} we
will provide more details on analysis of fermion masses. As
mentioned, $\om $ is responsible for the first stage of symmetry
breaking, while the second stage is due to $u$ and $v$; therefore,
the VEVs in this model satisfies the constraint: \be u^2, v^2 \ll
\om^2 . \label{vevcons} \ee

The Yukawa couplings in Eq. (\ref{y1}) possess an extra global
symmetry \cite{changlong,tujo} which is not broken by $ v,
\omega$, but by $u$. From these couplings, one can find the
following lepton symmetry $L$ as in Table \ref{lnumber} (only the
fields with nonzero $L$ are listed; all other fields have
vanishing $L$). Here $L$ is broken by $u$ which is behind
$L(\chi^0_1)=2$, i.e., $u$ {\it is a kind of the SLB scale}
\cite{major-models}.
\begin{table}\bc
 \caption{\label{lnumber}Nonzero lepton number $L$
 of the model particles.}
\begin{tabular}{|c|c|c|c|c|c|c|c|c|}
    \hline
  Field
&$\nu_{aL}$&$l_{aL,R}$&$\nu^c_{aR}$ & $\chi^0_1$&$\chi^-_2$ &
$\phi^+_3$ & $U_{L,R}$ & $D_{\alpha L,R}$\\
    \hline
        $L$ & $1$ & $1$ & $-1$ & $2$&$2$&$-2$&$-2$&$2$\\
    \hline
\end{tabular}
\ec
\end{table} It is interesting that the exotic quarks also carry the lepton
number (so-called leptoquarks); therefore, this $L$ obviously does
not commute with the gauge symmetry. One can then construct a new
conserved charge $\mathcal L$ through $L$ by making a linear
combination $L= xT_3 + yT_8 + {\mathcal L} I$. Applying $L$ on a
lepton triplet, the coefficients will be determined \be L =
\fr{4}{\sqrt{3}}T_8 + {\mathcal L} I \label{lepn}.\ee Another useful
conserved charge $\mathcal B$ which is exactly not broken by $u$,
$v$ and $\om$ is usual baryon number: $B ={\mathcal B} I$. Both the
charges $\mathcal{L}$ and $\mathcal{B}$ for the fermion and Higgs
multiplets are listed in Table~\ref{bcharge}.
\begin{table}\bc
\caption{\label{bcharge}${\mathcal B}$ and ${\mathcal L}$ charges of
the model multiplets.}
\begin{tabular}{|c|c|c|c|c|c|c|c|c|c|c|}
\hline
 Multiplet & $\chi$ & $\phi$ & $Q_{1L}$ & $Q_{\al L}$ &
$u_{aR}$&$d_{aR}$ &$U_R$ & $D_{\al R}$ & $\psi_{aL}$ & $l_{aR}$
\\ \hline $\mathcal B$-charge &$0$ & $ 0  $ &  $\fr 1 3  $ & $\fr 1 3
$& $\fr 1 3  $ &
 $\fr 1 3  $ &  $\fr 1 3  $&  $\fr 1 3  $&
 $0  $& $0$ \\ \hline
 $\mathcal L$-charge &$\fr 4 3$ & $-\fr 2 3  $ &
   $-\fr 2 3  $ & $\fr 2 3  $& 0 & 0 & $-2$& $2$&
 $\fr 1 3  $& $ 1   $\\
 \hline
\end{tabular}
\ec
\end{table}

Let us note that the Yukawa couplings of (\ref{y2}) conserve
$\mathcal{B}$, however, violate ${\mathcal L}$ with $\pm 2$ units
which implies that these interactions are much smaller than the
first ones \cite{dhhl}: \be s_a^u, \ s_{\al a}^d,\ s_\al^D, \
s_\al^U \ll h^U,\ h_{\al \bet}^D,\ h_a^d,\ h_{\al
a}^u.\label{dkhsyu}\ee In previous studies \cite{ponc1,violat}, the
LNV terms of this kind have often been excluded, commonly by the
adoption of an appropriate discrete symmetry. There is no reason
within the 3-3-1 models why such terms should not be present.

In this model, the most general Higgs potential has very simple form
\bea V(\chi,\phi) &=& \mu_1^2 \chi^\+\chi + \mu_2^2 \phi^\+\phi +
\la_1 (\chi^\+ \chi)^2 + \la_2 (\phi^\+\phi)^2\crn & & + \la_3
(\chi^\+\chi)(\phi^\+\phi) + \la_4 (\chi^\+\phi)(\phi^\+\chi).
\label{poten} \eea It is noteworthy that $V(\chi,\phi)$ does not
contain trilinear scalar couplings and conserves both the mentioned
global symmetries, this makes the Higgs potential much simpler and
discriminative from the previous ones of the 3-3-1 models
\cite{long98,changlong,tujo,ochoa2}. This potential is closer to
that of the standard model. In the next section we will show that
after spontaneous symmetry breaking, there are eight Goldstone
bosons---the needed number for massive gauge ones and three physical
scalar fields (one charged and two neutral). One of two physical
neutral scalars is the standard model Higgs boson.

To break the gauge symmetry spontaneously, the Higgs vacuums are not
$\mathrm{SU}(3)_L\otimes \mathrm{U}(1)_X$ singlets. Hence, non-zero
values of $\chi$ and $\phi$ at the minimum value of $V(\chi,\phi)$
can be easily obtained by (for details, see Section
\ref{hginteractions})\bea \chi^\+\chi
&\equiv&\fr{u^2+\om^2}{2}=\fr{\lambda_3\mu^2_2
-2\lambda_2\mu^2_1}{4\lambda_1\lambda_2-\lambda^2_3},\label{vev1}\\
\phi^\+\phi &\equiv&\fr{v^2}{2}=\fr{\lambda_3\mu^2_1
-2\lambda_1\mu^2_2}{4\lambda_1\lambda_2-\lambda^2_3}.\label{vev2}\eea
It is important noting that any other choice of $u,\om$ for the
vacuum value of $\chi$ satisfying (\ref{vev1}) gives the same
physics because it is related to (\ref{vevc}) by an
$\mathrm{SU}(3)_L\otimes \mathrm{U}(1)_X$ transformation. It is
worth noting that the assumed $u\neq 0$ is therefore given in a
general case. This model, however, does not lead to the formation of
Majoron \cite{ponc3,major-models}.

\subsection{\label{gaugeb}Gauge Bosons}

The covariant derivative of a triplet is given by \be D_\mu =
\pa_\mu-igT_iW_{i\mu}-ig_X T_9 X B_\mu
\equiv\pa_\mu-i\mathcal{P}_\mu,\label{coderi} \ee where the gauge
fields $W_i$ and $B$ transform as the adjoint representations of
$\mathrm{SU}(3)_L$ and $\mathrm{U}(1)_X$, respectively, and the
corresponding gauge coupling constants $g$, $g_X$. Moreover,
$T_9=\fr{1}{\sqrt{6}}\mathrm{diag}(1,1,1)$ is fixed so that the
relation $\mathrm{Tr}(T_{i} T_{j})=\fr{1}{2}\delta_{ij}$
$(i,j=1,2,...,9)$ is satisfied. The $\mathcal{P}_\mu$ matrix
appeared in the above covariant derivative is rewritten in a
convenient form \be \mathcal{P}_\mu= \fr{g}{2}\left(
\begin{array}{ccc}
  W_{3\mu}+\fr{W_{8\mu}}{\sqrt{3}}+t\sqrt{\fr 2 3}XB_\mu
  & \sqrt{2} W'^+_\mu & \sqrt{2}X'^0_\mu \\
  \sqrt{2}W'^-_\mu & -W_{3\mu}+\fr{W_{8\mu}}{\sqrt{3}}+
  t\sqrt{\fr 2 3}X B_\mu & \sqrt{2}Y'^-_\mu \\
  \sqrt{2}X'^{0*}_\mu & \sqrt{2}Y'^+_\mu &
  -\fr{2W_{8\mu}}{\sqrt{3}}+t\sqrt{\fr 2 3}X B_\mu \\
\end{array}
\right)\label{ccnnp}\ee where $t\equiv g_X/g$. Let us denote the
following combinations \be W'^{\pm} _\mu \equiv \fr{W_{1\mu}\mp
iW_{2\mu}}{\sqrt{2}},\hs Y'^\mp_\mu \equiv  \fr{W_{6\mu}\mp
iW_{7\mu}}{\sqrt{2}}, \hs X'^0_\mu \equiv
\fr{W_{4\mu}-iW_{5\mu}}{\sqrt{2}} \ee having defined charges under
the generators of the $\mathrm{SU}(3)_L$ group. For the sake of
convenience in  further reading, we note that, $W_4$ and $W_5$ are
pure real and imaginary parts of $ X'^0_\mu$ and $ X'^{0*}_\mu$,
respectively \be W_{4\mu}  =  \fr{1}{\sqrt{2}} ( X'^0_\mu +
X'^{0*}_\mu),\hs W_{5\mu} = \fr{i}{\sqrt{2}} ( X'^0_\mu -
X'^{0*}_\mu).\label{w4xx}\ee

The masses of the gauge bosons in this model are followed from
\bea {\mathcal
L}^{\mathrm{GB}}_{\mathrm{mass}}&=&(D_\mu\langle\phi\rangle)^\+
(D^\mu\langle\phi\rangle)+ (D_\mu\langle\chi\rangle)^\+
(D^\mu\langle\chi\rangle)\crn &=&\fr{g^2}{4}(u^2+v^2)W'^-_\mu
W'^{+\mu}+\fr{g^2}{4}(\om^2+v^2)Y'^-_\mu Y'^{+\mu}\crn
&&+\fr{g^2u\om}{4}(W'^-_\mu Y'^{+\mu}+Y'^-_\mu W'^{+\mu})\crn & &
+ \fr{g^2v^2}{8}\left(-W_{3\mu}+\fr{1}{\sqrt{3}}W_{8\mu}+t\fr 2 3
\sqrt{\fr 2 3} B_\mu\right)^2\crn & & +
\fr{g^2u^2}{8}\left(W_{3\mu}+\fr{1}{\sqrt{3}}W_{8\mu}-t\fr 1 3
\sqrt{\fr 2 3}B_\mu \right)^2\crn
&&+\fr{g^2\om^2}{8}\left(-\fr{2}{\sqrt{3}}W_{8\mu}-t\fr 1
3\sqrt{\fr 2 3} B_\mu\right)^2\crn & & + \fr{g^2 u
\om}{4\sqrt{2}}\left(W_{3\mu}+\fr{1}{\sqrt{3}}W_{8\mu}-t \fr 1 3
\sqrt{\fr 2 3}B_\mu\right)\left(X'^{0\mu}+X'^{0*\mu}\right)\crn &
& + \fr{g^2 u \om}{4\sqrt{2}}\left(-\fr{2}{\sqrt{3}}W_{8\mu}-t\fr
1 3 \sqrt{\fr 2 3}
B_\mu\right)\left(X'^{0\mu}+X'^{0*\mu}\right)\crn &  & +
\fr{g^2}{16}(u^2+\om^2)\left\{(X'^0_\mu+X'^{0*}_\mu)^2+
[i(X'^0_\mu-X'^{0*}_\mu)]^2\right\}.\label{ngbmassm}\eea

The combinations  $W'$ and $Y'$ are mixing via\bea {\mathcal
L}^{\mathrm{CG}}_{\mathrm{mass}}=\fr{g^2}{4}(W'^-_\mu,Y'^-_\mu)\left(%
\begin{array}{cc}
  u^2+v^2 & u\om \\
  u\om & \om^2+v^2 \\
\end{array}%
\right)\left(%
\begin{array}{c}
  W'^{+\mu} \\
  Y'^{+\mu} \\
\end{array}%
\right).\nn\eea Diagonalizing this mass matrix, we get {\it
physical} charged gauge bosons
 \be W_\mu = \cos\theta\
W'_\mu-\sin\theta\ Y'_\mu ,\hs Y_\mu = \sin\theta\
W'_\mu+\cos\theta\ Y'_\mu,\ee where the mixing angle is defined by
\be \tan\theta=\fr{u}{\om}.\ee The
mass eigenvalues are  \bea M^2_{W}&=&\fr{g^2v^2}{4},\label{massw}\\
M^2_{Y}&=&\fr{g^2}{4}(u^2+v^2+\om^2).\label{massy}\eea Because of
the constraints in (\ref{vevcons}), the following remarks are in
order: \ben\item $\theta$ should be very small, and then $W_\mu
\simeq W'_\mu, Y_\mu \simeq Y'_\mu$. \item  $v \simeq
v_{\mathrm{weak}} = 246$ GeV due to identification of $W$ as the
$W$ boson in the standard model.\een

Next, from (\ref{ngbmassm}), the $W_5$ gains mass as follows \be
M^2_{W_5}=\fr{g^2}{4}(\om^2+u^2).\ee

Finally, there is a mixing among $W_3, W_8, B, W_4$ components. In
the basis of these elements, the mass matrix is given by
\be M^2=\fr{g^2}{4}\left(%
\begin{array}{cccc}
  u^2+v^2 & \fr{u^2-v^2}{\sqrt{3}} & -\fr{2t}{3\sqrt{6}}(u^2+2v^2) & 2u\om \\
  \fr{u^2-v^2}{\sqrt{3}} & \fr{1}{3}(4\om^2+u^2+v^2) &
  \fr{\sqrt{2}t}{9}(2\om^2-u^2+2v^2)
   & -\fr{2}{\sqrt{3}}u\om \\
  -\fr{2t}{3\sqrt{6}}(u^2+2v^2) & \fr{\sqrt{2}t}{9}(2\om^2-u^2+2v^2)
  & \fr{2t^2}{27}(\om^2+u^2+4v^2)
  & -\fr{8t}{3\sqrt{6}}u\om \\
  2u\om & -\fr{2}{\sqrt{3}}u\om & -\fr{8t}{3\sqrt{6}}u\om & u^2+\om^2 \\
\end{array}%
\right). \label{nmass}\ee Note that the mass Lagrangian in this case
has the form \be {\mathcal L}^{\mathrm{NG}}_{\mathrm{mass}}=\fr 1 2
V^T M^2 V,\hs V^T \equiv (W_3, W_8, B, W_4).\ee In the limit
$u\rightarrow 0$, $W_{4}$ does not mix with $W_{3}, W_{8}, B$. In
the general case $u \neq 0$,  the  mass matrix in (\ref{nmass})
contains  two {\it exact eigenvalues} such as  \be M^2_\ga =0, \hs
M^2_{W'_4}=\fr{g^2}{4}(\om^2+u^2).\label{massx}\ee Thus the $W'_4$
and $W_5$ components have the same mass, and this conclusion {\it
contradicts the previous analysis in} Ref. \cite{ponce}. With this
result, we should  identify the combination of $W'_4$ and $W_5$: \be
\sqrt{2} X_\mu^0 = W'_{4\mu} - i W_{5\mu} \label{chbln} \ee as {\it
physical} neutral {\it non-Hermitian}  gauge boson. The subscript
$0$ denotes neutrality of gauge boson $X$. However, in the
following, this subscript may be dropped. This boson caries the
lepton number with two units, hence it is the bilepton like those in
the usual 3-3-1 model with right-handed neutrinos. From
(\ref{massw}), (\ref{massy}) and (\ref{massx}), it follows an
interesting relation between  the bilepton masses similar to the law
of Pythagoras \bea M^2_{Y}&=& M^2_{X}+M^2_{W}. \label{massrel} \eea
Thus the charged bilepton $Y$ is slightly heavier than the neutral
one $X$. Remind that the similar relation in the 3-3-1 model with
right-handed neutrinos is \cite{il}: $ | M_Y^2 - M_X^2 | \leq
m_W^2$.

Now we turn to the eigenstate question. The  eigenstates
corresponding to the two values in (\ref{massx}) are determined as
follows
\bea A_\mu=\fr{1}{\sqrt{18+4t^2}}\left(%
\begin{array}{c}
  \sqrt{3}t \\
  -t \\
  3\sqrt{2} \\
  0 \\
\end{array}%
\right), \hs W'_{4\mu}=\fr{1}{\sqrt{1+4\tan^22\theta}}\left(%
\begin{array}{c}
  \tan2\theta \\
  \sqrt{3}\tan2\theta \\
  0 \\
  1 \\
\end{array}%
\right).\eea To embed this model in the effective theory at the low
energy we follow an appropriate method in Ref. \cite{ld,mohapatra},
where the photon field couples with the lepton by strength \be
{\mathcal
L}^{\mathrm{EM}}_{\mathrm{int}}=-\fr{\sqrt{3}g_X}{\sqrt{18+4t^2}}\bar{l}\ga^\mu
l A_\mu.\ee Therefore the coefficient of the electromagnetic
coupling constant can be identified as \be
\fr{\sqrt{3}g_X}{\sqrt{18+4t^2}} =e\ee Using continuation of the
gauge coupling constant $g$ of $\mathrm{SU}(3)_L$ at the spontaneous
symmetry breaking point \be g=g[\mathrm{SU}(2)_L]=\fr{e}{s_W}\ee
from which it follows \be t=\fr{3\sqrt{2}s_W}{\sqrt{3-4s^2_W}}.\ee
The eigenstates are now rewritten as follows\bea A_\mu &=& s_W
W_{3\mu}+c_W\left(-\fr{t_W}{\sqrt{3}}
W_{8\mu}+\sqrt{1-\fr{t^2_W}{3}}B_\mu\right),\crn
W'_{4\mu}&=&\fr{t_{2\theta}}{\sqrt{1+4t^2_{2\theta}}}W_{3\mu}+
\fr{\sqrt{3}t_{2\theta}}{\sqrt{1+4t^2_{2\theta}}}W_{8\mu}
+\fr{1}{\sqrt{1+4t^2_{2\theta}}}W_{4\mu},\eea where we have denoted
$s_W\equiv\sin\theta_W$, $t_{2\theta}\equiv\tan2\theta$, and so
forth.

The diagonalization of the mass matrix is done via three steps. In
the first step, it is in the base of
$(A_\mu,Z_\mu,Z'_\mu,W_{4\mu})$, where the two remaining gauge
vectors are given by \bea Z_\mu&=& c_W
W_{3\mu}-s_W\left(-\fr{t_W}{\sqrt{3}}
W_{8\mu}+\sqrt{1-\fr{t^2_W}{3}}B_\mu\right),\crn Z'_\mu &=&
\sqrt{1-\fr{t^2_W}{3}} W_{8\mu}+\fr{t_W}{\sqrt{3}}B_\mu.\eea In
this basis, the mass matrix $M^2$ becomes \bea M'^2=\fr{g^2}{4}\left(%
\begin{array}{cccc}
  0 & 0 & 0 & 0 \\
  0 & \fr{u^2 + v^2}{c^2_W} & \fr{c_{2W}u^2-v^2}{c^2_W
  \sqrt{3-4s^2_W}} & \fr{2 u \om}{c_W} \\
  0 & \fr{c_{2W}u^2-v^2}{c^2_W\sqrt{3-4s^2_W}} &
  \fr{v^2+4c^4_W\om^2+c^2_{2W}u^2}{c^2_W(3-4s^2_W)} &
  -\fr{2 u \om}{c_W\sqrt{3-4s^2_W}} \\
  0 & \fr{2 u \om}{c_W} & -\fr{2 u \om}{c_W\sqrt{3-4s^2_W}} & u^2+\om^2 \\
\end{array}%
\right).\eea Also, in the limit $u\rightarrow 0$, $W_{4\mu}$ does
not mix with $Z_{\mu},Z'_{\mu}$. The eigenstate $W'_{4\mu}$ is now
defined by \bea
W'_{4\mu}=\fr{t_{2\theta}}{c_W\sqrt{1+4t^2_{2\theta}}}Z_\mu+
\fr{\sqrt{4c^2_W-1}t_{2\theta}}{c_W\sqrt{1+
4t^2_{2\theta}}}Z'_\mu+\fr{1}{\sqrt{1+4t^2_{2\theta}}}W_{4\mu}.\eea

We turn to the second step. To see explicitly that the following
basis is orthogonal and normalized, let us put\be
s_{\theta'}\equiv\fr{t_{2\theta}}{c_W\sqrt{1+4t^2_{2\theta}}},\label{htmix}\ee
which leads to \be W'_{4\mu}=s_{\theta'} Z_\mu+
c_{\theta'}\left[t_{\theta'}\sqrt{4c^2_W-1}Z'_\mu+
\sqrt{1-t^2_{\theta'}(4c^2_W-1)}W_{4\mu}\right].\label{thetapr}\ee
Note that the mixing angle in this step $\theta'$ is the same order
as  the mixing angle in the charged gauge boson sector. Taking into
account \cite{pdg} $s^2_W \simeq 0.231$, from (\ref{htmix})   we get
$s_{\theta'}\simeq 2.28 s_{\theta}$. It is now easy to choose two
remaining gauge vectors orthogonal to $W'_{4\mu}$:\bea
\mathcal{Z}_{\mu}&=&c_{\theta'} Z_\mu-
s_{\theta'}\left[t_{\theta'}\sqrt{4c^2_W-1}Z'_\mu+
\sqrt{1-t^2_{\theta'}(4c^2_W-1)}W_{4\mu}\right],\crn
\mathcal{Z}'_{\mu}&=&\sqrt{1-t^2_{\theta'}(4c^2_W-1)}Z'_\mu-
t_{\theta'}\sqrt{4c^2_W-1}W_{4\mu}.\eea Therefore, in the base of
$(A_\mu,\mathcal{Z}_{\mu},{\mathcal Z}'_{\mu}$,$W'_{4\mu})$  the
mass
matrix $M'^2$  has a quasi-diagonal form  \bea M''^2=\left(%
\begin{array}{cccc}
  0 & 0 & 0 & 0 \\
  0 & m^2_{\mathcal{Z}} & m^2_{\mathcal{Z}\mathcal{Z}'} & 0 \\
  0 & m^2_{\mathcal{Z}\mathcal{Z}'} & m^2_{\mathcal{Z}'} & 0 \\
  0 & 0 & 0 & \fr{g^2}{4}(u^2+\om^2) \\
\end{array}%
\right)\label{fmassmatrix}\eea with \bea
m^2_{\mathcal{Z}}&=&\fr{(1+3t^2_{2\theta})u^2+(1+4t^2_{2\theta})v^2-
t^2_{2\theta}\om^2}{4g^{-2}[c^2_W+(3-4s^2_W)t^2_{2\theta}]},\crn
m^2_{\mathcal{Z}\mathcal{Z}'}&=&\fr{\sqrt{1+4t^2_{2\theta}}
\left\{[c_{2W}+(3-4s^2_W)t^2_{2\theta}]u^2-v^2-(3-4s^2_W)t^2_{2\theta}
\om^2\right\}}{4g^{-2}\sqrt{3-4s^2_W}[c^2_W+(3-4s^2_W)t^2_{2\theta}]},\\
m^2_{\mathcal{Z}'}&=&\fr{[c^2_{2W}+(3-4s^2_{2W})t^2_{2\theta}
]u^2+v^2+[4c^4_W+(1+4c^2_W)(3-4s^2_W)t^2_{2\theta}]\om^2}
{4g^{-2}(3-4s^2_W)[c^2_W+(3-4s^2_W)t^2_{2\theta}]}.\nn\eea

In the last step, it is trivial to diagonalize the  mass matrix in
(\ref{fmassmatrix}). The two remaining mass eigenstates are given by
\be Z^1_\mu = c_\va {\mathcal Z}_{\mu}-s_\va \mathcal{Z}'_{\mu},\hs
Z^2_\mu = s_\va {\mathcal Z}_{\mu}+c_\va \mathcal{Z}'_{\mu},\ee
where the mixing angle $\va$ between $\mathcal{Z}$ and
$\mathcal{Z}'$ is defined by \bea
t_{2\va}&=&\left\{\left[\left(3-4s^2_W\right)\left(1+4t^2_{2\theta}
\right)\right]^{1/2}\left\{\left[c_{2W}+\left(3-4s^2_W\right)t^2_{2\theta}
\right]u^2-v^2\right.\right.\crn
&&\left.\left.-\left(3-4s^2_W\right)t^2_{2\theta}\om^2\right\}\right\}\left\{\left[2s^4_W-1
+\left(8s^4_W-2s^2_W-3\right) t^2_{2\theta}\right]u^2-
\left[c_{2W}\right.\right.\crn
&&\left.\left.+2\left(3-4s^2_W\right)t^2_{2\theta}\right]v^2+\left[2c^4_W+
\left(8s^4_W+9c_{2W}\right)t^2_{2\theta}\right]\om^2\right\}^{-1}.\eea
The physical mass eigenvalues are defined  by\bea
M^2_{Z^1}&=&[2g^{-2}(3-4s^2_W)]^{-1}\left\{c^2_W(u^2+\om^2)+v^2\right.
\crn &&-\left.\sqrt{[c^2_W (u^2+\om^2)+v^2]^2+(3-4s^2_W)(3u^2\om^2-
u^2v^2-v^2\om^2)}\right\},\crn
M^2_{Z^2}&=&[2g^{-2}(3-4s^2_W)]^{-1}\left.\{c^2_W(u^2+\om^2)+v^2\right.
\crn&& \left. +\sqrt{[c^2_W (u^2+\om^2)+v^2]^2+(3-4s^2_W)(3u^2\om^2
-u^2v^2-v^2\om^2)}\right\}.\nn\eea

Because of the condition (\ref{vevcons}), the angle $\va$ has to be
very small\be
t_{2\va}\simeq-\fr{\sqrt{3-4s^2_W}[v^2+(11-14s^2_W)u^2]}{2c^4_W\om^2}.\ee
In this approximation,  the above physical states have masses \bea
M^2_{Z^1}&\simeq& \fr{g^2}{4c^2_W}(v^2-3u^2),\label{massz1}\\
M^2_{Z^2}&\simeq& \fr{g^2c^2_W\om^2}{3-4s^2_W}.\eea Consequently,
$Z^1$ can be identified as the $Z$ boson in the standard model,
and $Z^2$ being the new neutral (Hermitian) gauge boson. It is
important to note that in the limit  $u\rightarrow 0$ the mixing
angle $\va$ between $\mathcal{Z}$ and $\mathcal{Z}'$ is always
non-vanishing. This differs from the mixing angle $\theta$ between
the $W$ boson of the standard model and the singly-charged
bilepton $Y$. Phenomenology of the mentioned mixing is quite
similar to the $W_L - W_R$ mixing in the left-right symmetric
model based on the $\textrm{SU}(2)_R \otimes \textrm{SU}(2)_L
\otimes \textrm{U}(1)_{B-L}$ group (the interested reader can find
in \cite{mohapatra}).

\subsection{\label{currents}Currents}

The interaction among fermions with gauge bosons arises in part from
\bea i \bar{\psi}\ga_\mu D^\mu \psi =  \textrm{kinematic terms} +
H^{\mathrm{CC}} + H^{\mathrm{NC}}. \label{lagcurrent} \eea

\subsubsection{Charged Currents} Despite neutrality, the gauge
bosons $X^0$, $X^{0*}$ belong to this section by their nature.
Because of the mixing among the standard model $W$ boson and the
charged bilepton $Y$ as well as among ($X^0 + X^{0*}$) with $(W_3,
W_8, B)$, the new interaction terms exist as follows \bea
H^{\mathrm{CC}}=\fr{g}{\sqrt{2}}\left(J^{\mu-}_W W^+_\mu +
J^{\mu-}_Y Y^+_\mu + J^{\mu 0*}_X X^{0}_\mu + \mathrm{H.c.}\right)
\eea where \bea J^{\mu-}_W&=&c_\theta \left(\bar{\nu}_{aL}\ga^\mu
l_{aL}+\bar{u}_{aL}\ga^\mu d_{aL}\right)\crn &&-s_\theta
\left(\bar{\nu}^c_{aL}\ga^\mu l_{aL}+\bar{U}_{L}\ga^\mu
d_{1L}+\bar{u}_{\al L}\ga^\mu
D_{\al L}\right),\label{dongw}\\
J^{\mu-}_Y&=&c_\theta \left(\bar{\nu}^c_{aL}\ga^\mu
l_{aL}+\bar{U}_{L}\ga^\mu d_{1L}+\bar{u}_{\al L}\ga^\mu D_{\al
L}\right)\crn &&+s_\theta \left(\bar{\nu}_{aL} \ga^\mu
l_{aL}+\bar{u}_{a L}\ga^\mu d_{a L}\right),\label{dongy}\\
J^{\mu 0*}_X &\simeq& (1-t^2_{2\theta})\left(\bar{\nu}_{aL}\ga^\mu
\nu^c_{a L}+\bar{u}_{1L}\ga^\mu U_{L}-\bar{D}_{\al L}\ga^\mu
d_{\al L}\right)\crn
&&-t^2_{2\theta}\left(\bar{\nu}^c_{aL}\ga^\mu\nu_{aL}+\bar{U}_L\ga^\mu
u_{1L}-\bar{d}_{\al L}\ga^\mu D_{\al
L}\right)+\fr{t_{2\theta}}{\sqrt{1+4t^2_{2\theta}}}\label{dongx}\\
&&\times\left(\bar{\nu}_a\ga^\mu \nu_a+\bar{u}_{1L}\ga^\mu
u_{1L}-\bar{U}_L\ga^\mu U_L-\bar{d}_{\al L}\ga^\mu d_{\al
L}+\bar{D}_{\al L}\ga^\mu D_{\al L}\right).\nn \eea

Comparing with the charged currents in the usual 3-3-1 model with
right-handed neutrinos \cite{long} we get the following discrepances
\ben \item The second term in (\ref{dongw}) \item The second term in
(\ref{dongy}) \item The second and the third terms in (\ref{dongx})
\een All mentioned above interactions are lepton-number violating
and weak (proportional to $\sin \theta$ or its square  $\sin^2
\theta$). However, these couplings lead to lepton-number violations
only in the neutrino sector.

\subsubsection{Neutral Currents}

As before, in this model, a real part of the non-Hermitian neutral
$X'^0$ mixes with the real neutral ones such as $Z$ and $Z'$. This
gives the {\it unusual} term as follows
 \bea H^{\mathrm{NC}} = e
A^\mu J^{\mathrm{EM}}_\mu + {\mathcal L}^{\mathrm{NC} } +
{\mathcal L}^{\mathrm{NC} }_{\mathrm{unnormal}}. \label{ncurrent}
\eea

Despite the mixing among $W_3, W_8, B, W_4$, the electromagnetic
interactions {\it remain} the same as in the standard model and the
usual 3-3-1 model with right-handed neutrinos, i.e. \be
J^{\mathrm{EM}}_\mu = \sum_f q_f \bar{f} \ga_\mu f,
 \label{ecurrent} \ee
where $f$ runs among all the fermions of the model.

Interactions of the neutral currents with  fermions have a common
form \bea {\mathcal L}^{\mathrm{NC}}=\fr{g}{2c_W}\bar{f}\ga^\mu
\left[g_{k V}(f)-g_{k A}(f)\ga^5\right] f Z^k_\mu,\mathrm{ } k= 1,
2,\label{normal}\eea where  \bea
g_{1V}(f)&=&\fr{c_\va\left\{T_3(f_L)-3t^2_{2\theta}
X(f_L)+[(3-8s^2_W)t^2_{2\theta}-2s^2_W]Q(f)\right\}
}{\sqrt{(1+4t^2_{2\theta})[1+(3-t^2_W)t^2_{2\theta}]}} \crn
&&-\fr{s_\va [(4c^2_W-1)T_3(f_L)+3c^2_W X(f_L)-(3-5s^2_W)Q(f)]}
{\sqrt{(4c^2_W-1)[1+(3-t^2_W)t^2_{2\theta}]}},\\
g_{1A}(f)&=&\fr{c_\va
[T_3(f_L)-3t^2_{2\theta}(X-Q)(f_L)]}{\sqrt{(1+4t^2_{2\theta})
[1+(3-t^2_W)t^2_{2\theta}]}}\crn &&-\fr{s_\va
[(4c^2_W-1)T_3(f_L)+3c^2_W(X-Q)(f_L)]}{
\sqrt{(4c^2_W-1)[1+(3-t^2_W)t^2_{2\theta}]}},\\
g_{2V}(f)&=&g_{1V}(f)(c_\va\rightarrow s_\va, s_\va\rightarrow
-c_\va),\\ g_{2A}(f)&=&g_{1A}(f)(c_\va\rightarrow s_\va,
s_\va\rightarrow -c_\va).\eea Here $T_3(f_L)$, $X(f_L)$ and $Q(f)$
are, respectively, the third component of the weak isospin, the
$\mathrm{U}(1)_X$ charge and the electric charge of the fermion
$f_L$. Note that the isospin for the $\mathrm{SU}(2)_L$ fermion
singlet (in the bottom of triplets) vanishes: $T_3(f_L)= 0$. The
values of $g_{1V}(f)$, $g_{1A}(f) $ and $g_{2V}(f)$, $g_{2A}(f)$
are listed in Table \ref{tab1} and Table \ref{tab2}.

\begin{table}
\caption{\label{tab1}The $Z^1_\mu \rightarrow \bar{f} f$
couplings.} \bc
\begin{tabular}{|l|c|c|}
\hline $f$  &  $g_{1V}(f)$  &  $g_{1A}(f)$ \\ \hline $\nu_a$ &
$\fr{c_\va -s_\va \sqrt{(4c^2_W-1)(1+4t^2_{2\theta})}}
{2\sqrt{(1+4t^2_{2\theta})[1+(3-t^2_W)t^2_{2\theta}]}}$&
$\fr{c_\va \sqrt{(4c^2_W-1)(1+4t^2_{2\theta})}+s_\va}
{2\sqrt{(4c^2_W-1)[1+(3-t^2_W)t^2_{2\theta}]}}$\\ \hline  $l_a$ &
$\fr{(3-4c^2_W)[c_\va\sqrt{(4c^2_W-1)(1+4t^2_{2\theta})}+s_\va]}
{2\sqrt{(4c^2_W-1)[1+(3-t^2_W)t^2_{2\theta}]}}$  &
$-\fr{c_\va\sqrt{(4c^2_W-1)(1+4t^2_{2\theta})}+s_\va}
{2\sqrt{(4c^2_W-1)[1+(3-t^2_W)t^2_{2\theta}]}}$\\ \hline  $u_1$ &
$\fr{c_\va
\sqrt{4c^2_W-1}[3(1+2t^2_{2\theta})-8s^2_W(1+4t^2_{2\theta})]
-s_\va(3+2s^2_W)\sqrt{1+4t^2_{2\theta}}}{6\sqrt{(4c^2_W-1)(1+4t^2_{2\theta})
[1+(3-t^2_W)t^2_{2\theta}]}}$ &
$\fr{c_\va\sqrt{4c^2_W-1}(1+2t^2_{2\theta})-s_\va
c_{2W}\sqrt{1+4t^2_{2\theta}}}{2\sqrt{(4c^2_W-1)(1+4t^2_{2\theta})
[1+(3-t^2_W)t^2_{2\theta}]}}$\\ \hline  $d_1$ &
$\fr{(1-4c^2_W)[c_\va\sqrt{(4c^2_W-1)(1+4t^2_{2\theta})}+s_\va]}
{6\sqrt{(4c^2_W-1)[1+(3-t^2_W)t^2_{2\theta}]}}$ &
$-\fr{c_\va\sqrt{(4c^2_W-1)(1+4t^2_{2\theta})}+s_\va}
{2\sqrt{(4c^2_W-1)[1+(3-t^2_W)t^2_{2\theta}]}}$\\ \hline $u_\al$ &
$\fr{(3-8s^2_W)[c_\va\sqrt{(4c^2_W-1)
(1+4t^2_{2\theta})}+s_\va]}{6\sqrt{(4c^2_W-1)
[1+(3-t^2_W)t^2_{2\theta}]}}$ &
$\fr{c_\va\sqrt{(4c^2_W-1)(1+4t^2_{2\theta})}+s_\va}
{2\sqrt{(4c^2_W-1)[1+(3-t^2_W)t^2_{2\theta}]}}$\\ \hline $d_\al$ &
$\fr{c_\va
\sqrt{4c^2_W-1}[(1-4c^2_W)(1+4t^2_{2\theta})+6t^2_{2\theta}]+s_\va(1+2c^2_W)
\sqrt{1+4t^2_{2\theta}}}{6\sqrt{(4c^2_W-1)
(1+4t^2_{2\theta})[1+(3-t^2_W)t^2_{2\theta}]}}$ & $-\fr{c_\va
\sqrt{4c^2_W-1}(1+2t^2_{2\theta})-s_\va
c_{2W}\sqrt{1+4t^2_{2\theta}}}
{2\sqrt{(4c^2_W-1)(1+4t^2_{2\theta})[1+(3-t^2_W)t^2_{2\theta}]}}$\\
\hline $U$  & $\fr{c_\va
\sqrt{4c^2_W-1}[3t^2_{2\theta}-4s^2_W(1+4t^2_{2\theta})] +s_\va
(3-7s^2_W)\sqrt{1+4t^2_{2\theta}}}{3\sqrt{(4c^2_W-1)(1+4t^2_{2\theta})
[1+(3-t^2_W)t^2_{2\theta}]}}$ & $\fr{c_\va
\sqrt{4c^2_W-1}t^2_{2\theta}+s_\va
c^2_W\sqrt{1+4t^2_{2\theta}}}{\sqrt{(4c^2_W-1)(1+4t^2_{2\theta})
[1+(3-t^2_W)t^2_{2\theta}]}}$\\ \hline $D_\al$ & $\fr{c_\va
\sqrt{4c^2_W-1}[2s^2_W(1+4t^2_{2\theta})-3t^2_{2\theta}]
-s_\va(3-5s^2_W)\sqrt{1+4t^2_{2\theta}}}{3\sqrt{(4c^2_W-1)(1+4t^2_{2\theta})
[1+(3-t^2_W)t^2_{2\theta}]}}$ & $-\fr{c_\va
\sqrt{4c^2_W-1}t^2_{2\theta}+s_\va
c^2_W\sqrt{1+4t^2_{2\theta}}}{\sqrt{(4c^2_W-1)(1+4t^2_{2\theta})
[1+(3-t^2_W)t^2_{2\theta}]}}$\\ \hline
\end{tabular}
\ec
\end{table}

\begin{table}
\caption{\label{tab2}The $Z^2_\mu \rightarrow \bar{f} f$
couplings.} \bc
\begin{tabular}{|l|c|c|}
\hline $f$ &  $g_{2V}(f)$ &  $g_{2A}(f)$ \\ \hline $\nu_a$ &
$\fr{s_\va+c_\va\sqrt{(4c^2_W-1)(1+4t^2_{2\theta})}}
{2\sqrt{(1+4t^2_{2\theta})[1+(3-t^2_W)t^2_{2\theta}]}}$&
$\fr{s_\va\sqrt{(4c^2_W-1)(1+4t^2_{2\theta})}-c_\va}
{2\sqrt{(4c^2_W-1)[1+(3-t^2_W)t^2_{2\theta}]}}$\\ \hline $l_a$ &
$\fr{(3-4c^2_W)[s_\va\sqrt{(4c^2_W-1)(1+4t^2_{2\theta})}-c_\va]}
{2\sqrt{(4c^2_W-1)[1+(3-t^2_W)t^2_{2\theta}]}}$  &
$-\fr{s_\va\sqrt{(4c^2_W-1)(1+4t^2_{2\theta})}-c_\va}
{2\sqrt{(4c^2_W-1)[1+(3-t^2_W)t^2_{2\theta}]}}$\\  \hline $u_1$
&$\fr{s_\va
\sqrt{4c^2_W-1}[3(1+2t^2_{2\theta})-8s^2_W(1+4t^2_{2\theta})]
+c_\va(3+2s^2_W)\sqrt{1+4t^2_{2\theta}}}{6\sqrt{(4c^2_W-1)(1+4t^2_{2\theta})
[1+(3-t^2_W)t^2_{2\theta}]}}$ &
$\fr{s_\va\sqrt{4c^2_W-1}(1+2t^2_{2\theta})+c_\va
c_{2W}\sqrt{1+4t^2_{2\theta}}}{2\sqrt{(4c^2_W-1)(1+4t^2_{2\theta})
[1+(3-t^2_W)t^2_{2\theta}]}}$\\ \hline $d_1$ &
$\fr{(1-4c^2_W)[s_\va\sqrt{(4c^2_W-1)(1+4t^2_{2\theta})}-c_\va]}
{6\sqrt{(4c^2_W-1)[1+(3-t^2_W)t^2_{2\theta}]}}$ &
$-\fr{s_\va\sqrt{(4c^2_W-1)(1+4t^2_{2\theta})}-c_\va}
{2\sqrt{(4c^2_W-1)[1+(3-t^2_W)t^2_{2\theta}]}}$\\ \hline $u_\al$ &
$\fr{(3-8s^2_W)[s_\va\sqrt{(4c^2_W-1)
(1+4t^2_{2\theta})}-c_\va]}{6\sqrt{(4c^2_W-1)
[1+(3-t^2_W)t^2_{2\theta}]}}$ &
$\fr{s_\va\sqrt{(4c^2_W-1)(1+4t^2_{2\theta})}-c_\va}
{2\sqrt{(4c^2_W-1)[1+(3-t^2_W)t^2_{2\theta}]}}$\\ \hline $d_\al$ &
$\fr{s_\va
\sqrt{4c^2_W-1}[(1-4c^2_W)(1+4t^2_{2\theta})+6t^2_{2\theta}]-
c_\va(1+2c^2_W)\sqrt{1+4t^2_{2\theta}}}{6\sqrt{(4c^2_W-1)
(1+4t^2_{2\theta})[1+(3-t^2_W)t^2_{2\theta}]}}$ & $-\fr{s_\va
\sqrt{4c^2_W-1}(1+2t^2_{2\theta})+c_\va
c_{2W}\sqrt{1+4t^2_{2\theta}}}
{2\sqrt{(4c^2_W-1)(1+4t^2_{2\theta})[1+(3-t^2_W)t^2_{2\theta}]}}$
\\ \hline
$U$ & $\fr{s_\va
\sqrt{4c^2_W-1}[3t^2_{2\theta}-4s^2_W(1+4t^2_{2\theta})]
-c_\va(3-7s^2_W)\sqrt{1+4t^2_{2\theta}}}{3\sqrt{(4c^2_W-1)(1+4t^2_{2\theta})
[1+(3-t^2_W)t^2_{2\theta}]}}$ & $\fr{s_\va
\sqrt{4c^2_W-1}t^2_{2\theta}-c_\va
c^2_W\sqrt{1+4t^2_{2\theta}}}{\sqrt{(4c^2_W-1)(1+4t^2_{2\theta})
[1+(3-t^2_W)t^2_{2\theta}]}}$\\ \hline  $D_\al$ & $\fr{s_\va
\sqrt{4c^2_W-1}[2s^2_W(1+4t^2_{2\theta})-3t^2_{2\theta}]
+c_\va(3-5s^2_W)\sqrt{1+4t^2_{2\theta}}}{3\sqrt{(4c^2_W-1)(1+4t^2_{2\theta})
[1+(3-t^2_W)t^2_{2\theta}]}}$ & $-\fr{s_\va
\sqrt{4c^2_W-1}t^2_{2\theta}-c_\va
c^2_W\sqrt{1+4t^2_{2\theta}}}{\sqrt{(4c^2_W-1)(1+4t^2_{2\theta})
[1+(3-t^2_W)t^2_{2\theta}]}}$ \\ \hline
\end{tabular}
\ec
\end{table}

Because of the above-mentioned mixing, the lepton-number violating
interactions mediated by neutral gauge bosons $Z^1$ and $Z^2$
exist in the {\it neutrino  and the exotic quark sectors}
 \bea {\mathcal L}^{\mathrm{NC} }_{\mathrm{unnormal}}&=&-\fr{g
 t_{2\theta}g_{kV}(\nu)}{2}\left(\bar{\nu}_{aL}\ga^\mu \nu^c_{a
L}+\bar{u}_{1L}\ga^\mu U_{L}-\bar{D}_{\al L}\ga^\mu d_{\al
L}\right) Z^k_\mu + \mathrm{H.c.} \label{un} \eea Again, these
interactions are very weak and proportional to  $\sin \theta$.
From (\ref{dongw}) - (\ref{dongx}) and (\ref{un}) we conclude that
all lepton-number violating interactions are expressed in the
terms dependent only  in  the mixing angle between the charged
gauge bosons.

\subsection{\label{phenomenology}Phenomenology}

\begin{figure}[htbp]
\begin{center}
\includegraphics[width=10cm,height=7cm]{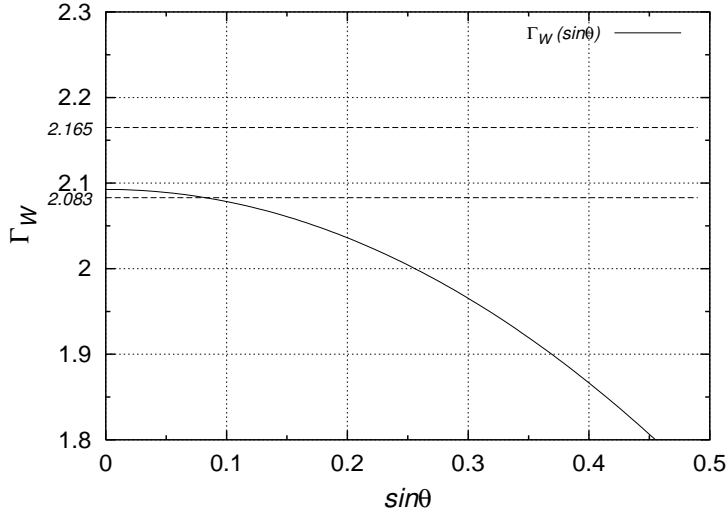}
\caption{\label{wsin}$W$ width as function of $\sin \theta$, and
the horizontal lines are an upper and a lower limit.}
\end{center}
\end{figure}

First of all we should  find some constraints on the parameters of
the model. There are many ways to get constraints on the mixing
angle $\theta$ and the charged bilepton mass $M_Y$. Below we
present a simple one. In our model, the $W$ boson has the
following {\it normal main} decay modes: \bea W^-  & \rightarrow &
l\  \tilde{\nu}_l\  (l = e,\mu,\tau),\crn
 &  \searrow & u^c
d, u^c s, u^c b,  (u\rightarrow c),\label{decaywn}\eea
 which are the same as
in the standard model and in the 3-3-1 model with right-handed
neutrinos. Beside the above modes, there are additional ones which
are lepton-number violating $(\De L = 2)$ - the model's  specific
feature   \be W^- \rightarrow l \ \nu_l \ (l = e,\mu,\tau).
\label{decaywan}\ee It is easy to compute the tree level decay
widths as follows \cite{bardin} \bea
\Ga^{\mathrm{Born}}(W\rightarrow l\ \tilde{\nu}_l)&=&\fr{g^2
c^2_{\theta}}{8}\fr{M_W}{6\pi}(1-x)(1-\fr{x}{2}-\fr{x^2}{2})\simeq
\fr{c^2_\theta \al M_W}{12 s^2_W},\crn
\Ga^{\mathrm{Born}}(W\rightarrow l\ \nu_l)&=&\fr{g^2
s^2_{\theta}}{8}\fr{M_W}{6\pi}(1-x)(1-\fr{x}{2}-\fr{x^2}{2})\simeq
\fr{s^2_\theta \al M_W}{12 s^2_W},\crn && x\equiv m^2_l/M^2_W,\crn
\sum_{\mathrm{color}}\Ga^{\mathrm{Born}}(W\rightarrow u^c_i
d_j)&=&\fr{3g^2
c^2_{\theta}}{8}\fr{M_W}{6\pi}|V_{ij}|^2\left[1-2(x+\bar{x})+(x-\bar{x})^2\right]^{\fr
{1}{2}} \label{qcd}\\
&&\times\left[1-\fr{x+\bar{x}}{2}-\fr{(x-\bar{x})^2}{2}\right]\simeq
\fr{c^2_\theta \al M_W}{4 s^2_W}|V_{ij}|^2,\crn && x\equiv
m^2_{d_j}/M^2_W,\hs \bar{x} \equiv m^2_{u^c_i}/M^2_W.\nn\eea Quantum
chromodynamics radiative corrections modify Eq.(\ref{qcd}) by a
multiplicative factor \cite{pdg,bardin}\bea \de_{\mathrm{QCQ}}&=&
1+\al_s(M_Z)/\pi+1.409\al^2_s/\pi^2-12.77\al^3_s/\pi^3 \simeq
1.04,\eea which is estimated from $\al_s(M_Z)\simeq 0.12138$. All
the state masses can be ignored, the predicted total width for $W$
decay into fermions is \bea \Ga^{\mathrm{tot}}_W=1.04\fr{\al
M_W}{2s^2_W}(1-s^2_{\theta})+\fr{\al M_W}{4s^2_W}.\eea Taking
$\al(M_Z) \simeq 1/128$, $M_W=80.425\mathrm{GeV}$, $s^2_W=0.2312$
and $\Ga^{\mathrm{tot}}_W=2.124\pm 0.041 \mathrm{GeV}$ \cite{pdg},
in Fig.\ref{wsin}, we have plotted   $ \Ga^{\mathrm{tot}}_W$ as
function of $s_{\theta}$.
From the figure we get an upper limit: \be \sin \theta \leq 0.08.
\label{uperlim} \ee It is important to note that this limit value on
the LNV parameter $u/\om$ is much larger than those in Refs.
\cite{tujo,plei}.

\begin{figure*}\bc
\includegraphics{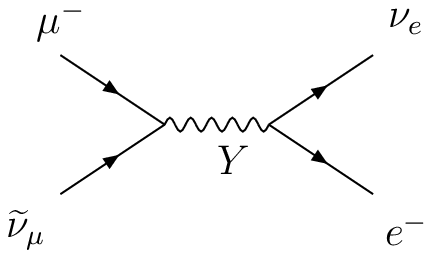}
\includegraphics{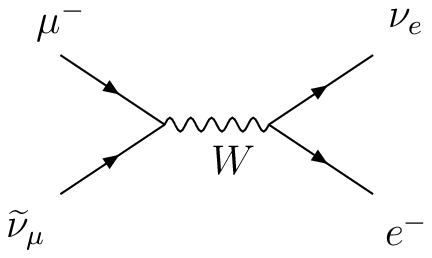}
\caption{\label{fig:fig1}Feynman diagram for the wrong muon decay
$\mu^-\rightarrow e^- \nu_e \tilde{\nu}_\mu$.}\ec
\end{figure*}

Since one of the VEVs is closely to the those in the standard
model:  $v \simeq v_{weak}= 246 $ GeV, therefore only two free
VEVs exist  in the considering model, namely $u$ and $\om$. The
bilepton mass limit can be obtained from the ``wrong" muon decay
 \be \mu^- \rightarrow e^- \nu_e \tilde{\nu}_\mu
\label{muondecayw} \ee  mediated,  at the tree level, by both the
standard model $W$ and the singly-charged  bilepton $Y$ (see
Fig.\ref{fig:fig1}). Remind that in the 3-3-1 model with
right-handed neutrinos, at the lowest order, this decay is mediated
only by the singly-charged bilepton $Y$. In our case, the second
diagram in Fig.\ref{fig:fig1} gives main contribution.
Taking into account of the famous experimental data \cite{pdg}
 \be R_{muon} \equiv
\fr{\Ga(\mu^- \rightarrow e^- \nu_e \tilde{\nu}_\mu)}{\Ga(\mu^-
\rightarrow e^- \tilde{\nu}_e \nu_\mu)}
 < 1.2 \% \hs  \textrm{90 \% \ CL} \label{wrdecayrat}
 \ee
 we get the constraint: $ R_{muon} \simeq \fr{M_W^4}{M_Y^4}$.
 Therefore, it follows that $M_Y \geq 230 $ GeV.

However, the stronger bilepton mass bound of 440 GeV has been
derived from consideration of experimental limit on  lepton-number
violating charged lepton decays \cite{tullyjoshi}.

In the case of $u \rightarrow 0$, analyzing the $Z$ decay width
\cite{ponc1,ponc2}, the $Z - Z'$ mixing angle is constrained by
$-0.0015 \leq \varphi \leq 0.001$. From atomic parity violation in
cesium, bounds for mass of the new exotic $Z'$ and the $Z - Z'$
mixing angles, again in the limit $u\rightarrow 0$, are given
\cite{ponc1,ponc2} \be -0.00156 \leq \varphi \leq 0.00105, \hs
M_{Z_2} \geq 2.1 \ \textrm{TeV} \label{boundz}\ee These values
coincide with the bounds in the usual 3-3-1 model with right-handed
neutrinos \cite{longtrung}. The interested reader can find in
\cite{dln} for the general case $u\neq0$ of the constraints.

For our purpose we consider the $\rho$ parameter - one of the most
important quantities of the standard model, having a leading
contribution in terms of the $T$ parameter, is very useful to get
the new-physics effects. It is well-known relation between $\rho$
and $T$ parameter \be \rho = 1 + \al T \label{rhot} \ee In the
usual 3-3-1 model with right-handed neutrinos, $T$ gets
contribution from the oblique correction and the $Z-Z'$ mixing
\cite{il} \bea T_{RHN} &= &T_{Z Z'} + T_{oblique},\eea where $
T_{Z Z'} \simeq \fr{\tan^2 \varphi}{\al}\left(
\fr{M^2_{Z_2}}{M^2_{Z_1}} -1 \right)$ is negligible for $M_{Z'}$
less than 1 TeV, $T_{oblique}$ depends on masses of  the top quark
and the standard model Higgs boson. Again at  the tree level and
the limit (\ref{vevcons}), from (\ref{massw}) and (\ref{massz1})
we get an expression for the $\rho$ parameter in the considering
model
 \be \rho = \fr{M^2_{W}}{c^2_W
M^2_{Z^1}}=\fr{v^2}{v^2-3u^2} \simeq 1 + \fr{3u^2}{v^2}.
\label{rhopar} \ee Note that formula (\ref{rhopar}) has only one
free parameter $u$, since   $v$ is very close to the VEV in the
standard model. Neglecting the contribution from the usual 3-3-1
model with right-handed neutrinos and taking into account the
experimental data \cite{pdg} $\rho = 0.9987 \pm 0.0016 $ we get
the constraint on $u$ parameter by $ \fr u v \leq 0.01$ which
leads to $ u \leq 2.46 $ GeV. This means  that $u$ is much smaller
than $v$, as expected.

It seems that the $\rho$ parameter, at the tree level, in this
model, is favorable to be bigger than one and this is similar to the
case of the models contained  heavy $Z'$ \cite{luo}.

\begin{figure*}\bc
\includegraphics{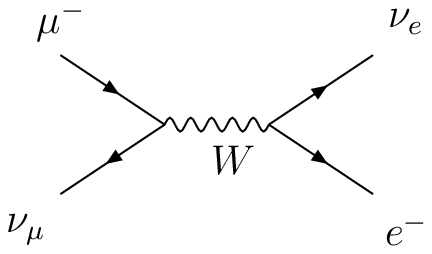}
\includegraphics{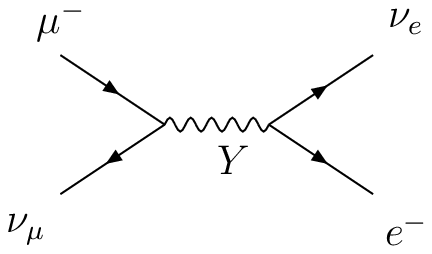}
\caption{\label{fig:fig2}Feynman diagram for $\mu^-\rightarrow e^-
\nu_e \nu_\mu$.}\ec
\end{figure*}

\begin{figure*}
\includegraphics{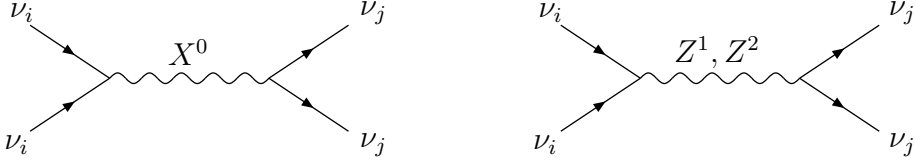}
\caption{\label{fig:fig3}Feynman diagram for
$\nu_i\nu_i\rightarrow \nu_j\nu_j \hs (i\neq j =e,\mu,\tau)$.}
\end{figure*}

The interesting new physics compared with other 3-3-1 models is
the neutrino physics. Due to lepton-number violating couplings we
have the following interesting consequences: \ben
\item {\it Processes with $\De L = \pm 2$}\\
From the charged currents we have the following lepton-number
violating $\De L = \pm 2$ decays such as \bea \mu^- &\rightarrow&
e^- \nu_e \nu_\mu, \crn \mu^- &\rightarrow& e^- \tilde{\nu}_e
\tilde{\nu}_\mu, \hs ( \mu \ \textrm{can be replaced by  } \ \tau)
\label{muondecay1} \eea in which both the standard model $W$ boson
and charged bilepton $Y_\mu^-$ are in intermediate states (see
Fig. \ref{fig:fig2}).
Here the main contribution arises from the first diagram. Note
that the wrong muon decay violates only {\it family}
lepton-number, i.e. $\De L = 0$, but not lepton-number at all as
in (\ref{muondecay1}). The decay rates are given by \bea R_{rare}
&\equiv& \fr{\Ga(\mu^- \rightarrow e^- \nu_e \nu_\mu)}{\Ga(\mu^-
\rightarrow e^- \tilde{\nu}_e \nu_\mu)} = \fr{\Ga(\mu^-
\rightarrow e^- \tilde{\nu}_e \tilde{\nu}_\mu )}{\Ga(\mu^-
\rightarrow e^- \tilde{\nu}_e \nu_\mu)} \simeq s_\theta^2.
\label{decayrat}
 \eea
Taking $s_\theta = 0.08$, we get $R_{rare} \simeq 6 \times
10^{-3}$. This rate is the same as the  wrong  muon decay one.
Interesting to note that, the family lepton-number violating
processes \be \nu_i \nu_i \rightarrow \nu_j \nu_j, \ (i \neq
j)\label{netross}\ee
 are mediated not only by the non-Hermitian
bilepton $X$ but also by the Hermitian neutral $Z^1, Z^2$ (see
Fig.\ref{fig:fig3}).

The first diagram in Fig.\ref{fig:fig3} exists also in the 3-3-1
model with right-handed neutrinos, but the second one does not
appear there.

\item {\it  Lepton-number violating kaon decays}\\
 Next, let us consider the  lepton-number violating   decay \cite{pdg}
\be K^+ \rightarrow \pi^0 + e^+ \tilde{\nu}_e\  < 3 \times 10^{-3}
\ \textrm{at} \  90 \% \ \textrm{CL}
 \ee
This decay can be explained in the considering model  as the
subprocess given below \be \tilde{s} \rightarrow \tilde{u} + e^+
\tilde{\nu}_e. \ee
 This process is mediated by the standard model $W$ boson
and the charged bilepton $Y$. Amplitude of the considered process is
proportional to $\sin\theta$ \be M (\tilde{s} \rightarrow \tilde{u}
+ e^+ \tilde{\nu}_e) \simeq \fr{\sin 2 \theta}{ 2 M_W^2} \left( 1 -
\fr{M_W^2}{M_Y^2}\right) \label{radecay}\ee Next, let us consider
the ``normal decay" \cite{pdg} \be K^+ \rightarrow \pi^0 + e^+
\nu_e\  \ (4.87 \pm 0.06)\ \% \ee with amplitude \be M (\tilde{s}
\rightarrow \tilde{u} + e^+ \nu_e) \simeq \fr{1}{ M_W^2}
\label{radecayn}\ee From (\ref{radecay}) and (\ref{radecayn}) we get
\be R_{kaon} \equiv \fr{\Ga (\tilde{s} \rightarrow \tilde{u} + e^+
\tilde{\nu}_e)}{\Ga (\tilde{s} \rightarrow \tilde{u} + e^+ \nu_e)}
\simeq \sin^2 \theta . \ee

In the framework of this model, we derive  the following decay
modes with rates \be  R_{kaon} =
 \fr{\Ga (K^+\rightarrow \pi^0 + e^+
\tilde{\nu}_e)}{\Ga (K^+ \rightarrow \pi^0 + e^+ \nu_e)} \simeq
\fr{\Ga (K^+ \rightarrow \pi^0 + \mu^+ \tilde{\nu}_\mu)}{\Ga (K^+
\rightarrow \pi^0 + \mu^+ \nu_\mu)}\simeq \sin^2 \theta \leq 6
\times 10^{-3}. \ee Note that the similar lepton-number violating
processes exist in the $\textrm{SU}(2)_R$ $\otimes\textrm{SU}(2)_L
\otimes \textrm{U}(1)_{B-L}$
 model (for details,
see Ref.\cite{mohapatra}).

\een

\subsection{\label{sec:conclusions}Summary}

In this section we have presented the 3-3-1 model with the minimal
scalar sector (only two Higgs triplets). This version belongs to the
3-3-1 model without exotic charges (charges of the exotic quarks are
$\fr 2 3$ and $-\fr 1 3$). The spontaneous symmetry breakdown is
achieved with only two Higgs triplets.  One of the VEVs $u$  is a
source of lepton-number violations and a reason for the mixing
between the charged gauge bosons - the standard model $W$ and the
singly-charged bilepton gauge bosons as well as between neutral
non-Hermitian $X^0$ and neutral gauge bosons: the $Z$ and the new
exotic $Z'$. At the tree level, masses of the charged gauge bosons
satisfy the law of Pythagoras $ M_Y^2 = M_X^2 + M_W^2$ and in the
limit $\om \gg u, v$, the $\rho$ parameter gets additional
contribution dependent only on $\fr u v$. Thus, this leads to $u \ll
v$, and there are three quite different scales for the VEVs of the
model: one is very small $u \simeq \emph{O}(1)  $ GeV - a
lepton-number violating parameter, the second $v$ is close to the
standard model one : $v\simeq v_{weak} = 246$ GeV and the last is in
the range of new physics scale about $\emph{O}(1)$ TeV.

In difference with  the usual 3-3-1 model with right-handed
neutrinos, in this model the first family of quarks should be
distinctive of the two others.

The exact diagonalization of the neutral gauge boson sector  is
derived. Because of the parameter $u$, the lepton-number violation
 happens only in neutrino
but not in charged lepton sector. It is interesting to note that
despite the mentioned above mixing, the electromagnetic current
remains unchanged.  In this model, the lepton-number changing
($\De L = \pm 2$) processes exist but only in the neutrino sector.

It is worth  mentioning on  the advantage of the considered model:
the new mixing angle between the charged gauge bosons $\theta$ is
connected with one of the VEVs $u$ - the parameter of
lepton-number violations. There is no new  parameter, but it
contains very simple Higgs sector, hence the significant number of
free parameters is reduced.

The model contains new kinds of interactions in the neutrino sector.
Hence neutrino physics in this model is very rich. We will turn to
further studies on neutrino masses and mixing in Section
\ref{fermionmasses}.

\section{\label{hginteractions}Higgs-Gauge Boson Interactions}

We first obtain the scalar fields and mass spectra. The couplings of
the scalar fields with the ordinary gauge bosons are presented then.
Cross section for the production of the charged Higgs boson at LHC
are calculated.

\subsection{\label{higgs}Higgs Potential}

The Higgs potential in the model under consideration is given by
Eq. (\ref{poten}). Let us first shift the Higgs fields into
physical ones:
\be \chi=\left(%
\begin{array}{c}
  \chi^{P 0}_1 + \fr{u}{\sqrt{2}}  \\
  \chi^-_2 \\
  \chi^{P 0}_3 + \fr{\om}{\sqrt{2}}  \\
\end{array}%
\right),  \hs  \phi=\left(%
\begin{array}{c}
  \phi^{+}_1   \\
  \phi^{P 0}_2 + \fr{v}{\sqrt{2}}\\
  \phi^{+}_3   \\
\end{array}%
\right).\label{higgsshipt} \ee The subscript $P$ denotes {\it
physical} fields as in the usual treatment. However, in the
following, this subscript will be  dropped. By substitution of
(\ref{higgsshipt}) into  (\ref{poten}), the potential becomes \be
V(\chi,\phi) = \mu_1^2 \left[ \left(\chi^{0 *}_1 +
\fr{u}{\sqrt{2}}\right)\left(\chi^{0 }_1 + \fr{u}{\sqrt{2}}\right) +
\chi_2^+ \chi_2^- + \left(\chi^{0 *}_3 +
\fr{\om}{\sqrt{2}}\right)\left(\chi^{ 0 }_3 +
\fr{\om}{\sqrt{2}}\right)\right]\nn\ee\be
 +
 \mu_2^2 \left[ \phi_1^- \phi_1^+ + \left(\phi^{ 0 *}_2 +
\fr{v}{\sqrt{2}}\right)\left(\phi^{ 0 }_2 + \fr{v}{\sqrt{2}}\right)
+ \phi_3^- \phi_3^+\right]\nn\ee \be +
  \la_1 \left[ \left(\chi^{ 0 *}_1 +
\fr{u}{\sqrt{2}}\right)\left(\chi^{ 0 }_1 + \fr{u}{\sqrt{2}}\right)
+ \chi_2^+ \chi_2^- + \left(\chi^{ 0 *}_3 +
\fr{\om}{\sqrt{2}}\right)\left(\chi^{ 0 }_3 +
\fr{\om}{\sqrt{2}}\right)\right]^2 \nn\ee\bea &&
   + \la_2\left[ \phi_1^- \phi_1^+ + \left(\phi^{ 0 *}_2 +
\fr{v}{\sqrt{2}}\right)\left(\phi^{ 0 }_2 +
\fr{v}{\sqrt{2}}\right) + \phi_3^- \phi_3^+\right]^2\crn &&
 + \la_3 \left[ \left(\chi^{ 0 *}_1 +
\fr{u}{\sqrt{2}}\right)\left(\chi^{ 0 }_1 +
\fr{u}{\sqrt{2}}\right) + \chi_2^+ \chi_2^- + \left(\chi^{ 0 *}_3
+ \fr{\om}{\sqrt{2}}\right)\left(\chi^{ 0 }_3 +
\fr{\om}{\sqrt{2}}\right)\right]\crn && \times \left[ \phi_1^-
\phi_1^+ +\left(\phi^{ 0 *}_2 +
\fr{v}{\sqrt{2}}\right)\left(\phi^{ 0 }_2 +
\fr{v}{\sqrt{2}}\right) + \phi_3^- \phi_3^+\right]\crn &&
 + \la_4\left[ \left(\chi^{ 0 *}_1 +
\fr{u}{\sqrt{2}}\right)\phi_1^+ + \chi_2^+ \left(\phi^{ 0 }_2 +
\fr{v}{\sqrt{2}}\right)+\left(\chi^{ 0 *}_3 +
\fr{\om}{\sqrt{2}}\right)\phi_3^+\right]\crn&& \times \left[
\phi_1^-\left(\chi^{ 0 }_1 + \fr{u}{\sqrt{2}}\right) +
\left(\phi^{ 0 *}_2 + \fr{v}{\sqrt{2}}\right)\chi_2^-
+\phi_3^-\left(\chi^{ 0 }_3 + \fr{\om}{\sqrt{2}}\right)\right].
\label{potennew} \eea  From the above expression, we get
constraint equations at the tree level \bea \mu_1^2 + \la_1 (u^2 +
\om^2) + \la_3 \fr{v^2}{2}
 & = & 0,\label{potn1}\\
 \mu_2^2 +  \la_2 v^2   + \la_3 \fr{(u^2 + \om^2)}{2} & = &
0.\label{potenn2} \eea The nonzero values of $\chi$ and $\phi$ at
the potential minimum as mentioned can be easily derived from these
equations to yield the given (\ref{vev1}) and (\ref{vev2}).

Since $u$ is a parameter of lepton-number violation, therefore the
terms linear in $u$ violate the latter.  Applying the constraint
equations (\ref{potn1}) and (\ref{potenn2}) we get the minimum
value, mass terms, lepton-number conserving and violating
interactions as follows \bea V(\chi,\phi)
&=&V_{\mathrm{min}}+V^{\mathrm{N}}_{\mathrm{mass}}
+V^{\mathrm{C}}_{\mathrm{mass}}+V_{\mathrm{LNC}} + V_{\mathrm{LNV}},
\label{potenn2a}\eea where \bea V_{\mathrm{min}}&=&- \fr{\la_2}{4}
v^4 - \fr 1 4 (u^2+\om^2)[\la_1(u^2 + \om^2) + \la_3 v^2],\crn
V^{\mathrm{N}}_{\mathrm{mass}}&=& \la_1 (uS_1+\om S_3)^2+\la_2 v^2
S^2_2 +\la_3 v (uS_1+\om S_3)S_2,
\label{potenn8}\\
V^{\mathrm{C}}_{\mathrm{mass}}&=&\fr{\la_4}{2}(u\phi^+_1+v\chi^+_2+\om
\phi^+_3)(u\phi^-_1+v\chi^-_2+\om \phi^-_3),
\label{potenn12}\\
V_{\mathrm{LNC}} &= &\la_1
(\chi^\+\chi)^2+\la_2(\phi^\+\phi)^2+\la_3
(\chi^\+\chi)(\phi^\+\phi)+\la_4 (\chi^\+\phi)(\phi^\+\chi)\crn &&
+ 2\la_1\om S_3(\chi^\+\chi)+2\la_2 v S_2(\phi^\+\phi)+\la_3 v
S_2(\chi^\+\chi)+\la_3\om S_3(\phi^\+\phi) \crn
&&+\fr{\la_4}{\sqrt{2}}(v\chi^-_2+\om
\phi^-_3)(\chi^\+\phi)+\fr{\la_4}{\sqrt{2}}(v\chi^+_2+\om
\phi^+_3)(\phi^\+\chi),  \label{potenn3} \\ V_{\mathrm{LNV}} &= &
2 \la_1 u S_1(\chi^\+\chi)+\la_3u
S_1(\phi^\+\phi)+\fr{\la_4}{\sqrt{2}}u
\left[\phi^-_1(\chi^\+\phi)+\phi^+_1(\phi^\+\chi)\right].
\label{potenn4} \eea In the above equations, we have dropped the
subscript $P$ and used $\chi=(\chi^{0}_1,\chi^-_2,\chi^{0}_3)^T$,
$\phi=(\phi^{+}_1,\phi^{0}_2,\phi^{+}_3)^T$. Moreover, we have
expanded the neutral Higgs fields as \be \chi^0_1  =  \fr{S_1 +i
A_1}{\sqrt{2}},\hs \chi^0_3  =  \fr{S_3 + i A_3}{\sqrt{2}},\hs
\phi^0_2 = \fr{S_2 + i A_2}{\sqrt{2}}.\label{potenn5}\ee In the
literature, the real parts $(S_i, i=1,2,3)$  are also called
CP-even scalar and the imaginary part $(A_i, i=1,2,3)$ -- CP-odd
scalar. In this paper, for short, we call them scalar and
pseudoscalar field, respectively. As expected, the lepton-number
violating part $V_{\mathrm{LNC}}$ is linear in $u$ and  trilinear
in scalar fields. These couplings will be also a source for
lepton-number violations such as the mass spectra of quarks
including exotic ones as well as neutrino Majorana masses, but
given at higher-order corrections.

In the pseudoscalar sector, all the fields are Goldstone bosons:
$G_1= A_1$, $G_2= A_2$ and $G_3 = A_3$ (cl.  Eq.(\ref{potenn8})).
The scalar fields $S_1$, $S_2$ and $S_3$ gain masses via
(\ref{potenn8}), thus we get one Goldstone boson $G_4$ and two
neutral physical fields---the standard model $H^0$ and the new
$H^0_1$ with masses  \bea m^2_{H^0}&=&\la_2 v^2+
\la_1(u^2+\om^2)-\sqrt{[\la_2 v^2-\la_1(u^2+\om^2)]^2+\la^2_3 v^2
(u^2+\om^2)}\crn
&\simeq&\fr{4\la_1\la_2-\la^2_3}{2\la_1}v^2,\label{potenn10a}\eea
\bea M^2_{H^0_1}&=&\la_2 v^2+ \la_1(u^2+\om^2)+\sqrt{[\la_2
v^2-\la_1(u^2+\om^2)]^2+\la^2_3 v^2 (u^2+\om^2)}\crn &\simeq& 2\la_1
\om^2.\label{potenn10}\eea  In term of original fields, the
Goldstone and Higgs fields are given by \bea G_4 &=&
\fr{1}{\sqrt{1+t^2_{\theta}}}(S_1 -t_\theta
S_3),\\
H^0&=&c_\zeta S_2
-\fr{s_\zeta}{\sqrt{1+t^2_{\theta}}}(t_{\theta}S_1 +S_3),
\label{potenn11a}\\
H^0_1&=&s_\zeta S_2
+\fr{c_\zeta}{\sqrt{1+t^2_{\theta}}}(t_{\theta}S_1
+S_3),\label{potenn11b}\eea where \bea t_{2\zeta}&\equiv &
\fr{\la_3 M_W M_X}{\la_1M^2_X-\la_2 M^2_W}.\label{potenn11}\eea
From Eq.(\ref{potenn10}), it follows that mass of the new Higgs
boson $M_{H^0_1}$ is related to mass of the bilepton gauge $X^0$
(or $Y^\pm$ via the law of Pythagoras) through \bea M^2_{H^0_1}&
=& \fr{8\la_1}{g^2} M_X^2 \left[1 +
\mathcal{O}\left(\fr{M_W^2}{M_X^2}\right)\right]\crn &=& \fr{2
\la_1 s^2_W}{\pi \al} M_X^2  \left[1 +
\mathcal{O}\left(\fr{M_W^2}{M_X^2}\right)\right] \approx 18.8
\la_1 M_X^2. \label{potenn11mass}\eea Here, we have used $\al =
\fr{1}{128}$ and $s^2_W = 0.231$.

In the charged Higgs sector, the mass terms for
$(\phi_1,\chi_2,\phi_3)$ are given by (\ref{potenn12}), thus there
are two Goldstone bosons and one physical scalar field:\be
H^+_2\equiv \fr{1}{\sqrt{u^2+v^2+\om^2}}(u\phi^+_1+v\chi^+_2+\om
\phi^+_3)\label{potenn13}\ee with mass \bea M^2_{H^+_2}&
=&\fr{\la_4}{2}(u^2+v^2+\om^2) = 2 \la_4 \fr{M_Y^2}{g^2} = \fr{s_W^2
\la_4}{2 \pi \al} M_Y^2 \simeq 4.7 \la_4 M_Y^2.\label{potenn14}\eea
The two remaining  Goldstone bosons are \bea
G^+_5&=&\fr{1}{\sqrt{1+t^2_{\theta}}}(\phi^+_1-t_{\theta}\phi^+_3),\\
G^+_6
&=&\fr{1}{\sqrt{(1+t^2_{\theta})(u^2+v^2+\om^2)}}\left[v(t_\theta
\phi^+_1+\phi^+_3)- \om
(1+t^2_\theta)\chi^+_2\right].\label{potenn15}\eea

Thus, all the pseudoscalars are eigenstates and massless
(Goldstone). Other fields are related to the scalars in the weak
basis by the linear transformations: \be \left(
\begin{array}{ccc} H^0
\\ H^0_1 \\G_4
\end{array}\right) = \left( \begin{array}{ccc} -s_\zeta
s_\theta & c_\zeta
&  -s_\zeta c_\theta\\
c_\zeta s_\theta &
 s_\zeta & c_\zeta c_\theta\\
c_\theta & 0 & -s_\theta \end{array}\right) \left(
\begin{array}{ccc} S_1
\\ S_2  \\ S_3 \end{array}\right), \label{potenn16}\ee \be
 \left(%
\begin{array}{c}
  H^+_2 \\
  G^+_5 \\
  G^+_6
\end{array}%
\right)=\fr{1}{\sqrt{\om^2+c^2_\theta v^2}}\left(%
\begin{array}{ccc}
  \om s_\theta & vc_\theta & \om c_\theta\\
  c_\theta \sqrt{\om^2+c^2_\theta v^2} & 0 & -
  s_\theta \sqrt{\om^2+c^2_\theta v^2}\\
  \fr{v s_{2\theta}}{2} & -\om & v c^2_{\theta}
\end{array}%
\right)\left(%
\begin{array}{c}
  \phi^+_1 \\
  \chi^+_2 \\
  \phi^+_3
\end{array}%
\right).\label{potenn18}\ee

With the two Higgs triplets of the model, there are twelve real
scalar components. Eight of the gauge symmetries of
$\mathrm{SU}(3)_L \otimes \mathrm{U}(1)_X$ are spontaneously broken,
which eliminates just eight Goldstone bosons associated with these
fields. It leaves over just four massive scalar particles as
obtained (one charged and two natural). There is no Majoron field in
this model which contrasts to the 3-3-1 model with right-handed
neutrinos \cite{majo331}. Let us remind the reader that among the
Goldstone bosons there are four fields carrying the lepton number
but they can be gauged away by an unitary transformation
\cite{ponc3}.

From (\ref{potenn10a}) and (\ref{potenn10}), we come to the
previous result in Ref.\cite{ponce} \be \la_1
> 0, \ \la_2 > 0, \hs 4 \la_1 \la_2 > \la_3^2. \label{potenn20}\ee
Eq.(\ref{potenn14}) shows that the mass of the charged Higgs boson
$H^\pm_2$ is proportional to those of the charged bilepton $Y$
through a coefficient of Higgs self-interaction $\la_4>0$.
Analogously, this happens for the standard-model-like Higgs boson
$H^0$ $(M_{H^0} \sim M_W)$ and the new $H^0_1$ $(M_{H^0_1}\sim
M_X)$. Combining (\ref{potenn20}) with the constraint equations
(\ref{potn1}), (\ref{potenn2}) we get a consequence: $\la_3$ is
negative ($\la_3 < 0$). Let us remind the reader that the couplings
$\la_{4,1,2}$ are fixed by the Higgs boson masses and $\la_3$, where
the $\la_3$ defines the splitting $\Delta
m^2_{H}\simeq-[\la^2_3/(2\la_1)]v^2$ from the standard model
prediction.

To finish this section, let us comment on our physical Higgs
bosons. In the effective approximation $w \gg v, u$, from Eqs
(\ref{potenn16}),
 and (\ref{potenn18}) it follows that
 \bea H^0 &\sim &
S_2,\hs H_1^0 \sim S_3, \hs G_4 \sim S_1, \crn H^+_2 &\sim &
\phi_3^+,\hs G^+_5 \sim \phi^+_1, \hs G^+_6 \sim
\chi^+_2.\label{potenn20a} \eea
 This means that, in the
effective approximation, the charged boson $H^-_2$ is a scalar
bilepton (with lepton number $L=2$), while the neutral scalar
bosons $H^0$ and $H^0_1 $ do not carry lepton number (with $L=0$).

\subsection{\label{smhiggs}Higgs--Standard Model Gauge Couplings}

There are a total of 9 gauge bosons in the $ \mathrm {SU}(3)_L
\otimes {\mathrm U}(1)_X$  group and 8 of them are massive. As
shown in the previous section, we have got just 8 massless
Goldstone bosons---the justified number for the model. One of the
neutral scalars is identified with the standard model Higgs boson,
therefore  its couplings to ordinary gauge bosons such as the
photon, the $Z$ and the $W^\pm$ bosons have to have, in the
effective limit, usual known forms. To search Higgs bosons at
future high energy colliders, one needs their couplings with
ordinary particles, specially with the gauge bosons in the
standard model.

The interactions among the gauge bosons and the Higgs bosons arise
in part from \be \sum_{Y=\chi,\ \phi}\left(D_\mu Y\right)^\+\left(
D^\mu Y\right).\nn \ee In the following the summation over
$\emph{Y}$ is default and only the terms giving interested couplings
are explicitly displayed. The covariant derivative is given by Eq.
(\ref{coderi}),\be D_\mu =\pa_\mu-i\mathcal{P}_\mu \equiv \pa_\mu
-i\mathcal{P}^{\mathrm{NC}}_\mu - i\mathcal{P}_\mu^{\mathrm{CC}},
\label{gau1} \ee where the matrices $\mathcal{P}^{\mathrm{NC}}_\mu$
and $\mathcal{P}^{\mathrm{CC}}_\mu$ are written as
 \be \mathcal{P}_\mu^{\mathrm{NC}} =
\fr{g}{2}\left(
\begin{array}{ccc}
  W_{3\mu}+\fr{W_{8\mu}}{\sqrt{3}}+t\sqrt{\fr 2 3}XB_\mu
  & 0 & y_\mu  \\
  0 & -W_{3\mu}+\fr{W_{8\mu}}{\sqrt{3}}+
  t\sqrt{\fr 2 3}X B_\mu & 0 \\
  y_\mu &0 &
  -\fr{2W_{8\mu}}{\sqrt{3}}+t\sqrt{\fr 2 3}X B_\mu \\
\end{array}
\right)\ee and \be \mathcal{P}_\mu^{\mathrm{CC}} =
\fr{g}{\sqrt{2}}\left(%
\begin{array}{ccc}
  0
  & c_\theta W^+_\mu + s_\theta Y_\mu^+ & X^0_\mu \\
 c_\theta W^-_\mu + s_\theta Y_\mu^-&
  0 & c_\theta Y^-_\mu - s_\theta W_\mu^- \\
  X^{0*}_\mu & c_\theta Y^+_\mu - s_\theta W_\mu^+  &
  0 \\
\end{array}%
\right).\label{pcc}\ee Let us recall that $t = g_X/g =
3\sqrt{2}s_W /\sqrt{3-4s_W^2}$, $\tan
 \theta =u/\om$,
 and $W^\pm_\mu, Y_\mu^\pm$ and $X^0_\mu$ are  the physical fields.
  The existence of $y^\mu$ is a consequence of mixing among the
real part $(X^{0*}_\mu + X^0_\mu)$ with $W_{3\mu}, W_{8 \mu}$ and
$B_\mu$; and its expression is determined from the mixing matrix
$U$ given in Appendix \ref{matrantron}: \bea y_\mu &\equiv& U_{42}
Z_\mu + U_{43} Z'_\mu + (U_{44}-1)\fr{(X^{0*}_\mu + X^0_\mu)}{
  \sqrt{2}}\label{potenn28},\eea where \bea  U_{42}&=&
-t_{\theta'}\left(c_\va\sqrt{1-4s^2_{\theta'}c^2_W}
  -s_\va\sqrt{4c^2_W-1}\right),\crn
 U_{43} &=&  -t_{\theta'}\left(
  s_\va\sqrt{1-4s^2_{\theta'}c^2_W}
  +c_\va\sqrt{4c^2_W-1}\right),\label{potenn30}\\
 U_{44}  &=&
  \sqrt{1-4s^2_{\theta'}c^2_W}.\nn
\eea

First, we consider the relevant couplings of the standard model
$W$ boson with the Higgs and Goldstone bosons. The trilinear
couplings of the pair $W^+ W^-$ with the neutral scalars are given
by \be (\mathcal{P}^{\mathrm{CC}}_\mu \langle \chi
\rangle)^\+(\mathcal{P}^{\mathrm{CC}\mu}
\chi)+(\mathcal{P}^{\mathrm{CC}}_\mu \langle \phi
\rangle)^\+(\mathcal{P}^{\mathrm{CC}\mu}
\phi)+\mathrm{H.c.}=\fr{g^2 v}{2} W^+_\mu W^{-\mu}S_2.\ee Because
of $S_2$ is a combination of only $H$ and $H_1^0$, therefore,
there are two couplings which are given in Table \ref{tab3}.
\begin{table}
\caption{Trilinear coupling constants of  $W^+W^-$ with
 neutral Higgs bosons.}
 \bc
\begin{tabular}{|c|c|}
\hline Vertex  &   Coupling \\ \hline $W^+W^- H$ & $\fr{g^2}{2}v
c_\zeta $\\ \hline $W^+W^- H_1^0$ & $\fr{g^2}{2}v s_\zeta $\\
\hline
\end{tabular}
\label{tab3} \ec
\end{table}

Couplings of the single $W$ with two Higgs bosons exist in  \bea i
\left(Y^\+ \mathcal{P}_\mu^{\mathrm{CC}} \pa^\mu Y - \pa^\mu Y^\+
\mathcal{P}_\mu^{\mathrm{CC}} Y \right)&=&\fr{ig}{\sqrt{2}}
W^-_\mu \left[ Y^*_2 (c_\theta \pa^\mu Y_1 - s_\theta \pa^\mu Y_3)
\right.\crn &&\left.- \pa^\mu Y^*_2 (c_\theta  Y_1 - s_\theta
Y_3)\right] + \mathrm{H.c.} \label{potenn25}\eea \bea
&=&\fr{ig}{\sqrt{2}} W^-_\mu \left[ \chi_2^+ (c_\theta \pa^\mu
\chi^0_1 - s_\theta \pa^\mu \chi^0_3) -\pa^\mu \chi_2^+ (c_\theta
\chi^0_1 - s_\theta \chi^0_3)\right.\crn &+&\left.\phi^{0*}_2
(c_\theta \pa^\mu \phi^+_1 - s_\theta \pa^\mu \phi^+_3) -\pa^\mu
\phi^{0*}_2 (c_\theta \phi^+_1 - s_\theta
\phi^+_3)\right]+\mathrm{H.c.} \label{potenn26}\eea
 The resulting couplings of the single
$W$ boson with two scalar fields are listed in Table \ref{tab4},
\begin{table}
\caption{\label{tab4}Trilinear coupling constants of  $W^-$ with
 two Higgs bosons.}
\bc
\begin{tabular}{|c|c|c|c|}
\hline Vertex  &   Coupling & Vertex & Coupling \\ \hline  $W^{\mu
-} H^+_2\overleftrightarrow{\pa_\mu}G_4  $ & $\fr{ig v c_\theta}{2
\sqrt{\om^2 + c_\theta^2 v^2}}$&  $W^{\mu -} G^+_6
\overleftrightarrow{\pa_\mu}G_1 $ & $\fr{g c_\theta
\om}{2\sqrt{\om^2 + c_\theta^2 v^2} }$\\ \hline $W^{\mu -}G^+_5
\overleftrightarrow{\pa_\mu}H  $ & $-\fr{ig c_\zeta}{2 }$& $W^{\mu
-}G^+_5\overleftrightarrow{\pa_\mu} G_2  $ & $-\fr{g}{2}$\\ \hline
$W^{\mu -} G^+_6\overleftrightarrow{\pa_\mu}G_4   $ &
$\fr{ig\om}{2\sqrt{\om^2 + c_\theta^2 v^2} }$&  $W^{\mu -}G^+_5
\overleftrightarrow{\pa_\mu}H_1^0  $ & $-\fr{i g }{2}s_{\zeta}$
\\ \hline
$W^{\mu -} H^+_2 \overleftrightarrow{\pa_\mu}G_1 $ & $-\fr{g v
c^2_\theta}{2\sqrt{\om^2 + c_\theta^2 v^2} }$ & $W^{\mu -}
G^+_6\overleftrightarrow{\pa_\mu} G^0_3  $ & $-\fr{g s_\theta \om
}{2\sqrt{\om^2+c^2_\theta v^2}}$\\ \hline $W^{\mu -} H^+_2
\overleftrightarrow{\pa_\mu} G_3 $ & $\fr{g v
s_{2\theta}}{4\sqrt{\om^2 + c_\theta^2 v^2} }$ & &\\ \hline
\end{tabular}
\ec
\end{table}
where we have used a notation $ A\overleftrightarrow{\pa_\mu} B =
A(\pa_\mu B) - (\pa_\mu A) B$. Vanishing couplings are
 \bea && \mathcal{V}(W^-H^+_2 H^0) = \mathcal{V}(W^-H^+_2 H^0_1) =
 \mathcal{V}(W^-H^0 G^+_6)\crn && = \mathcal{V}(W^-H^0_1 G^+_6)=
 \mathcal{V}(W^-H^+_2 G_2)=  \mathcal{V}(W^-G^+_6 G_2)=
 0\nn. \eea

Quartic couplings of $W^+W^-$ with two scalar fields arise in part
from  \bea && (\mathcal{P}_\mu^{\mathrm{CC}}\emph{Y}
)^+(\mathcal{P}^{ \mathrm{CC}\mu} \emph{Y})
 = \fr{g^2}{2}W^+_\mu W^{-\mu}\left[\chi^+_2\chi^-_2+c^2_\theta
\chi^{0*}_1 \chi^0_1 \right. \crn && +s^2_\theta \chi^{0*}_3
\chi^0_3 -c_\theta s_\theta
(\chi^{0*}_1\chi^0_3+\chi^0_1\chi^{0*}_3)+\phi^{0*}_2\phi^0_2\crn
&& \left.+c^2_\theta \phi^-_1\phi^+_1+s^2_\theta
\phi^-_3\phi^+_3-c_\theta
s_\theta(\phi^+_1\phi^-_3+\phi^-_1\phi^+_3) \right].
\label{potenn24}\eea
 With the help of
(\ref{potenn17}) and (\ref{potenn19}), we get the interested
couplings of $W^+ W^-$ with two scalars which are listed in Table
\ref{tab5}.
\begin{table}
\caption{Nonzero quartic coupling constants of $W^+W^-$ with
 Higgs bosons.}
 \bc
\begin{tabular}{|c|c|c|c|}
\hline Vertex  & Coupling &  Vertex & Coupling \\ \hline  $W^+W^-
H_2^+ H_2^-$ & $\fr{g^2 c_\theta^2 v^2}{2(\om^2+v^2 c_\theta^2)}$&
$W^+W^- G_1^0 G_1^0 $ & $ \fr{g^2c_{\theta}^2}{2} $\\ \hline
 $W^+W^- G_5^+ G_5^-$ & $ \fr{g^2}{2}
$ &  $W^+W^- G_3^0 G_3^0 $ &  $ \fr{g^2s_{\theta}^2}{2} $
\\ \hline
$W^+W^- G_6^+ G_6^-$ & $ \fr{g^2\om^2}{2(\om^2 + c_\theta^2 v^2)}
$&  $W^+W^- G_4^0 G_4^0 $ & $ \fr{g^2}{2} $
\\ \hline
$W^+W^- H_2^+ G_6^-$ & $-\fr{g^2 c_\theta v \om}{2 (\om^2 +
c_\theta^2 v^2)}$&  $W^+W^- H H_1^0 $ & $ \fr{g^2 s_{2\zeta}}{4} $
\\ \hline
$W^+W^- H H $ & $ \fr{g^2 c^2_\zeta}{2}$&  $W^+W^- G_1^0 G_3^0 $ &
$ -\fr{g^2s_{2\theta}}{4} $
\\ \hline
$W^+W^- H_1^0 H_1^0 $ & $ \fr{g^2s^2_\zeta}{2}$&
$W^+W^-G^0_2G^0_2$ & $\fr{g^2}{2}$ \\ \hline
\end{tabular}
\label{tab5} \ec
\end{table}
Our calculation give following vanishing couplings \bea &&
\mathcal{V}(W^+W^- H_2^+ G_5^-) = \mathcal{V}(W^+W^- G_5^+
G_6^-)\crn && = \mathcal{V}(W^+W^- H^0 G_4^0) = \mathcal{V}(W^+W^-
H^0_1 G_4^0 )= 0. \eea

Now we turn to the couplings of neutral gauge bosons with Higgs
bosons.  In this case, the interested couplings exist in  \bea &&i
\left(Y^\+\mathcal{P}_\mu^{\mathrm{NC}} \pa^\mu Y - \pa^\mu Y^\+
\mathcal{P}_\mu^{\mathrm{NC}} Y \right)\crn
 &=& -\fr{ig}{2}\left\{ W_3^\mu \left(\pa_\mu \chi^{0*}_1
\chi^{0}_1 - \pa_\mu \chi^{+}_2 \chi^{-}_2 + \pa_\mu \phi^{-}_1
\phi^{+}_1 - \pa_\mu \phi^{0*}_2 \phi^{0}_2\right)\right. \crn &&
\left.+ \fr{W_8^\mu}{\sqrt{3}} \left(\pa_\mu \chi^{0*}_1
\chi^{0}_1 + \pa_\mu \chi^{+}_2 \chi^{-}_2 + \pa_\mu \phi^{-}_1
\phi^{+}_1 + \pa_\mu \phi^{0*}_2 \phi^{0}_2 - 2 \pa_\mu
\chi^{0*}_3 \chi^{0}_3 - 2 \pa_\mu \phi^{-}_3
\phi^{+}_3\right)\right.\crn && \left.+
 t\sqrt{\fr 2 3} B^\mu \left[ -\fr 1 3
 \left(\pa_\mu \chi^{0*}_1 \chi^{0}_1 + \pa_\mu
\chi^{+}_2 \chi^{-}_2 +  \pa_\mu \chi^{0*}_3 \chi^{0}_3\right)+
\fr 2 3 \left( \pa_\mu \phi^{-}_1 \phi^{+}_1 + \pa_\mu \phi^{0*}_2
\phi^{0}_2 \right. \right. \right. \crn &&\left. \left. \left.+
 \pa_\mu \phi^{-}_3 \phi^{+}_3\right)\right]
 + y^\mu (\pa_\mu \chi^{0*}_1 \chi^{0}_3 +
 \pa_\mu \chi^{0*}_3 \chi^{0}_1 +
\pa_\mu \phi^{-}_1 \phi^{+}_3 + \pa_\mu \phi^{-}_3 \phi^{+}_1)
 \right\} + \mathrm{H.c.}
 \label{potenn32}\eea It can be checked that, as expected,  the photon $A_\mu$ does
not interact with neutral Higgs bosons. Other vanishing couplings
are
 \bea \mathcal{V}(A H^+_2 G^-_5) = \mathcal{V}(A H^+_2 G^-_6) =
 \mathcal{V}(A G^+_6 G^-_5) =
 0 \eea and
 \bea
 \mathcal{V}(AAH^0)&=&\mathcal{V}(AAH^0_1)=\mathcal{V}(AAG_4)=0,\crn
 \mathcal{V}(AZH^0)&=&\mathcal{V}(AZH^0_1)=\mathcal{V}(AZG_4)=0,\crn
 \mathcal{V}(AZ'H^0)&=&\mathcal{V}(AZ'H^0_1)=\mathcal{V}(AZ'G_4)=0.\nn
 \eea
The nonzero electromagnetic couplings are listed in Table
\ref{tab6}.
\begin{table}[h]
\caption{Trilinear electromagnetic coupling constants of $A^\mu$
with two Higgs bosons.} \bc
\begin{tabular}{|c|c|c|c|}
\hline Vertex  & $A^\mu H^-_2\overleftrightarrow{\pa_\mu}H^+_2$ &
$A^\mu G^-_5\overleftrightarrow{\pa_\mu}G^+_5$ & $A^\mu
G^-_6\overleftrightarrow{\pa_\mu}G^+_6$\\ \hline  Coupling & $i e$
& $i e$ & $i e$\\ \hline
\end{tabular}
 \label{tab6}
\ec
\end{table}
It should be noticed that the electromagnetic interaction is
diagonal, i.e., the non-zero couplings in this model always have a
form\be ie q_H A^\mu H^*\overleftrightarrow{\pa_\mu}H. \ee

For the $Z$ bosons, the following observation is useful \bea
W_3^\mu & = & U_{12} Z^\mu + \cdots, \hs  W_8^\mu  =  U_{22} Z^\mu
+ \cdots,\crn
  B^\mu & = & U_{32} Z^\mu + \cdots, \hs \ \
  y^\mu  =  U_{42} Z^\mu + \cdots.
\label{potenn33}\eea  Here  \bea
 U_{12}&=&  c_\va c_{\theta'}c_W , \hs  U_{22} =
 \fr{c_\va(s^2_W-3c^2_Ws^2_{\theta'})
  -s_\va\sqrt{(1-4s^2_{\theta'}c^2_W)(4c^2_W-
  1)}}{\sqrt{3}c_Wc_{\theta'}},\\
 U_{32}&=& -\fr{t_W(c_\va\sqrt{4c^2_W-1}
  +s_\va\sqrt{1-4s^2_{\theta'}c^2_W})}{\sqrt{3}c_{\theta'}}
\label{potenn34}\eea  are elements in the mixing matrix of the
neutral gauge bosons given in Appendix \ref{matrantron}. From
(\ref{potenn32}) and (\ref{potenn33}), it follows that the
 trilinear couplings of the single $Z$ with charged
Higgs bosons exist in part from the Lagrangian terms
 \bea &&
-\fr{ig}{2} Z^\mu \left[ \left(U_{12} - \fr{U_{22}}{\sqrt{3}}
+\fr{t}{3}\sqrt{\fr 2 3} U_{32}\right) \pa_\mu \chi^{-}_2
\chi^{+}_2 + \left(U_{12} + \fr{U_{22}}{\sqrt{3}} + \fr{2 t}{ 3}
\sqrt{\fr 2 3} U_{32}\right) \pa_\mu \phi^{-}_1
\phi^{+}_1\right.\crn &&\left. +\left(- \fr{2}{\sqrt{3}} U_{22}  +
\fr{2 t}{ 3} \sqrt{\fr 2 3} U_{32}\right) \pa_\mu \phi^{-}_3
\phi^{+}_3 + U_{42} \left(\pa_\mu \phi^{-}_1 \phi^{+}_3 + \pa_\mu
\phi^{-}_3
\phi^{+}_1\right)\right]+\mathrm{H.c.}\label{potenn35}\eea  From
(\ref{potenn35}) we get trilinear couplings of the $Z$ with the
charged Higgs bosons which are listed in Table \ref{tab7}.
\begin{table}
\caption{Trilinear coupling constants of $Z^\mu$ with two charged
Higgs bosons.}\bc
\begin{tabular}{|c|c|}
\hline Vertex  &   Coupling \\ \hline $Z^\mu  H^-_2
\overleftrightarrow{\pa_\mu} H^+_2  $ & $\fr{ig}{2(\om^2 +
v^2c_\theta^2 )} \left\{(v^2 c^2_\theta + \om^2 s^2_\theta)U_{12}
+ [\om^2(1-3c^2_\theta) -
v^2c^2_\theta]\fr{U_{22}}{\sqrt{3}}\right.$\\
& $\left.+(v^2 c^2_\theta+2\om^2)\fr{t}{3}\sqrt{\fr 2 3}U_{32} +
\om^2 s_{2\theta}U_{42}\right\} \longrightarrow -ig s_W t_W
 $\\ \hline $Z^\mu  G^-_5\overleftrightarrow{\pa_\mu} G^+_5  $ & $\fr{ig}{2}
\left[ c^2_\theta U_{12} +(1-3s^2_\theta)\fr{U_{22}}{\sqrt{3}}
+\fr{2t}{3}\sqrt{\fr 2 3}U_{32} - s_{2\theta} U_{42}\right]
\longrightarrow \fr{ig}{2 c_W}(1-2 s^2_W) $\\ \hline $Z^\mu
G^-_6\overleftrightarrow{\pa_\mu} G^+_6  $ & $ \fr{ig}{2 (\om^2 +
c^2_\theta v^2) }\left\{ (\om^2+v^2s^2_\theta
c^2_\theta)U_{12}+[v^2c^2_\theta(1-3c^2_\theta)-\om^2]
\fr{U_{22}}{\sqrt{3}}\right.$\\ & $\left.+\fr t 3 \sqrt{\fr 2
3}(\om^2+2v^2c^2_\theta)U_{32}+2 v^2 s_\theta c^3_{\theta} U_{42}
\right\}\longrightarrow \fr{ig}{2 c_W}(1-2 s^2_W)$ \\
\hline $Z^\mu  H^-_2\overleftrightarrow{\pa_\mu} G^+_5  $ & $
\fr{ig\om}{4\sqrt{\om^2 + c^2_\theta v^2} }(s_{2\theta}U_{12}
+\sqrt{3} s_{2\theta}U_{22} + 2c_{2\theta}U_{42})\longrightarrow 0
 $\\ \hline
$Z^\mu  H^-_2\overleftrightarrow{\pa_\mu} G^+_6  $ & $
 \fr{ig \om v c_{\theta}}{2 (\om^2 + c^2_\theta v^2) }
 \left[-c^2_\theta U_{12}+
(2-3c^2_\theta)\fr{U_{22}}{\sqrt{3}}+\fr t 3 \sqrt{\fr 2 3}U_{32}
+s_{2\theta} U_{42}\right]\longrightarrow 0
  $\\ \hline
$Z^\mu  G^-_5\overleftrightarrow{\pa_\mu} G^+_6  $ & $ \fr{ig v
c_{\theta}}{4 \sqrt{\om^2 + c^2_\theta v^2}}
\left(s_{2\theta}U_{12} + \sqrt{3}s_{2\theta} U_{22}
+2c_{2\theta}U_{42}\right) \longrightarrow 0$\\ \hline
\end{tabular}
\ec \label{tab7}
\end{table}
The limit sign ($\longrightarrow$) in the Tables is the effective
one.

In the effective limit, the  $Z G_5 G_5$  vertex gets an exact
expression  as in the standard model. Hence $G_5$ can be
identified with the charged Goldstone boson in the standard model
$(G_{W^+})$.

Now we search couplings of the single $Z_\mu$ boson with neutral
scalar fields. With the help of the following equations  \bea
\chi^{0}_1\overleftrightarrow{\pa_\mu}\chi^{0*}_1& = &i
G_1\overleftrightarrow{\pa_\mu}S_1,\hs
\chi^{0}_3\overleftrightarrow{\pa_\mu}\chi^{0*}_3   = i
G_3\overleftrightarrow{\pa_\mu}S_3  ,\hs
\phi^{0}_2\overleftrightarrow{\pa_\mu}\phi^{0*}_2  = i
G_2\overleftrightarrow{\pa_\mu}S_2,\crn \pa_\mu \chi^{0*}_1
\chi^{0}_3 + \pa_\mu \chi^{0*}_3 \chi^{0}_1 &=& \fr 1 2
\left[\pa_\mu S_1 S_3 + \pa_\mu S_3 S_1 + \pa_\mu G_1 G_3 +
\pa_\mu G_3 G_1 + iG_3\overleftrightarrow{\pa_\mu}S_1 \right.\crn
&&\left. + i G_1\overleftrightarrow{\pa_\mu}S_3\right],\nn\eea the
necessary parts of Lagrangian are \bea
&&\fr{g}{2}\left[\left(U_{12}+\fr{U_{22}}{\sqrt{3}}-\fr t 3
\sqrt{\fr 2 3}U_{32}\right) G_1\overleftrightarrow{\pa_\mu}
S_1+U_{42}G_1\overleftrightarrow{\pa_\mu}S_3
+\left(-\fr{2}{\sqrt{3}}U_{22}-\fr t 3 \sqrt{\fr 2 3
}U_{32}\right)\right.\crn &&\left. \times
G_3\overleftrightarrow{\pa_\mu}S_3 +U_{42}
G_3\overleftrightarrow{\pa_\mu}S_1
+\left(-U_{12}+\fr{U_{22}}{\sqrt{3}}+\fr{2t}{3}\sqrt{\fr 2
3}U_{32}\right) G_2\overleftrightarrow{\pa_\mu}S_2 \right].\nn\eea
The resulting couplings are listed in Table \ref{tab8}.
\begin{table}
\caption{Trilinear coupling constants of $Z_\mu$ with two neutral
Higgs bosons.}\bc
\begin{tabular}{|c|c|}
\hline Vertex  &   Coupling \\ \hline $Z^\mu  G_1
\overleftrightarrow{\pa_\mu}H  $ & $-\fr{g s_\zeta}{2}
\left[\left( U_{12} + \fr{U_{22}}{\sqrt{3}} - \fr t 3 \sqrt{\fr 2
3} U_{32}\right)s_\theta  + U_{42}c_{\theta}\right]
\longrightarrow 0
 $\\ \hline
$Z^\mu  G_2\overleftrightarrow{\pa_\mu} H  $
 & $ \fr{g}{2} \left(-U_{12} +
\fr{U_{22}}{\sqrt{3}} + \fr{2 t}{ 3} \sqrt{\fr 2 3} U_{32}\right)
 c_\zeta \longrightarrow -\fr{g}{2 c_W}
$\\ \hline $Z^\mu G_3\overleftrightarrow{\pa_\mu} H   $
 & $ \fr{g s_\zeta}{2}\left[\left(\fr{2}{\sqrt{3}}U_{22}
 + \fr{t}{3} \sqrt{\fr 2 3}
 U_{32}\right)c_\theta - U_{42}s_{\theta}\right]
 \longrightarrow 0
$ \\ \hline $Z^\mu G_1\overleftrightarrow{\pa_\mu} H^0_1   $ &
$\fr{gc_\zeta}{2}\left[\left(U_{12} + \fr{U_{22}}{\sqrt{3}} - \fr
t 3 \sqrt{\fr 2 3} U_{32}\right)s_\theta + U_{42}c_{\theta}\right]
\longrightarrow 0
 $\\ \hline
$Z^\mu  G_2 \overleftrightarrow{\pa_\mu} H^0_1 $
 & $ \fr{g}{2} \left(-U_{12} +
\fr{U_{22}}{\sqrt{3}} + \fr{2 t}{ 3} \sqrt{\fr 2 3} U_{32}\right)
 s_\zeta \longrightarrow 0
$\\ \hline $Z^\mu G_3\overleftrightarrow{\pa_\mu}  H^0_1  $
 & $ -\fr{gc_\zeta}{2}\left[\left(\fr{2}{\sqrt{3}}U_{22}
 + \fr{t}{3}\sqrt{\fr 2 3}
 U_{32}\right)c_\theta - U_{42}s_{\theta}\right]
 \longrightarrow 0
$ \\ \hline $Z^\mu G_1\overleftrightarrow{\pa_\mu} G_4 $ &
$\fr{g}{2}\left[\left(U_{12}+\fr{U_{22}}{\sqrt{3}}-\fr t 3
\sqrt{\fr 2 3} U_{32}\right)c_\theta-U_{42}s_\theta\right]
\longrightarrow \fr{g}{2 c_W} $ \\ \hline $Z^\mu G_2
\overleftrightarrow{\pa_\mu}G_4  $ & $0$ \\
\hline  $Z^\mu  G_3 \overleftrightarrow{\pa_\mu}G_4  $ &
$\fr{g}{2}\left[\left(\fr{2}{\sqrt{3}}U_{22}+\fr t 3 \sqrt{\fr 2
3}U_{32}\right)s_\theta+U_{42}c_\theta\right]\longrightarrow 0$\\
\hline
\end{tabular}
\ec \label{tab8}
\end{table} From Table \ref{tab8}, we conclude that $G_2$
should be identified to $G_Z$ in the standard model. For the $Z'$
boson, the following remark is again helpful  \bea
 W_3^\mu & = & U_{13} Z^{'\mu} + \cdots,
 \hs  W_8^\mu  =  U_{23} Z^{'\mu} +
 \cdots,\crn
  B^\mu & = & U_{33} Z^{'\mu} + \cdots, \hs \ \
  y^\mu  =  U_{43} Z^{'\mu} + \cdots,
\label{potenn33a}\eea where  \bea
 U_{13}&=&  s_\va c_{\theta'}c_W , \hs  U_{23} =
\fr{s_\va(s^2_W-3c^2_Ws^2_{\theta'})+
  c_\va\sqrt{(1-4s^2_{\theta'}c^2_W)(4c^2_W-1)}}{\sqrt{3}
  c_Wc_{\theta'}},\\
 U_{33}&=&  -\fr{t_W(s_\va\sqrt{4c^2_W-1}
 -c_\va\sqrt{1-4s^2_{\theta'}c^2_W})}{\sqrt{3}c_{\theta'}}.
\label{potenn34}\eea
 Thus, with the replacement $Z
\rightarrow Z'$ one just replaces column  $2$ by $3$, for example,
trilinear coupling constants of the $Z'_\mu$ with two neutral
Higgs bosons are given in Table \ref{tab9}.
\begin{table}
\caption{Trilinear coupling constants of $Z'_\mu$ with two neutral
Higgs bosons.}\bc
\begin{tabular}{|c|c|}
\hline Vertex  &   Coupling \\ \hline $Z'^\mu G_1
\overleftrightarrow{\pa_\mu} H   $ & $-\fr{g s_\zeta}{2}
\left[\left( U_{13} + \fr{U_{23}}{\sqrt{3}} - \fr t 3 \sqrt{\fr 2
3} U_{33}\right)s_\theta  + U_{43}c_{\theta}\right]
\longrightarrow 0
 $\\ \hline
$Z'^\mu G_2 \overleftrightarrow{\pa_\mu} H $
 & $ \fr{g}{2} \left(-U_{13} +
\fr{U_{23}}{\sqrt{3}} + \fr{2 t}{ 3} \sqrt{\fr 2 3} U_{33}\right)
 c_\zeta \longrightarrow \fr{g}{2 c_W\sqrt{4c^2_W-1}}
$\\ \hline $Z'^\mu G_3\overleftrightarrow{\pa_\mu}H  $
 & $ \fr{g s_\zeta}{2}\left[\left(\fr{2}{\sqrt{3}}U_{23}
 + \fr{t}{3} \sqrt{\fr 2 3}
 U_{33}\right)c_\theta - U_{43}s_{\theta}\right]
 \longrightarrow 0
$ \\ \hline $Z'^\mu G_1\overleftrightarrow{\pa_\mu}  H^0_1  $ &
$\fr{gc_\zeta}{2}\left[\left(U_{13} + \fr{U_{23}}{\sqrt{3}} - \fr
t 3 \sqrt{\fr 2 3} U_{33}\right)s_\theta + U_{43}c_{\theta}\right]
\longrightarrow 0
 $\\ \hline
$Z'^\mu G_2\overleftrightarrow{\pa_\mu}  H^0_1  $
 & $ \fr{g}{2} \left(-U_{13} +
\fr{U_{23}}{\sqrt{3}} + \fr{2 t}{ 3} \sqrt{\fr 2 3} U_{33}\right)
 s_\zeta \longrightarrow 0
$\\ \hline $Z'^\mu  G_3\overleftrightarrow{\pa_\mu}  H^0_1 $
 & $ -\fr{gc_\zeta}{2}\left[\left(\fr{2}{\sqrt{3}}U_{23}
 + \fr{t}{3}\sqrt{\fr 2 3}
 U_{33}\right)c_\theta - U_{43}s_{\theta}\right]
 \longrightarrow -\fr{gc_W}{\sqrt{4c^2_W-1}}
$ \\ \hline $Z'^\mu G_1\overleftrightarrow{\pa_\mu} G_4 $ &
$\fr{g}{2}\left[\left(U_{13}+\fr{U_{23}}{\sqrt{3}}-\fr t 3
\sqrt{\fr 2 3}
U_{33}\right)c_\theta-U_{43}s_\theta\right]\longrightarrow \fr{g
c_{2W}}{2 c_W \sqrt{4c_W^2 - 1}}$ \\ \hline $Z'^\mu
G_2\overleftrightarrow{\pa_\mu} G_4 $ & $0$ \\ \hline $Z'^\mu G_3
\overleftrightarrow{\pa_\mu} G_4  $ &
$\fr{g}{2}\left[\left(\fr{2}{\sqrt{3}}U_{23}+\fr t 3 \sqrt{\fr 2
3}U_{33}\right)s_\theta+U_{43}c_\theta\right] \longrightarrow 0$\\
\hline
\end{tabular}
\ec \label{tab9}
\end{table}

Next, we search  couplings of two neutral gauge bosons with scalar
fields which arise in part from  \bea
Y^+\mathcal{P}^{\mathrm{NC}}_\mu
\mathcal{P}^{\mathrm{NC}\mu}Y&=&\fr{g^2}{4}\left\{\left[Y^*_1(A_{11}^\mu
A_{11 \mu} + y_\mu y^\mu) + Y^*_3(A_{11 \mu} y^\mu + A_{33 \mu}
y^\mu)\right]Y_1 + A_{22}^\mu A_{22 \mu}\right.\crn &&\left.\times
Y^*_2 Y_2+ \left[Y^*_1(A_{11 \mu} y^\mu + A_{33 \mu} y^\mu)+ Y^*_3
(A_{33}^\mu A_{33 \mu} + y_\mu y^\mu)\right]Y_3\right\}, \crn
&=&\fr{g^2}{4}\left\{\left[\chi^{0*}_1\left(A_{11}^{\mu \chi}
A_{11 \mu}^\chi + y_\mu y^\mu\right)  +
\chi^{0*}_3\left(A_{11\mu}^{\chi} y^\mu + A_{33 \mu}^\chi
y^\mu\right)\right]\chi^0_1 \right.\crn
&&+\left[\chi^{0*}_1\left(A_{11 \mu }^{ \chi} y^\mu + A_{33 \mu
}^{ \chi} y^\mu \right) + \chi^{0*}_3\left(A_{33}^{\mu \chi} A_{33
\mu}^\chi + y_\mu y^\mu\right)\right] \chi^0_3\crn &&+
\left[\phi^-_1\left(A_{11}^{\mu \phi} A_{11 \mu}^\phi + y_\mu
y^\mu\right) + \phi^-_3\left(A_{11\mu}^{\phi} y^\mu + A_{33
\mu}^\phi y^\mu\right)\right]\phi^+_1\crn &&+
\left[\phi^-_1\left(A_{11\mu}^{\phi} y^\mu + A_{33\mu}^{\phi}
y^\mu\right)+ \phi^-_3 \left(A_{33}^{\mu \phi} A_{33 \mu}^\phi +
y_\mu y^\mu\right)\right]\phi^+_3\crn &&+\left. \left(A_{22}^{\mu
\chi} A_{22 \mu}^\chi\right) \chi^+_2 \chi_2^- + \left(A_{22}^{\mu
\phi} A_{22 \mu}^\phi\right) \phi^{0*}_2
\phi^0_2\right\}.\label{potenn37}\eea Here $A_{ii}^\mu$
$(i=1,2,3)$ is a diagonal element in the matrix $\fr 2 g
\mathcal{P}^{\mathrm{NC}}_\mu $ which is dependent on the $U(1)_X$
charge: \bea A_{11}^{\mu \chi} &=& W_3^\mu +
\fr{W_8^\mu}{\sqrt{3}} - \fr t 3 \sqrt{\fr 2 3 } B^\mu, \hs
A_{11}^{\mu \phi} = W_3^\mu + \fr{W_8^\mu}{\sqrt{3}} + \fr{2 t}{
3} \sqrt{\fr 2 3 } B^\mu, \crn
 A_{22}^{\mu \chi} &=& - W_3^\mu +
\fr{W_8^\mu}{\sqrt{3}} -  \fr{ t}{ 3} \sqrt{\fr 2 3 } B^\mu,\hs
A_{22}^{\mu \phi} = - W_3^\mu + \fr{W_8^\mu}{\sqrt{3}} +  \fr{2
t}{ 3} \sqrt{\fr 2 3 } B^\mu,\label{potenn38} \\ A_{33}^{\mu \chi}
&=& - 2\fr{W_8^\mu}{\sqrt{3}} - \fr t 3 \sqrt{\fr 2 3 } B^\mu,\hs
A_{33}^{\mu \phi} = - 2\fr{W_8^\mu}{\sqrt{3}} + \fr{2 t}{ 3}
\sqrt{\fr 2 3 } B^\mu.\nn
  \eea
Quartic couplings of two  $Z$ with neutral scalar fields are given
by
 \bea &&
\fr{g^2}{4}\left\{\left[\chi^{0*}_1\left(A_{11}^{\mu \chi}  A_{11
\mu}^\chi + y_\mu y^\mu\right) + \chi^{0*}_3\left(A_{11\mu}^{\chi}
y^\mu + A_{33 \mu}^\chi y^\mu\right)\right]\chi^0_1 \right.\crn
&&+\left.\left[\chi^{0*}_1\left(A_{11 \mu }^{ \chi} y^\mu + A_{33
\mu }^{ \chi} y^\mu \right) + \chi^{0*}_3\left(A_{33}^{\mu \chi}
A_{33 \mu}^\chi + y_\mu y^\mu\right)\right] \chi^0_3+
 \left(A_{22}^{\mu \phi} A_{22 \mu}^\phi\right) \phi^{0*}_2
\phi^0_2 \right\}\crn &&= \fr{g^2}{4}\left\{\left(A_{11}^{\mu
\chi} A_{11 \mu}^\chi + y_\mu y^\mu\right)\chi^{0*}_1\chi^{0}_1
+\left(A_{33}^{\mu \chi} A_{33 \mu}^\chi + y_\mu
y^\mu\right)\chi^{0*}_3\chi^0_3\right. \crn &&\left.+\left(A_{11
\mu}^{\chi} y^\mu + A_{33 \mu}^\chi
y^\mu\right)(\chi^{0*}_1\chi^{0}_3  + \chi^{0*}_3\chi^{0}_1 )+
\left(A_{22}^{\mu \phi} A_{22 \mu}^\phi\right) \phi^{0*}_2
\phi^0_2 \right\} \label{potenn37}.\eea  In this case, the
couplings are listed in Table \ref{tab10}.
\begin{table}
\caption{Quartic coupling constants of $ZZ$ with
 two scalar  bosons.}\bc
\begin{tabular}{|c|c|}
\hline Vertex  & Coupling \\ \hline $ZZG_1G_1$ &
$\fr{g^2}{2}\left[\left(U_{12}+\fr{U_{22}}{\sqrt{3}}-\fr t 3
\sqrt{\fr 2 3 }U_{32}\right)^2+U_{42}^2\right]\longrightarrow
\fr{g^2}{2 c_W^2} $ \\ \hline $ZZG_2G_2$ &
$\fr{g^2}{2}\left(-U_{12}+\fr{U_{22}}{\sqrt{3}}+\fr{2t}{3}\sqrt{\fr
2 3 }U_{32}\right)^2\longrightarrow \fr{g^2}{2 c_W^2} $ \\ \hline
$ZZG_3G_3$ & $\fr{g^2}{2}\left[\left(\fr{2}{\sqrt{3}}U_{22}+\fr t
3 \sqrt{\fr 2 3 }U_{32}\right)^2+U^2_{42}\right]\longrightarrow 0
$ \\ \hline $ZZG_1G_3$ &
$\fr{g^2}{2}\left(U_{12}-\fr{U_{22}}{\sqrt{3}}-\fr{2t}{3}\sqrt{\fr
2 3}U_{32}\right)U_{42}\longrightarrow 0 $ \\ \hline $Z Z H H $ &
$
\fr{g^2}{2}\left\{s^2_\zeta\left[s^2_\theta\left(U_{12}+\fr{U_{22}}{\sqrt{3}}-\fr
t 3 \sqrt{\fr 2 3
}U_{32}\right)^2+c^2_\theta\left(\fr{2}{\sqrt{3}}U_{22}+\fr t 3
\sqrt{\fr 2 3 }U_{32}\right)^2+U_{42}^2\right.\right.$\\
&
$\left.\left.+s_{2\theta}U_{42}\left(U_{12}-\fr{U_{22}}{\sqrt{3}}-\fr{2
t}{3}\sqrt{\fr 2 3
}U_{32}\right)\right]+c^2_\zeta\left(U_{12}-\fr{U_{22}}{\sqrt{3}}-\fr{2t}{3}
\sqrt{\fr 2 3 }U_{32}\right)^2\right\} \longrightarrow \fr{g^2}{2
c_W^2}
 $\\ \hline
$Z Z H^0_1 H^0_1$ & $
\fr{g^2}{2}\left\{c^2_\zeta\left[s^2_\theta\left(U_{12}+\fr{U_{22}}{\sqrt{3}}-\fr
t 3 \sqrt{\fr 2 3
}U_{32}\right)^2+c^2_\theta\left(\fr{2}{\sqrt{3}}U_{22}+\fr t 3
\sqrt{\fr 2 3 }U_{32}\right)^2+U_{42}^2\right.\right.$\\
&$\left.\left.+s_{2\theta}U_{42}\left(U_{12}-\fr{U_{22}}{\sqrt{3}}-\fr{2
t}{3}\sqrt{\fr 2 3
}U_{32}\right)\right]+s^2_\zeta\left(U_{12}-\fr{U_{22}}{\sqrt{3}}-\fr{2t}{3}
\sqrt{\fr 2 3 }U_{32}\right)^2\right\}
  \longrightarrow 0$\\ \hline
 $Z Z G_4 G_4$ & $ \fr{g^2}{2}\left[c^2_\theta
 \left(U_{12}+\fr{U_{22}}{\sqrt{3}}-\fr
t 3 \sqrt{\fr 2 3
}U_{32}\right)^2+s^2_\theta\left(\fr{2}{\sqrt{3}}U_{22}+\fr t 3
\sqrt{\fr 2 3 }U_{32}\right)^2\right.$\\
&$\left.-s_{2\theta}\left(U_{12}-\fr{U_{22}}{\sqrt{3}}-\fr{2t}{3}
\sqrt{\fr 2 3 }U_{32}\right)U_{42}+U_{42}^2\right]
  \longrightarrow \fr{g^2}{2c^2_W} $
 \\ \hline
$Z Z H H_1$ & $ -\fr{g^2 s_{2\zeta}}{4}\left[s^2_\theta
 \left(U_{12}+\fr{U_{22}}{\sqrt{3}}-\fr
t 3 \sqrt{\fr 2 3
}U_{32}\right)^2+c^2_\theta\left(\fr{2}{\sqrt{3}}U_{22}+\fr t 3
\sqrt{\fr 2 3 }U_{32}\right)^2+U_{42}^2\right.$
\\ & $\left.-\left(U_{12}-\fr{U_{22}}{\sqrt{3}}-\fr{2t}{3} \sqrt{\fr 2 3
}U_{32}\right)^2+s_{2\theta}\left(U_{12}-\fr{U_{22}}{\sqrt{3}}-\fr{2t}{3}
\sqrt{\fr 2 3 }U_{32}\right)U_{42}\right] \longrightarrow 0 $
\\ \hline
$Z Z H G_4$ & $
-\fr{g^2s_\zeta}{4}\left(U_{12}-\fr{U_{22}}{\sqrt{3}}
-\fr{2t}{3}\sqrt{\fr 2
3}U_{32}\right)\left[2c_{2\theta}U_{42}+s_{2\theta}\left(U_{12}
+\sqrt{3}U_{22}\right)\right] \longrightarrow 0 $\\ \hline $Z Z
H_1 G_4$ & $ \fr{g^2 c_\zeta}{4}\left(U_{12} -
\fr{U_{22}}{\sqrt{3}} - \fr{2t}{3}\sqrt{\fr 2 3 }
U_{32}\right)\left[2c_{2\theta}U_{42}+s_{2\theta}\left(U_{12}+\sqrt{3}U_{22}\right)
\right]\longrightarrow 0 $\\ \hline
\end{tabular}
\ec \label{tab10}
\end{table}

Trilinear couplings of the pair $ZZ$ with one scalar field are
obtained via the following terms: \bea && \fr{g^2}{4}\left[vS_2
A_{22\mu}^\phi A_{22}^{\mu \phi}+u S_1 A_{11\mu}^\chi A_{11}^{\mu
\chi}+\om S_3 A_{33\mu}^\chi A_{33}^{\mu \chi}\right. \crn &&
\left.+(u S_1+\om S_3)y_\mu y^\mu -(\om S_1+u S_3)y^\mu
A_{22\mu}^\phi \right].\eea The obtained couplings are given in
Table \ref{tab11}.
\begin{table} \caption{Trilinear coupling
constants of $ZZ$ with one scalar bosons.}
 \bc
\begin{tabular}{|c|c|}
\hline Vertex  & Coupling \\ \hline  $Z Z H$ & $
\fr{g^2}{2}\left[v c_\zeta \left(U_{12} -\fr{U_{22}}{\sqrt{3}} -
\fr{2t}{3}\sqrt{\fr 2 3 } U_{32}\right)^2 -u s_\zeta s_\theta
\left(U_{12} + \fr{U_{22}}{\sqrt{3}} - \fr{t}{3} \sqrt{\fr 2 3
}U_{32}\right)^2-\om\fr{s_\zeta}{c_\theta}U^2_{42}\right.$\\
&$\left.-\om s_\zeta c_\theta\left(\fr{2}{\sqrt{3}} U_{22}+\fr t 3
\sqrt{\fr 2 3}U_{32}\right)^2-2\om s_\zeta s_\theta\left(U_{12}
-\fr{U_{22}}{\sqrt{3}} - \fr{2t}{3}\sqrt{\fr 2 3 }
U_{32}\right)U_{42} \right] \longrightarrow \fr{g^2 v}{2 c_W^2}$\\
\hline $Z Z H^0_1$ & $ \fr{g^2}{2}\left[v s_\zeta \left(U_{12}
-\fr{U_{22}}{\sqrt{3}} - \fr{2t}{3}\sqrt{\fr 2 3 } U_{32}\right)^2
+u c_\zeta s_\theta \left(U_{12} + \fr{U_{22}}{\sqrt{3}} -
\fr{t}{3} \sqrt{\fr 2 3
}U_{32}\right)^2+\om\fr{c_\zeta}{c_\theta}U^2_{42}\right.$\\ &
$\left.+\om c_\zeta c_\theta\left(\fr{2}{\sqrt{3}} U_{22}+\fr t 3
\sqrt{\fr 2 3}U_{32}\right)^2+2\om c_\zeta s_\theta\left(U_{12}
-\fr{U_{22}}{\sqrt{3}} - \fr{2t}{3}\sqrt{\fr 2 3 }
U_{32}\right)U_{42} \right]
 \longrightarrow 0$\\ \hline
 $Z Z G_4$ & $ \fr{g^2\om}{2}\left[s_\theta\left(U_{12} +
\sqrt{3}U_{22}\right) +
\fr{c_{2\theta}}{c_\theta}U_{42}\right]\left[
U_{12}-\fr{U_{22}}{\sqrt{3}}-\fr{2t}{3}\sqrt{\fr 2 3
}U_{32}\right] \longrightarrow 0$\\ \hline
\end{tabular}
\ec \label{tab11}
\end{table}

Because of (\ref{potenn33a}), for the $Z Z'$ couplings with scalar
fields, the above manipulation is  good enough. For example, Table
\ref{tab10} is replaced by Table \ref{tab12}.
\begin{table}
\caption{Trilinear coupling constants of $ZZ'$ with one scalar
bosons.}\bc
\begin{tabular}{|c|c|}
\hline Vertex  & Coupling \\ \hline $Z Z' H$ & $
\fr{g^2}{2}\left[v c_\zeta \left(U_{12} -\fr{U_{22}}{\sqrt{3}} -
\fr{2t}{3}\sqrt{\fr 2 3 } U_{32}\right)\left(U_{13}
-\fr{U_{23}}{\sqrt{3}} - \fr{2t}{3}\sqrt{\fr 2 3 } U_{33}\right)
-u s_\zeta s_\theta \right.$\\ & $\left. \times\left(U_{12} +
\fr{U_{22}}{\sqrt{3}} - \fr{t}{3} \sqrt{\fr 2 3
}U_{32}\right)\left(U_{13} + \fr{U_{23}}{\sqrt{3}}-\fr{t}{3}
\sqrt{\fr 2 3 }U_{33}\right)-\om s_\zeta
c_\theta\left(\fr{2}{\sqrt{3}} U_{22}+\fr t 3 \sqrt{\fr 2
3}U_{32}\right)\right.$\\ & $\left.\times\left(\fr{2}{\sqrt{3}}
U_{23}+\fr t 3 \sqrt{\fr 2
3}U_{33}\right)-\om\fr{s_\zeta}{c_\theta}U_{42}U_{43}-\om s_\zeta
s_\theta\left(U_{12} -\fr{U_{22}}{\sqrt{3}} - \fr{2t}{3}\sqrt{\fr
2 3 } U_{32}\right)U_{43}\right.$\\
&$\left.-\om s_\zeta s_\theta\left(U_{13} -\fr{U_{23}}{\sqrt{3}} -
\fr{2t}{3}\sqrt{\fr 2 3 } U_{33}\right)U_{42}
\right] \longrightarrow \fr{g^2 v c_{2W}}{2c_W\sqrt{4c^2_W-1}}$\\
\hline $Z Z' H^0_1$ & $ \fr{g^2}{2}\left[v s_\zeta \left(U_{12}
-\fr{U_{22}}{\sqrt{3}} - \fr{2t}{3}\sqrt{\fr 2 3 }
U_{32}\right)\left(U_{13} -\fr{U_{23}}{\sqrt{3}} -
\fr{2t}{3}\sqrt{\fr 2 3 } U_{33}\right) +u c_\zeta s_\theta
\right.$\\
&$\left.\times\left(U_{12} + \fr{U_{22}}{\sqrt{3}} - \fr{t}{3}
\sqrt{\fr 2 3 }U_{32}\right)\left(U_{13} + \fr{U_{23}}{\sqrt{3}} -
\fr{t}{3} \sqrt{\fr 2 3 }U_{33}\right)+\om c_\zeta
c_\theta\left(\fr{2}{\sqrt{3}} U_{22}+\fr t 3 \sqrt{\fr 2
3}U_{32}\right)\right.$\\ & $\left.\times\left(\fr{2}{\sqrt{3}}
U_{23}+\fr t 3 \sqrt{\fr 2
3}U_{33}\right)+\om\fr{c_\zeta}{c_\theta}U_{42}U_{43}+\om c_\zeta
s_\theta\left(U_{12} -\fr{U_{22}}{\sqrt{3}} - \fr{2t}{3}\sqrt{\fr
2 3 } U_{32}\right)U_{43}\right.$\\ & $\left.+\om c_\zeta
s_\theta\left(U_{13} -\fr{U_{23}}{\sqrt{3}} - \fr{2t}{3}\sqrt{\fr
2 3 } U_{33}\right)U_{42} \right] \longrightarrow 0$\\ \hline $Z
Z' G_4$ & $ \fr{g^2\om s_\theta}{2}\left[\left(U_{12} +
\fr{U_{22}}{\sqrt{3}} -\fr t 3 \sqrt{\fr 2 3
}U_{32}\right)\left(U_{13} + \fr{U_{23}}{\sqrt{3}} -\fr t 3
\sqrt{\fr 2 3 }U_{33}\right)\right.$\\ &
$\left.-\left(\fr{2}{\sqrt{3}}U_{22}+\fr{t}{3}\sqrt{\fr 2 3
}U_{32}\right)\left(\fr{2}{\sqrt{3}}U_{23}+\fr{t}{3}\sqrt{\fr 2 3
}U_{33}\right)+\cot_{2\theta}U_{42}\right.$\\ &
$\left.\times\left(U_{13} -
\fr{U_{23}}{\sqrt{3}}-\fr{2t}{3}\sqrt{\fr 2
3}U_{33}\right)+\cot_{2\theta}U_{43}\left(U_{12} -
\fr{U_{22}}{\sqrt{3}}-\fr{2t}{3}\sqrt{\fr 2 3}U_{32}
\right)\right] \longrightarrow 0$\\ \hline
\end{tabular}
\ec \label{tab12}
\end{table}

Now we turn to the interested coupling $ZW^\pm H^\mp_2$ arisen in
part from \bea && Y^+ \mathcal{P}^{\mathrm{NC}}_\mu
\mathcal{P}^{\mathrm{CC} \mu} Y + \mathrm{H.c.} =
\fr{g^2}{2\sqrt{2}}\left\{W^-_\mu A^\mu_{22} Y^*_2\left(c_\theta
Y_1 - s_\theta Y_3\right)\right.\crn &&
+\left.W^+_\mu\left[\left(c_\theta A^\mu_{11} - s_\theta
y^\mu\right)Y^*_1 +\left(c_\theta y^\mu - s_\theta
A^\mu_{33}\right)Y^*_3\right]Y_2\right\}+\mathrm{H.c.}\label{potenn38}\eea
For our Higgs triplets, one gets  \bea &&
\fr{g^2}{2\sqrt{2}}\left\{W^-_\mu \left[A^{\chi \mu}_{22}\chi^+_2
\left(c_\theta \chi^0_1 - s_\theta \chi^0_3\right)+ A^{\phi
\mu}_{22}\phi^{0*}_2 \left(c_\theta \phi^+_1 - s_\theta
\phi^+_3\right)\right]\right.\crn &&+ W^+_\mu \chi^-_2
\left[\left(c_\theta A^{\chi \mu}_{11} - s_\theta
y^\mu\right)\chi^{0*}_1 +\left(c_\theta y^\mu - s_\theta A^{\chi
\mu}_{33}\right)\chi^{0*}_3\right]\crn &&+
 \left.W^+_\mu \phi^0_2\left[\left(c_\theta  A^{\phi \mu}_{11}
 - s_\theta y^\mu\right)\phi^-_1 + \left(c_\theta  y^\mu -  s_\theta
 A^{\phi \mu}_{33}\right)\phi^-_3\right] \right\}+ \mathrm{H.c.}
 \label{potenn39}\eea

 From Eq. (\ref{potenn39}),  the trilinear couplings of the W
boson with one scalar and one neutral gauge bosons exist in a part
\bea &&\fr{g^2}{4}W^+_\mu\left\{v
\phi^-_1\left[c_\theta\left(\fr{2}{\sqrt{3}}W^\mu_8+\fr{4t}{3}\sqrt{\fr
2 3}B^\mu\right)-s_\theta y^\mu\right]\right.\crn &&\left.+v
\phi^-_3\left[c_\theta
y^\mu-s_\theta\left(-W_3^\mu-\fr{W^\mu_8}{\sqrt{3}}+\fr{4t}{3}\sqrt{\fr
2 3}B^\mu\right)\right]\right.\crn
&&\left.+\om\chi^-_2\left[s_\theta(W^\mu_3+\sqrt{3}W^\mu_8)+\fr{c_{2\theta}}
{c_\theta}y^\mu\right]\right\}+\mathrm{H.c.}
 \label{potenn51}\eea
From the above equation, we get necessary nonzero couplings, which
are listed in Table \ref{tab13}.
\begin{table}
\caption{Trilinear coupling constants of neutral gauge bosons with
$W^+$ and the charged scalar boson.} \bc
\begin{tabular}{|c|c|}
\hline Vertex  & Coupling \\ \hline $A W^+G^-_5$ &
$\fr{g^2}{2}vs_W$
\\ \hline
$Z W^+ H^-_2$ & $\fr{g^2 v \om}{2\sqrt{\om^2+c^2_\theta
v^2}}\left[s_\theta c_\theta
(U_{12}+\sqrt{3}U_{22})+c_{2\theta}U_{42}\right]$\\ \hline $Z' W^+
H^-_2$ & $\fr{g^2 v \om}{2\sqrt{\om^2+c^2_\theta
v^2}}\left[s_\theta c_\theta
(U_{13}+\sqrt{3}U_{23})+c_{2\theta}U_{43}\right]\longrightarrow 0$
\\ \hline
$ZW^+G^-_5$ & $\fr{g^2v}{4}\left[-s^2_\theta
U_{12}+(2-3s^2_\theta)\fr{U_{22}}{\sqrt{3}}+\fr{4t}{3}\sqrt{\fr 2
3} U_{32}-s_{2\theta}U_{42}\right]\longrightarrow -\fr{g^2}{2}vs_W
t_W$
 \\ \hline
$ZW^+G^-_6$ &
$\fr{g^2(v^2c^2_\theta-\om^2)}{8c_\theta\sqrt{\om^2+c^2_\theta
v^2}}\left[s_{2\theta}(U_{12}+\sqrt{3}U_{22})+2c_{2\theta}U_{42}\right]\longrightarrow
0$ \\ \hline
\end{tabular}
\ec \label{tab13}
\end{table} Vanishing couplings are \be
\mathcal{V}(A W^+H^-_2)= \mathcal{V}(A W^+G^-_6)=0.
\label{the1}\ee Eq. (\ref{the1}) is consistent with an evaluation
in Ref.~\cite{kame}, where authors neglected the diagrams with the
$\ga W^\pm H^\mp$ vertex.

From (\ref{pcc}), it follows that, to get couplings of the
bilepton gauge boson $Y^+$ with $Z H^-_2$, one just makes in
(\ref{potenn51}) the replacement: $ c_\theta \rightarrow -
s_\theta,\hs  s_\theta \rightarrow c_\theta$.

Finally, we can identify the scalar fields in the considered model
with that in the standard model as follows: \be H
\longleftrightarrow h, \hs G^+_5 \longleftrightarrow G_{W^+}, \hs
G_2 \longleftrightarrow G_Z.\ee In the effective limit $\om \gg v,
u$ our Higgs can be represented as
\be \chi=\left(%
\begin{array}{c}
 \fr{1}{\sqrt{2}}u +  G_{X^0}\\
   G_{Y^-} \\
  \fr{1}{\sqrt{2}}(\om + H_1^0 + i G_{Z'}) \\
\end{array}%
\right), \
\phi=\left(%
\begin{array}{c}
  G_{W^+} \\
  \fr{1}{\sqrt{2}}(v + h +  iG_Z) \\
  H_2^+ \\
\end{array}%
\right)\label{potenn35aa} \ee where $  G_3 \sim G_{Z'}, \hs G_6^-
\sim G_{Y^-}$ and \be G_4 + i\ G_1 \sim \sqrt{2}\ G_{X^0}\label{dnt}
\ee are the Goldstone boson of the massive gauge bosons $Z'$, $Y^-$
and $X^0$, respectively. Note that identification in (\ref{dnt}) is
possible due to the fact that both scalar and pseudoscalar parts of
$\chi_1^0$ are massless. In addition, the pseudoscalar part is
decoupled from others, while its scalar part mixes in the same as in
the gauge boson sector.

We emphasize again, in the effective approximation, all
Higgs-gauge boson couplings in the standard model are {\it
recovered} (see Table \ref{tab14}).
\begin{table} \caption{The standard model coupling constants
in the effective limit.} \bc
\begin{tabular}{|c|c|c|c|}
\hline Vertex  & Coupling  & Vertex & Coupling \\ \hline $WWhh$ &
$\fr{g^2}{2}$  & $G_WG_WA$ & $ie$
\\ \hline
$WWh$ & $\fr{g^2}{2}v$  & $WWG_ZG_Z$ & $\fr{g^2}{2}$
\\ \hline
$WG_W h$ & $-\fr{ig}{2} $  &$WWG_WG_W$ & $\fr{g^2}{2}$
\\ \hline
$WG_W G_Z$ & $\fr{g}{2}$  & $ZZh$ & $\fr{g^2}{2c^2_W}v$\\ \hline
$ZZhh$ & $\fr{g^2}{2c^2_W}$ & $ZZG_Z G_Z$ & $\fr{g^2}{2c^2_W}$ \\
\hline $AWG_W$ & $\fr{g^2}{2}v s_W$ & $Z W G_W$ &
$-\fr{g^2}{2}vs_W t_W$
\\ \hline
$ZG_Z h$ & $-\fr{g}{2c_W}$ & $ZG_W G_W$ &
$\fr{ig}{2c_W}(1-2s^2_W)$ \\ \hline
\end{tabular}
\ec \label{tab14}
\end{table} In contradiction with the
previous analysis in Ref.~\cite{ponce}, the condition $u \sim v$
or introduction of the third triplet are not necessary.

\subsection{\label{charged}Production of $H^\pm_2$ via $WZ$ Fusion
at LHC}

The possibility to detect the neutral Higgs boson in the minimal
version  at $e^+ e^-$ colliders was considered in~\cite{mal} and
production of the standard model-like neutral Higgs boson at LHC
was considered in Ref.\cite{ninhlong}. This section is devoted to
production of the charged $H^\pm_2$ at the CERN LHC.

Let us firstly discuss on the mass of this Higgs boson. Eq.
(\ref{potenn14}) gives us a connection between its mass and those of
the singly-charged bilepton $Y$ through the coefficient of Higgs
self-coupling $\la_4$. Note that in the considered model, the
neutrino Majorana masses exist only in the loop-levels. To keep
these masses in the experimental range, the mass of $M_{H^\pm_2}$
can be taken in the electroweak scale with $\la_4\sim 0.01$ (see the
next section). From (\ref{potenn14}), taking the lower limit for
$M_Y$ to be 1 TeV, the mass of $H^\pm_2$ is in range of 200 GeV.

Taking into account that, in the effective approximation, $H^-_2$
is the bilepton, we get the dominant decay channels as follows
\bea H^-_2 & \rightarrow &  l  \nu_l, \hs \tilde{U} d_a,\hs D_\al
\tilde{u}_a,\crn &\searrow& Z W^-,\hs Z^{\prime}W^{-},\hs XW^-,\hs
ZY^-. \label{modes}\eea Assuming that masses of the exotic quarks
$(U, D_\al)$ are larger than $M_{H^\pm_2}$, we come to the fact
that, the hadron modes are absent in  decay of the  charged Higgs
boson. Due to that the Yukawa couplings of $H_2^\pm l^\mp \nu$ are
very small, the main decay modes of the $H_2^\pm$ are in the
second line of (\ref{modes}). Note that the charged Higgs bosons
in doublet models such as two-Higgs doublet model or minimal
supersymmetric standard model, has both hadronic and leptonic
modes~\cite{roy}. This is a specific feature of the model under
consideration.

Because of the exotic $X,Y,Z'$ gauge bosons are heavy, the coupling
of a singly-charged Higgs boson ($H^\pm_2$) with the weak gauge
bosons, $H^\pm_2 W^\mp Z$, may dominate. Here, it is of particular
importance for the electroweak symmetry breaking. Its magnitude is
directly related to the structure of the extended Higgs sector under
global symmetries~\cite{glob}. This coupling can appear at the tree
level in models with scalar triplets, while it is induced at the
loop level in multi scalar doublet models. The coupling, in our
model, differs from zero at the tree level due to the fact that the
$H^\pm_2$ belongs to a triplet.

Thus, for the charged Higgs boson $H^\pm_2$, it is important to
study the couplings given by the interaction Lagrangian \be
\mathcal{L}_{int} =  f_{ZWH} H_2^\pm W_\mu^\mp Z^\mu,
\label{potenn50} \ee where $f_{ZWH}$, at tree level,  is given in
Table \ref{tab13}. The same as in~\cite{kame}, the dominant rate
is due to the diagram connected with the $W$ and $Z$ bosons.
Putting necessary matrix elements in Table \ref{tab13} , we get
\bea f_{ZWH}&=&-\fr{g^2 v \om
s_{2\theta}}{4\sqrt{\om^2+c^2_{\theta} v^2}}\fr{c_\va-s_\va
\sqrt{(4c^2_W-1)(1+4t^2_{2\theta})}}{\sqrt{(1+4t^2_{2\theta})
[c^2_W+(4c^2_W-1)t^2_{2\theta}]}}\nn
 \eea Thus, the form factor,
at the tree-level, is obtained by \bea
F&\equiv&\fr{f_{ZWH}}{gM_W}=-\fr{\om s_{2\theta}\left[c_\va-s_\va
\sqrt{(4c^2_W-1)(1+4t^2_{2\theta})}\right]}{2\sqrt{(\om^2+c^2_{\theta}
v^2)(1+4t^2_{2\theta})
[c^2_W+(4c^2_W-1)t^2_{2\theta}]}}.\label{fe}\eea The decay width
of $H_2^\pm\rightarrow W^\pm_iZ_i$, where $i=L,\ T$ represent
respectively the longitudinal and transverse polarizations, is
given by~\cite{kame}
 \bea \Gamma(H_2^\pm\to W^\pm_i Z_i) = M_{H_2^\pm}
 \frac{\la^{1/2}(1, w, z)}{16\pi} |M_{ii}|^2,\eea where
$\la(1,w,z)=(1-w-z)^2-4wz$, $w=M^2_W/M^2_{H^\pm_2}$ and
$z=M^2_Z/M^2_{H^\pm_2}$. The longitudinal and transverse
contributions are given in terms of $F$ by \bea |M_{LL}|^2 &=&
\frac{g^2}{4 {z}}
      (1-{w}-{z})^2 \left|F
             \right|^2, \\
|M_{TT}|^2 &=& 2 g^2
     {w} |F|^2. \eea For the case of $M_{H_2^\pm} \gg M_{Z}$, we have
$|M_{TT}|^2/|M_{LL}|^2 \sim 8 M_W^2 M_Z^2/M_{H_2^\pm}^4$ which
implies that the decay into a longitudinally polarized weak boson
pair dominates that into a transversely polarized one. The form
factor $F$ and mixing angle $t_\va$ are presented in Table
\ref{values}, where we have used: $s_W^2 = 0.2312,\ \ v = 246\
\textrm{GeV},\ \om = 3\ \textrm{TeV} \ ( \textrm{or} \ M_Y = 1
\textrm{TeV})$ as the typical values to get five cases corresponding
with the $s_\theta$ values under the constraint (\ref{uperlim}).
\begin{table}
\caption{Values of $F$, $t_\va$ and $M^{\mathrm{max}}_{H^\pm_2}$
for given $s_{\theta}$.} \bc
\begin{tabular}{|c|c|c|c|c|c|}
\hline $s_{\theta}$  & $0.08$ & $0.05$ & $0.02$ & $0.009$ &
$0.005$ \\ \hline $t_\va$ & $-0.0329698$ & $-0.0156778$ &
$-0.00598729$ & $-0.00449063$ & $-0.00422721$ \\ \hline $F$ &
$-0.087481$ & $-0.0561693$ & $-0.022803$ & $-0.0102847$ &
$-0.00571598$ \\ \hline $M^{\mathrm{max}}_{H^\pm_2}[\mathrm{GeV}]$
& $1700$ & $1300$ & $700$& $420$ & 320\\ \hline
\end{tabular}
\label{values} \ec
\end{table}

\begin{figure}[htbp]
\begin{center}
\includegraphics[width=12cm,height=9cm]{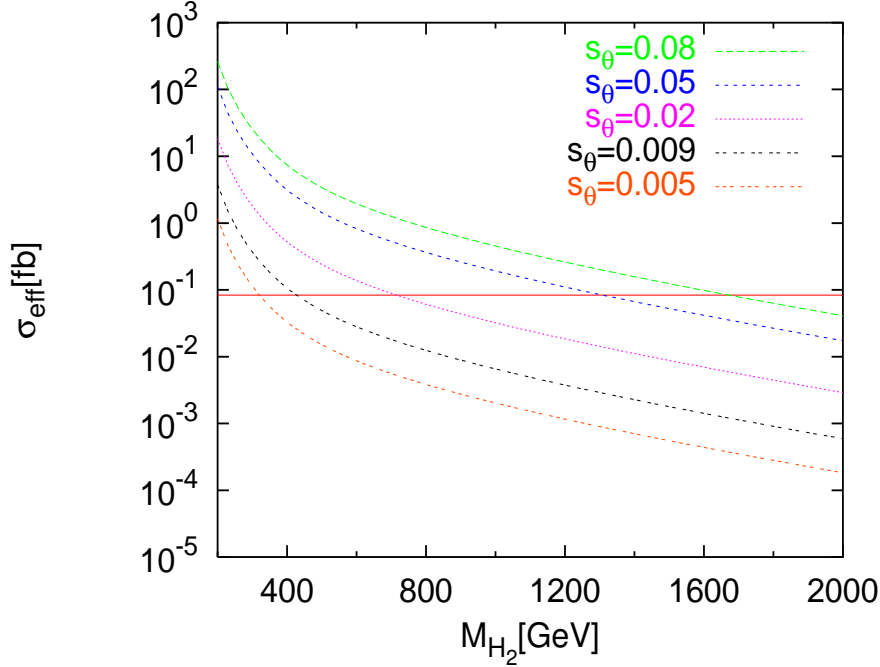}
\caption{\label{plseff}Hadronic cross section of $W^\pm Z$ fusion
process as a function of the charged Higgs boson mass for five
cases of $\sin\theta$. Horizontal line is discovery limit (25
events)}
\end{center}
\end{figure}

Next, let us study the impact of the $H^\pm_2 W^\mp Z$ vertex on
the production cross section of $pp \rightarrow W^{\pm*} Z^* X
\rightarrow H^\pm_2 X$ which is a pure electroweak process with
high $p_T$ jets going into the forward and backward directions
from the decay of the produced scalar boson without color flow in
the central region. The hadronic cross section for $pp \to H_2^\pm
X$ via $W^\pm Z$ fusion is expressed in the effective vector boson
approximation~\cite{kane} by 
\bea \sigma_{\rm eff}(s,M_{H_2^\pm}^2) \simeq \frac{16\pi^2 }{
\lambda(1,w,z) M_{H_2^\pm}^3} \sum_{\lambda=T,L} \Gamma(H_2^\pm
\to W^\pm_\lambda Z_\lambda)
       \tau \left.\frac{d {\mathcal L}}{d \tau}
      \right|_{pp/W^\pm_\lambda Z_\lambda},
\eea where $\tau=M_{H_2^\pm}^{2}/s$, and \bea
 \left.\frac{d {\mathcal
L}}{d \tau}\right|_{pp/W^\pm_\lambda Z_\lambda} = \sum_{ij}
\int_\tau^{1} \frac{d\tau'}{\tau'} \int_{\tau'}^{1} \frac{d x}{x}
f_i(x) f_j(\tau'/x) \left.\frac{d {\mathcal L}}{d
\xi}\right|_{q_iq_j/W^\pm_\lambda Z_\lambda}, \eea
 with
$\tau'=\hat{s}/s$ and $\xi=\tau/\tau'$. Here $f_i(x)$ is the
parton structure function for the $i$-th quark, and \bea \left.
\frac{d{\mathcal L}}{d\xi} \right|_{q_iq_j/W_T^\pm Z_T}
&=&\frac{c}{64\pi^4} \frac{1}{\xi} \ln \left(
\frac{\hat{s}}{M_W^2} \right) \ln \left( \frac{\hat{s}}{M_Z^2}
\right) \left[ (2+\xi)^2 \ln (1/\xi)-2(1-\xi)(3+\xi) \right],\crn
\left. \frac{d{\mathcal L}}{d\xi} \right|_{q_iq_j/W_L^\pm Z_L}
&=&\frac{c}{16\pi^4} \frac{1}{\xi} \left[ (1+\xi) \ln
(1/\xi)+2(\xi-1) \right],\nn \eea
 where
$c=\fr{g^4c^2_\theta}{16c^2_W}\left[g_{1V}^{2}(q_j)+g_{1A}^{2}(q_j)\right]$
with $g_{1V}(q_j)$, $g_{1A}(q_j)$ for quark $q_j$ are given in
Table I of Ref.~\cite{dlns}. Using CTEQ6L~\cite{cteq6}, in Fig.
\ref{plseff}, we have plotted $\sigma_{\rm eff}(s,M_{H_2^\pm}^2)$
at $\sqrt{s} = 14 \ \mathrm{TeV} $, as a function of the Higgs
boson mass corresponding five cases in Table \ref{values}.

Assuming discovery limit of 25 events corresponding to the
horizontal line, and taking the integrated luminosity of $300\
fb^{-1}$ ~\cite{lumino}, from the figure, we come to  conclusion
that, for $s_\theta = 0.08$ (the line on top), the charged Higgs
boson $H_2^\pm$ with mass larger than 1700 GeV, cannot be seen at
the LHC. These limiting masses are denoted by
$M^{\mathrm{max}}_{H^\pm_2}$ and listed in Table \ref{values}. If
the mass of the above mentioned Higgs boson is in range of 200 GeV
and $s_\theta = 0.08$, the cross section can exeed 260 $fb$: i.e.,
78000 of $H_2^\pm$ can be produced at the integrated LHC
luminosity of $300\ fb^{-1}$.
 This production rate is about ten times larger than those
in Ref.~\cite{kame}. The cross-sections decrease rapidly as mass
of the Higgs boson increases from 200 GeV to 400 GeV.

\subsection{Summary}

In this section we have considered the scalar sector in the
economical 3-3-1 model. The model contains eight Goldstone bosons -
the justified number of the massless ones eaten by the massive gauge
bosons. Couplings of the standard model-like gauge bosons such as of
the photon, the $Z$ and the new $Z'$ gauge bosons with physical
Higgs ones are also given. From these couplings, the standard
model-like Higgs boson as well as Goldstone ones are identified. In
the effective approximation, full content of scalar sector can be
recognized. The CP-odd part of Goldstone associated with the neutral
non-Hermitian bilepton gauge bosons $G_{X^0}$ is decoupled, while
its CP-even counterpart has the mixing by the same way in the gauge
boson sector. Despite the mixing among the photon with the
non-Hermitian neutral bilepton $X^0$ as well as with the $Z$ and the
$Z'$ gauge bosons, the electromagnetic couplings remain unchanged.

It is worth mentioning that, masses of all physical Higgs bosons are
related to that of gauge bosons through the coefficients of Higgs
self-interactions. All gauge-scalar boson couplings in the standard
model are recovered. The coupling of the photon with the Higgs
bosons are diagonal.

It should be mentioned that in Ref.\cite{ponce}, to get nonzero
coupling $ZZh$ at the tree level, the authors suggested the
following solution: (i) $u \sim v$ or (ii) by introducing the
third Higgs scalar with VEV ($\sim v$). This problem does not
happen in our consideration.

After all we focused attention to the singly-charged Higgs boson
$H^\pm_2$ with mass proportional to the bilepton mass $M_Y$ through
the coefficient $\la_4$. Mass of the $H^\pm_2$  is estimated in a
range of 200 GeV. This boson, in difference with those arisen in the
Higgs doublet models, does not have the hadronic and leptonic decay
modes. The trilinear coupling $ZW^\pm H_2^\mp$ which differs, at the
tree level, while the similar coupling of the photon $\ga W^\pm
H^\mp_2$ as expected, vanishes. In the model under consideration,
the charged Higgs boson $H_2^\pm$ with mass larger than 1700 GeV,
cannot be seen at the LHC.  If the mass of the above mentioned Higgs
boson is in range of 200 GeV, however, the cross section can exceed
260 fb: i.e., 78000 of $H_2^\pm$ can be produced at the LHC for the
luminosity of $300\ fb^{-1}$. By measuring this process we can
obtain useful information to determine the structure of the Higgs
sector.

\section{\label{fermionmasses}Fermion Masses} We first give some comments
on the charged lepton masses and set conventions. The neutrino and
quark masses are correspondingly considered.

\subsection{\label{chargeleptons}Charged-Lepton Masses} The charged
leptons $(l= e, \mu, \tau)$ gain masses via the following couplings
\bea {\mathcal L}^l_Y= h^l_{ab}\bar{\psi}_{aL} \phi l_{bR}+
\mathrm{H.c.}\eea The mass matrix is therefore followed by \bea
M_{l}= -\fr{v}{\sqrt{2}}\left(%
\begin{array}{ccc}
  h^l_{11} & h^l_{12} & h^l_{13} \\
  h^l_{21} & h^l_{22} & h^l_{23} \\
  h^l_{31} & h^l_{32} & h^l_{33} \\
\end{array}%
\right),\eea which of course is the same as in the standard model
and thus gives consistent masses for the charged leptons
\cite{ponc1}.

For the sake of simplicity, in the following, we can suppose that
the Yukawa coupling of charged leptons $h^l$ is flavor diagonal,
thus $l_a$ $(a=1,2,3)$ are mass eigenstates respective to the mass
eigenvalues $m_a=-\fr{v}{\sqrt{2}}h^l_{aa}$.

\begin{figure}
\includegraphics{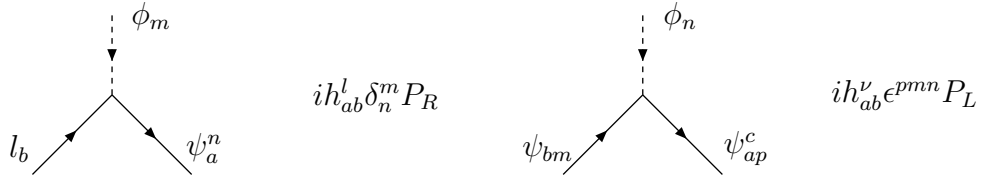}
\caption{\label{figh2}Lepton Yukawa couplings.}
\end{figure}

\begin{figure}
\includegraphics{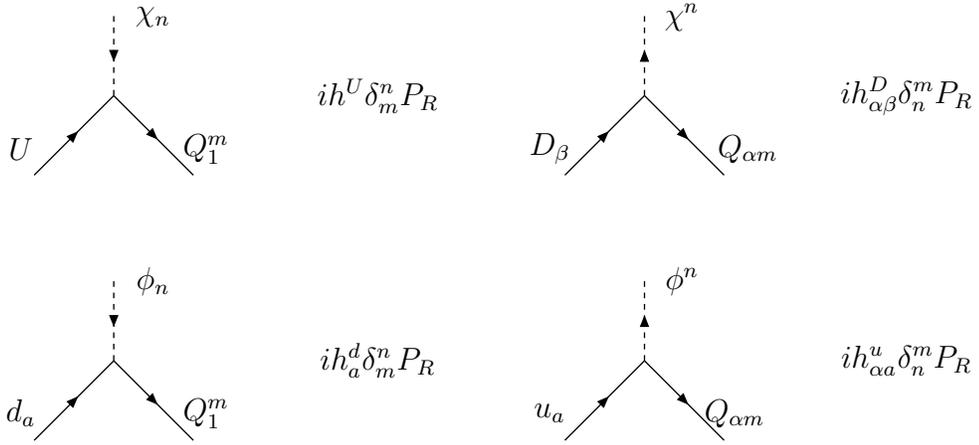}
 \caption{\label{fighh1}Relevant lepton-number conserving
quark Yukawa couplings}
\end{figure}

\begin{figure}
\includegraphics{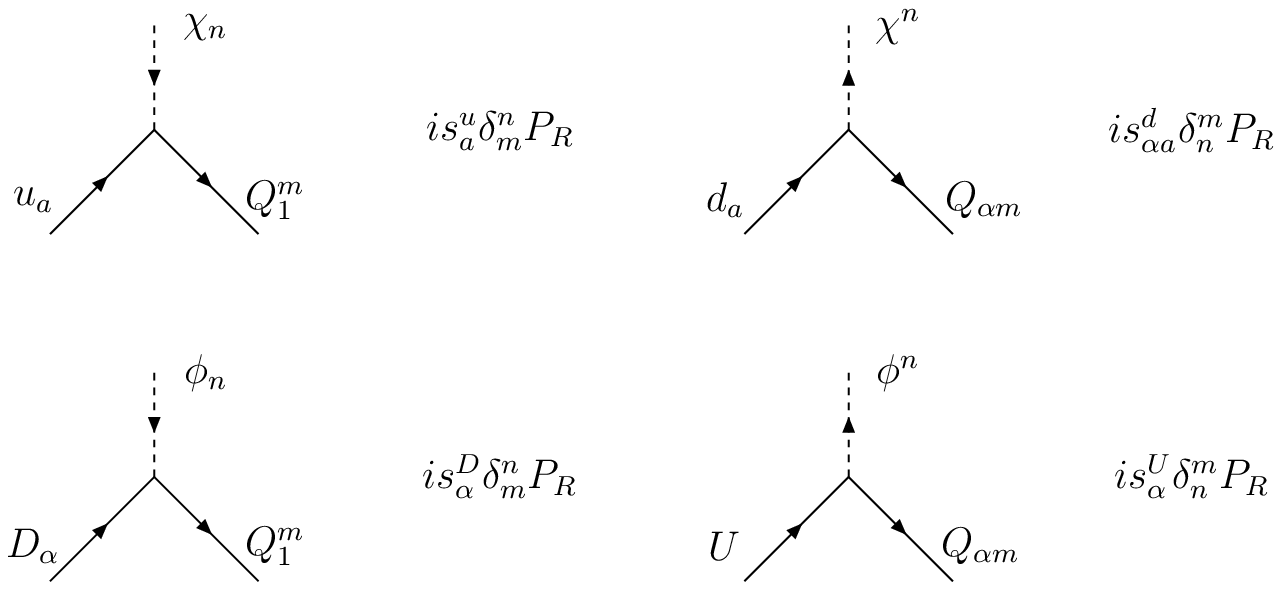}
 \caption{\label{fighh2}Lepton-number violating
 quark Yukawa couplings}
\end{figure}

\begin{figure}
\includegraphics[width=15cm,height=7cm]{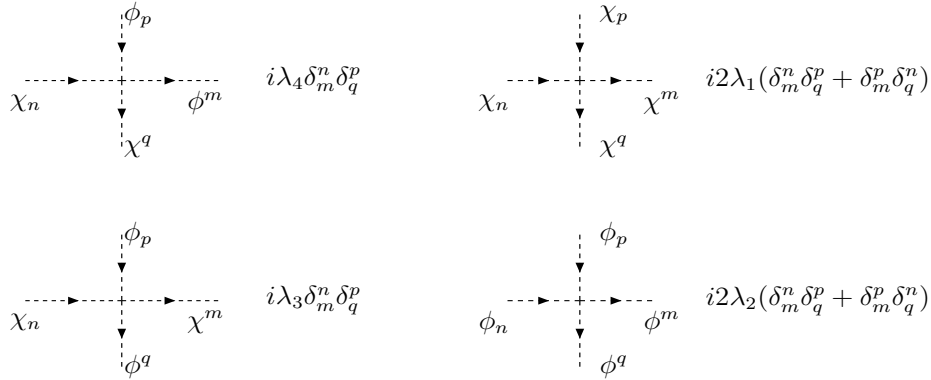}
\caption{Higgs boson self-couplings} \label{figh1}
\end{figure}

For convenience in further reading, we present the Yukawa
interactions of (\ref{y1}) and (\ref{y2}) in terms by Feynman
diagrams in Figures (\ref{figh2}),
(\ref{fighh1}),
and (\ref{fighh2}),
where the Hermitian adjoint ones are not displayed. The Higgs
boson self-couplings are depicted in Figure (\ref{figh1}).

\subsection{\label{neutrinomasses}Neutrino Masses} First we present
mass mechanisms for the neutrinos. Next, detailed calculations and
analysis of the neutrino mass spectrum are given. The experimental
constraints on the coupling $h^\nu$ are also considered.

\subsubsection{\label{model1}Neutrino Mass Mechanisms}

In the considering model, the possible different mass-mechanisms
for the neutrinos can be summarized through the three dominant
$\mathrm{SU}(3)_C\otimes \mathrm{SU}(3)_L \otimes
\mathrm{U}(1)_X$-invariant effective operators as follows
\cite{wbg}: \bea O^{\mathrm{LNC}}_{ab}
&=&\bar{\psi}^c_{aL}\psi_{bL}\phi, \label{yka}\\
O^{\mathrm{LNV}}_{ab}
&=&(\chi^*\bar{\psi}^c_{aL})(\chi^*\psi_{bL}),
\label{heavyparticles}\\
O^{\mathrm{SLB}}_{ab}
&=&(\chi^*\bar{\psi}^c_{aL})(\psi_{bL}\phi\chi),\label{loops}\eea
where the Hermitian adjoint operators are not displayed. It is
worth noting that they are also all the performable operators with
the mass dimensionality $d\leq6$ responsible for the neutrino
masses. The difference among the mass-mechanisms can be verified
through the operators. Both (\ref{yka}) and (\ref{loops}) conserve
$\mathcal{L}$, while (\ref{heavyparticles}) violates this charge
with two units. Since $d(O^{\mathrm{LNC}})=4$ and
$L\langle\phi\rangle=0$, (\ref{yka}) provides only Dirac masses
for the neutrinos which can be obtained at the tree level through
the Yukawa couplings in (\ref{y1}). Since $d(O^{\mathrm{SLB}})=6$
and $(L\langle\chi\rangle)_p \neq 0$ for $p=1$, vanishes for other
cases, (\ref{loops}) provides both Dirac and Majorana masses for
the neutrinos through radiative corrections mediated by the model
particles. The masses induced by (\ref{yka}) are given by the
standard $\mathrm{SU}(2)_L\otimes \mathrm{U}(1)_Y$ symmetry
breaking via the VEV $v$. However, those by (\ref{loops}) are
obtained from both the stages of $\mathrm{SU}(3)_L\otimes
\mathrm{U}(1)_X$ breaking achieved by the VEVs $u,\ \om$ and $v$.

Note that, the LNV interactions in (\ref{y2}) are due to quarks.
Hence, they do not give contribution to LNV of the leptons such as
of the neutrinos. Except, the LNV couplings of (\ref{y2}), all the
remaining interactions of the model (lepton Yukawa couplings
(\ref{y1}), Higgs self-couplings (\ref{poten}), and etc.) conserve
$\mathcal{L}$. This means that the operator (\ref{heavyparticles})
of LNV cannot be  mediated by particles of the model, in other
words, it must be introduced by hands. As a fact, the economical
3-3-1 model including the alternative versions \cite{flt,ppf} are
only extensions beyond the standard model in the scales of orders
of TeV \cite{dln,ochoa}. Hence, it is expected that the operator
in (\ref{heavyparticles}) has to be mediated by heavy particles of
an underlined new physics at a scale $\mathcal{M}$ much greater
than $\om$ which have been followed in various of grand unified
theories (GUTs) \cite{wbg,331GUTs,GUTs}. Thus, in this model the
neutrinos can get mass from three very different sources widely
ranging over the mass scales: $u\sim \mathcal{O}(1)\
\mathrm{GeV}$, $v \approx 246\ \mathrm{GeV}$, $\om \sim
\mathcal{O}(1)\ \mathrm{TeV}$, and $\mathcal{M}\sim
\mathcal{O}(10^{16})\ \mathrm{GeV}$.

We remind that, in the former version \cite{flt}, the authors in
\cite{diasalex} have considered operators of the type
(\ref{heavyparticles}), however, under a discrete symmetry
\cite{discr,ponc1}. As shown in Section \ref{fermionmasses}, the
current model is realistic, and such a discrete symmetry is not
needed, because, as a fact that the model will fail if it is
enforced. In addition, if such discrete symmetries are not
discarded, the important mass contributions for the neutrinos
mediated by model particles are then suppressed; for example, in
this case the remaining operators (\ref{yka}) and (\ref{loops}) will
be removed. With the only operator (\ref{heavyparticles}) the three
active neutrinos will get effective zero-masses under a type II
seesaw \cite{seesaw} (see below); however, this operator occupies a
particular importance in this version.

Alternatively, in such model, the authors in \cite{changlong} have
examined two-loop corrections to (\ref{heavyparticles}) by the aid
of explicit LNV Higgs self-couplings, and using a fine-tuning for
the tree-level Dirac masses of (\ref{yka}) down to current values.
However, as mentioned, this is not the case in the considering
model, because our Higgs potential (\ref{poten}) conserves
$\mathcal{L}$. We know that one of the problems of the 3-3-1 model
with RH neutrinos is associated with the Dirac mass term of
neutrinos. In the following, we will show that, if such a
fine-tuning is done to get  small values for these terms, then the
mass generation of neutrinos mediated by model particles is not
able, or the results will be trivial.  This is in contradiction
with \cite{changlong}. In the next, the large bare Dirac masses
for the neutrinos, which are as of charged fermions of a natural
result from standard symmetry breaking, will be studied.

\subsubsection{\label{neumass}Neutrino Mass Matrix} The operators
$O^{\mathrm{LNC}}$, $O^{\mathrm{SLB}}$ and $O^{\mathrm{LNV}}$
(including their Hermitian adjoint) will provide the masses for
the neutrinos: the first responsible for tree-level masses, the
second for one-loop corrections, and the third for contributions
of heavy particles.

\bc{\it Tree-Level Dirac Masses}\ec From the Yukawa couplings in
(\ref{y1}), the tree-level mass Lagrangian for the neutrinos is
obtained by \cite{feynmanrules}\bea
\mathcal{L}^\mathrm{LNC}_\mathrm{mass}&=&
h^\nu_{ab}\bar{\nu}_{aR}\nu_{bL}\langle
\phi^0_2\rangle-h^\nu_{ab}\bar{\nu}^c_{aL}
\nu^c_{bR}\langle\phi^0_2\rangle+\mathrm{H.c.}\crn &=&
2\langle\phi^0_2\rangle
h^\nu_{ab}\bar{\nu}_{aR}\nu_{bL}+\mathrm{H.c.}
=-(M_D)_{ab}\bar{\nu}_{aR}\nu_{bL}+\mathrm{H.c.}\crn
&=&-\fr{1}{2}\left(\bar{\nu}^c_{aL},\bar{\nu}_{aR}\right) \left(
\begin{array}{cc}
 0 & (M^T_D)_{ab} \\
(M_D)_{ab} & 0 \\
 \end{array}
\right)\left(
\begin{array}{c}
\nu_{bL} \\
\nu^c_{bR} \\
\end{array}
\right)+\mathrm{H.c.} \crn &=& -\fr{1}{2}\bar{X}^c_L M_\nu X_L
+\mathrm{H.c.}, \eea where $h^\nu_{ab}=-h^\nu_{ba}$ is due to
Fermi statistics. The $M_D$ is the mass matrix for the Dirac
neutrinos: \bea (M_D)_{ab}&\equiv&-\sqrt{2}v
h^\nu_{ab}=(-M^T_D)_{ab}=\left(
\begin{array}{ccc}
0 & -A & -B \\
A & 0 & -C \\
B & C & 0 \\
\end{array}
\right), \label{dirac1} \eea where \be  A \equiv
\sqrt{2}h^\nu_{e\mu}v,\hs B \equiv \sqrt{2}h^\nu_{e\tau}v,\hs
C\equiv \sqrt{2}h^\nu_{\mu\tau}v.\nn\ee This mass matrix has been
rewritten in a general basis $X^T_L \equiv(\nu_{eL},\nu_{\mu
L},\nu_{\tau L}, \nu^c_{e R},\nu^c_{\mu R},\nu^c_{\tau R})$: \be
M_\nu \equiv \left(
\begin{array}{cc}
0 & M_D^T \\
M_D & 0 \\
\end{array}
\right).\label{treedirac} \ee

The tree-level neutrino spectrum therefore consists of only Dirac
fermions. Since $h^\nu_{ab}$ is antisymmetric in $a$ and $b$, the
mass matrix $M_D$ gives one neutrino massless and two others
degenerate in mass: $0,\ -m_D,\ m_D$, where $m_D\equiv
(A^2+B^2+C^2)^{1/2}$. This mass spectrum is not realistic under
the data, however, it will be severely changed by the quantum
corrections, the most general mass matrix can then be written as
follows
\bea M_{\nu}=\left(%
\begin{array}{cc}
  M_{L} & M_D^T \\
  M_D & M_{R} \\
\end{array}%
\right),\label{matran}\eea where $M_{L,R}$ (vanish at the
tree-level) and $M_D$ get possible corrections.

If such a tree-level contribution dominates the resulting mass
matrix (after corrections), the model will provide an explanation
about a large splitting either $\Delta m^2_\mathrm{atm} \gg \Delta
m^2_\mathrm{sol}$ or $\Delta m^2_\mathrm{LSND} \gg \Delta
m^2_\mathrm{atm,sol}$ \cite{pdg} (see also \cite{changlong}).
Hence, we need a fine-tuning at the tree-level \cite{changlong}
either $m_D \sim (\Delta m^2_\mathrm{atm})^{1/2}$ $(\sim
5\times10^{-2}\ \mathrm{eV})$ or $m_D \sim (\Delta
m^2_\mathrm{LSND})^{1/2}$ $(\sim \ \mathrm{eV})$ \cite{pdg}.
Without loss of generality, assuming that $h^\nu_{e\mu}\sim
h^\nu_{e\tau}\sim h^\nu_{\mu\tau}$ we get then $h^\nu\sim
10^{-13}\ (\mathrm{or}\ 10^{-12})$. The coupling $h^\nu$ in this
case is so small and therefore this fine-tuning is not natural
\cite{pecsmr}. Indeed, as shown below, since $h^\nu$ enter the
dominant corrections from (\ref{loops}) for $M_{L,R}$, these terms
$M_{L,R}$ get very small values which are not large enough to
split the degenerate neutrino masses into a realistic spectrum.
(The largest degenerate splitting in squared-mass is still much
smaller than $\Delta m^2_\mathrm{sol}\sim 8\times10^{-5}\
\mathrm{eV}^2$ \cite{pdg}.) In addition, in this case, the Dirac
masses get corrections trivially.

The above problem can be solved {\it just} by the LNV  operator
(\ref{heavyparticles}); and then the operator (\ref{loops})
obtaining the contributions from particles in the model is
suppressed (for details, see \cite{diasalex}). However, we do not
consider the above solution in this work. This implies that the
tree-level Dirac mass term for the neutrinos by its naturalness
should be treated as those as of the usual charged fermions
resulted of the standard symmetry breaking, say, $h^\nu\sim h^{e}
\ (\sim 10^{-6})$ \cite{pecsmr}. It turns out that this term is
regarded as a large bare quantity and unphysical. Under the
interactions, they will of course change to physical masses. In
the following we will obtain such {\it finite renormalizations}
(for more details, see \cite{chengli}) in the masses of neutrinos.

\begin{figure}
\includegraphics{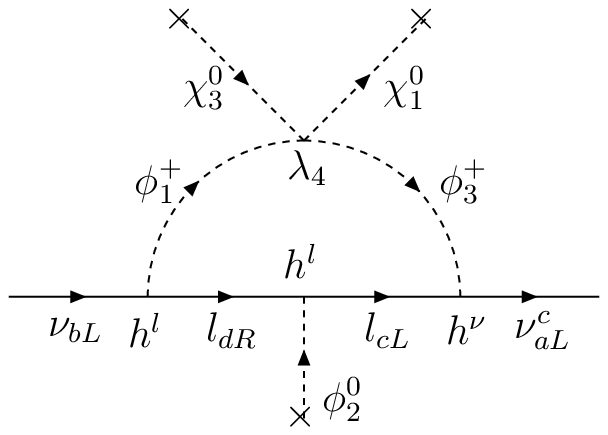}
\includegraphics{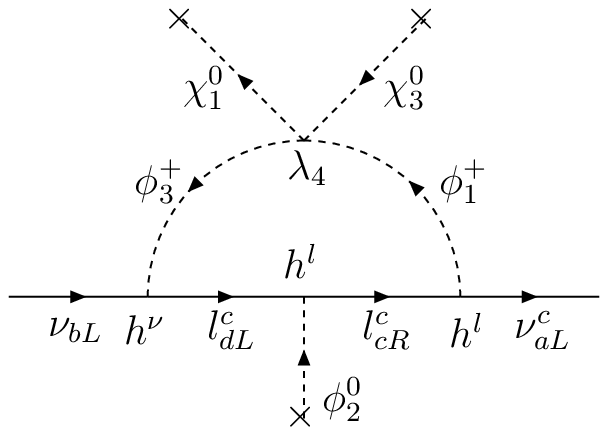}
\caption{\label{hinh12}The one-loop corrections for the mass
matrix $M_L$.}
\end{figure}

\begin{figure}
\includegraphics{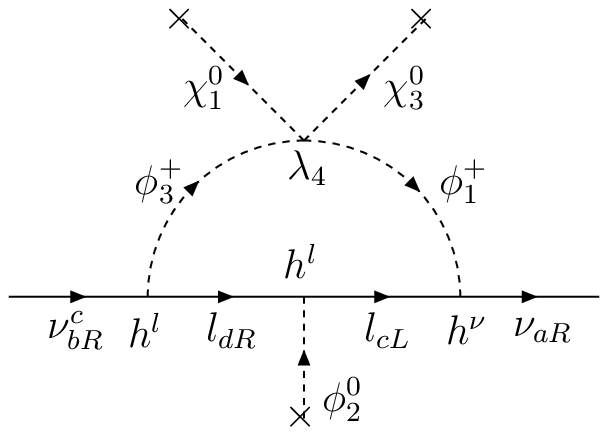}
\includegraphics{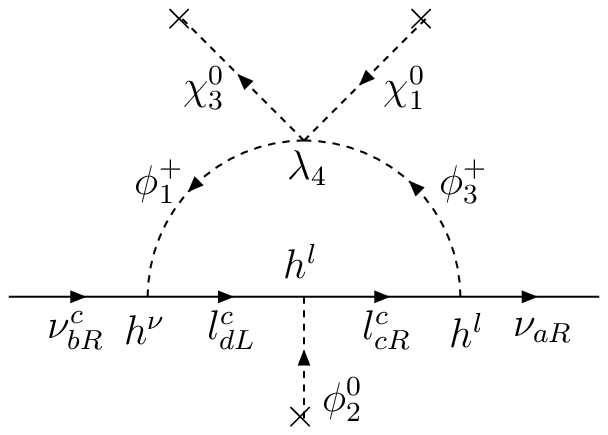}
\caption{\label{hinh34}The one-loop corrections for the mass
matrix $M_R$.}
\end{figure}

\begin{figure}
\includegraphics{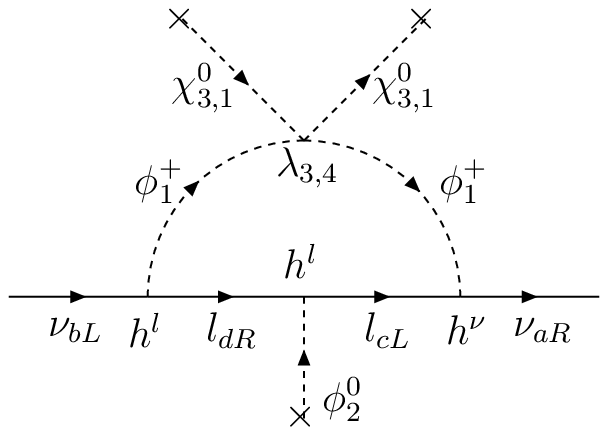}
\includegraphics{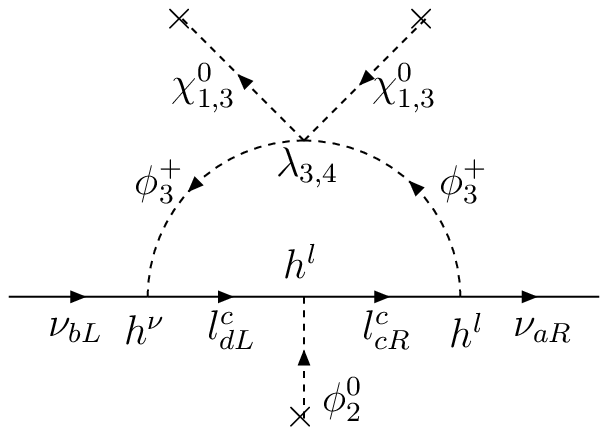}
\caption{\label{hinh56}The one-loop corrections for the mass
matrix $M_D$.}
\end{figure}

\bc{\it \label{radia}One-Loop Level Dirac and Majorana masses}\ec
The operator (\ref{loops}) and its Hermitian adjoint arise from
the radiative corrections mediated by the model particles, and
give contributions to Majorana and Dirac mass terms $M_L$, $M_R$
and $M_D$ for the neutrinos. The Yukawa couplings of the leptons
in (\ref{y1}) and the relevant Higgs self-couplings in
(\ref{poten}) are explicitly rewritten as follows \bea
\mathcal{L}^\mathrm{lept}_\mathrm{Y}&=&
2h^\nu_{ab}\bar{\nu}^c_{aL}l_{bL}\phi^+_3
-2h^\nu_{ab}\bar{\nu}_{aR}l_{bL}\phi^+_1+h^l_{ab}
\bar{\nu}_{aL}l_{bR}\phi^+_1
+h^l_{ab}\bar{\nu}^c_{aR}l_{bR}\phi^+_3+h^l_{ab}\bar{l}_{aL}l_{bR}\phi^0_2\crn&&+\mathrm{H.c.},\\
\mathcal{L}^\mathrm{relv}_\mathrm{H}&=&\la_3\phi^-_1
\phi^+_1(\chi^{0*}_1\chi^{0}_1
+\chi^{0*}_3\chi^{0}_3)+\la_3\phi^-_3\phi^+_3
(\chi^{0*}_1\chi^{0}_1+\chi^{0*}_3\chi^{0}_3) \crn
&&+\la_4\phi^-_1
\phi^+_1\chi^{0*}_1\chi^0_1+\la_4\phi^-_3\phi^+_3\chi^{0*}_3\chi^0_3
+\la_4\phi^-_3\phi^+_1\chi^{0*}_1\chi^0_3 +\la_4\phi^-_1
\phi^+_3\chi^{0*}_3\chi^0_1.\eea The one-loop corrections to the
mass matrices $M_L$ of $\nu_L$, $M_R$ of $\nu_R$ and $M_D$ of
$\nu$ are therefore given in Figs. (\ref{hinh12}), (\ref{hinh34})
and (\ref{hinh56}), respectively.

\centerline{{\it Radiative Corrections to $M_L$ and $M_R$}}

With the Feynman rules at hand \cite{feynmanrules}, $M_L$ is
obtained by \bea
-i(M_L)_{ab}P_L&=&\int\fr{d^4p}{(2\pi)^4}\left(i2h^\nu_{ac}P_L\right)
\fr{i(p\!\!\!/+m_c)}{p^2-m^2_c}\left(ih^l_{cd}\fr{v}{\sqrt{2}}P_R\right)
\fr{i(p\!\!\!/+m_d)}{p^2-m^2_d}\crn
&\times&(ih^{l*}_{bd}P_L)\fr{-1}{(p^2-m^2_{\phi_1})
(p^2-m^2_{\phi_3})}\left(i\la_4\fr{u\om}{2}\right)\crn
&+&\int\fr{d^4p}{(2\pi)^4}\left(ih^{l*}_{ac}P_L\right)
\fr{i(-p\!\!\!/+m_c)}{p^2-m^2_c}\left(ih^l_{dc}\fr{v}{\sqrt{2}}P_R\right)
\fr{i(-p\!\!\!/+m_d)}{p^2-m^2_d}\crn
&\times&(i2h^{\nu}_{bd}P_L)\fr{-1}{(p^2-m^2_{\phi_1})
(p^2-m^2_{\phi_3})}\left(i\la_4\fr{u\om}{2}\right).\label{eq1}\eea
Because the Yukawa couplings of the charged leptons are flavor
diagonal, the equation (\ref{eq1}) becomes \bea
(M_L)_{ab}&=&\fr{i\sqrt{2}\la_4u\om}{v} h^\nu_{ab}\left[m^2_b
I(m^2_b,m^2_{\phi_3},m^2_{\phi_1})-m^2_a
I(m^2_a,m^2_{\phi_3},m^2_{\phi_1})\right],\crn && (a,b \mbox{ not
summed}),\label{ap2}\eea where the integral $I(a,b,c)$ is given in
Appendix \ref{ap1}.

In the effective approximation (\ref{vevcons}), identifications
are given by $\phi^{\pm}_3 \sim H^{\pm}_2$ and $\phi^{\pm}_1 \sim
G^{\pm}_{W}$ \cite{dls}, where $H^{\pm}_2$ and $G^{\pm}_{W}$ as
above mentioned, are the charged bilepton Higgs boson and the
Goldstone boson associated with $W^\pm$ boson, respectively. For
the masses, we have also $m^2_{\phi_3}\simeq m^2_{H_2}\
(\simeq\fr{\la_4}{2}\om^2)$ and $m^2_{\phi_1}\simeq 0$. Using
(\ref{ketqua}), the integrals are given by \bea
I(m^2_a,m^2_{\phi_3},m^2_{\phi_1})\simeq
-\fr{i}{16\pi^2}\fr{1}{m^2_a-m^2_{H_2}}\left[1-\fr{m^2_{H_2}
}{m^2_a-m^2_{H_2}}\ln\fr{m^2_a}{m^2_{H_2}}\right],\hs
a=e,\mu,\tau.\eea Consequently, the mass matrix (\ref{ap2})
becomes\bea (M_L)_{ab} &\simeq& \fr{\sqrt{2}\la_4 u \om
h^\nu_{ab}}{16\pi^2
v}\left[\fr{m^2_{H_2}(m^2_a-m^2_b)}{(m^2_b-m^2_{H_2})(m^2_a-m^2_{H_2})}+\fr{m^2_a
m^2_{H_2}}{(m^2_a-m^2_{H_2})^2}\ln\fr{m^2_a}{m^2_{H_2}}\right.\crn
&&\left.-\fr{m^2_b
m^2_{H_2}}{(m^2_b-m^2_{H_2})^2}\ln\fr{m^2_b}{m^2_{H_2}}\right]\crn
&\simeq& \fr{\sqrt{2}\la_4 u \om h^\nu_{ab}}{16\pi^2 v
m^2_{H_2}}\left[m^2_a\left(1+\ln\fr{m^2_a}{m^2_{H_2}}\right)-
m^2_b\left(1+\ln\fr{m^2_b}{m^2_{H_2}}\right)\right],\label{tdr}\eea
where the last approximation (\ref{tdr}) is kept in the orders up
to $\mathcal{O}[(m^2_{a,b}/m^2_{H_2})^2]$. Since $m^2_{H_2}\simeq
\fr{\la_4}{2} \om^2$, it is worth noting that the resulting $M_L$
is not explicitly dependent on $\la_4$, however, proportional to
$t_{\theta}=u/\om$ (the mixing angle between the $W$ boson and the
singly-charged bilepton gauge boson $Y$ \cite{dlns}), $\sqrt{2}v
h^\nu_{ab}$ (the tree-level Dirac mass term of neutrinos), and
$m_{H_2}$ in the logarithm scale. Here the VEV $v\approx
v_{\mathrm{weak}}$, and the charged-lepton masses $m_a$
$(a=e,\mu,\tau)$ have the well-known values. Let us note that
$M_L$ is symmetric and has vanishing diagonal elements.

For the corrections to $M_R$, it is easily to check that the
relationship $(M_R)_{ab}=-(M_L)_{ab}$ is exact at the one-loop
level. (This result can be derived from Fig. (\ref{hinh34}) in a
general case without imposing any additional condition on $h^l$,
$h^\nu$, and further.) Combining this result with (\ref{tdr}), the
mass matrices are explicitly rewritten as follows \bea
(M_L)_{ab}=-(M_R)_{ab}\simeq \left(
\begin{array}{ccc}
0 & f & r \\
f & 0 & t \\
r & t & 0 \\
\end{array}\right),\label{majo} \eea where the elements are obtained by
\bea f&\equiv&\left(\sqrt{2}vh^\nu_{e\mu}\right)
\left\{\left(\fr{t_{\theta}}{8\pi^2v^2}\right)
\left[m^2_e\left(1+\ln\fr{m^2_e}{m^2_{H_2}}\right)
-m^2_\mu\left(1+\ln\fr{m^2_\mu}{m^2_{H_2}}
\right)\right]\right\},\crn
r&\equiv&\left(\sqrt{2}vh^{\nu}_{e\tau}\right)
\left\{\left(\fr{t_{\theta}}{8\pi^2v^2}\right)
\left[m^2_e\left(1+\ln\fr{m^2_e}{m^2_{H_2}}\right)-m^2_\tau
\left(1+\ln\fr{m^2_\tau}{m^2_{H_2}}\right)\right]\right\},\crn
t&\equiv&\left(\sqrt{2}vh^\nu_{\mu\tau}\right)
\left\{\left(\fr{t_{\theta}}{8 \pi^2v^2}\right)\left[m^2_\mu
\left(1+\ln\fr{m^2_\mu}{m^2_{H_2}}\right)
-m^2_\tau\left(1+\ln\fr{m^2_\tau}
{m^2_{H_2}}\right)\right]\right\}.\label{majo1}\eea

It can be checked that $f,r,t$ are much smaller than those of
$M_D$. To see this, we can take $m_e\simeq 0.51099\ \mathrm{MeV}$,
$m_\mu \simeq 105.65835\ \mathrm{MeV}$, $m_\tau \simeq 1777\
\mathrm{MeV}$, $v \simeq 246\ \mathrm{GeV}$, $u \simeq 2.46\
\mathrm{GeV}$, $\om \simeq 3000\ \mathrm{GeV}$, and $m_{H_2}\simeq
700\ \mathrm{GeV}\ (\la_4 \sim 0.11)$ \cite{dlns,dls,dln}, which
give us then \bea
f&\simeq&\left(\sqrt{2}vh^\nu_{e\mu}\right)\left(3.18\times
10^{-11}\right), \hs
r\simeq\left(\sqrt{2}vh^{\nu}_{e\tau}\right)\left(5.93\times
10^{-9}\right), \crn
t&\simeq&\left(\sqrt{2}vh^\nu_{\mu\tau}\right)\left(5.90\times
10^{-9}\right),\label{majo2}\eea where the second factors rescale
negligibly with $\om \sim 1-10\ \mathrm{TeV}$ and $m_{H_{2}}\sim
200-2000\ \mathrm{GeV}$. This thus implies that \be
|M_{L,R}|/|M_D|\sim 10^{-9},\label{cont1}\ee which can be checked
with the help of $|M|\equiv (M^\dag M)^{1/2}$. In other words, the
constraint is given as follows \be |M_{L,R}|\ll |M_D|.\label{b12}
\ee

With the above results at hand, we can then get the masses by
studying diagonalization of the mass matrix (\ref{matran}), in
which, the submatrices $M_{L,R}$ and $M_D$ satisfying the
constraint (\ref{b12}), are given by (\ref{majo}) and
(\ref{dirac1}), respectively. In calculation, let us note that,
since $M_D$ has one vanishing eigenvalue, $M_\nu$ does not possess
the pseudo-Dirac property in all three generations \cite{kobcsl},
however, is very close to those because the remaining eigenvalues
do. As a fact, we will see that $M_\nu$ contains a combined
framework of the seesaw \cite{seesaw} and the pseudo-Dirac
\cite{pseudo-Dirac}. To get mass, we can suppose that $h^\nu$ is
real, and therefore the matrix $iM_D$ is Hermitian:
$(iM_D)^\dag=iM_D$ (\ref{dirac1}). The Hermitianity for $M_{L,R}$
is also followed by (\ref{majo}). Because the dominant matrix is
$M_D$ (\ref{b12}), we first diagonalize it by biunitary
transformation \cite{chengli}: \bea
\bar{\nu}_{aR}&=&\bar{\nu}_{iR}(-iU)^\dag_{ia},\hs
\nu_{bL}= U_{bj}\nu_{jL},\hs (i,j=1,2,3),\label{csl}\\
M_\mathrm{diag}&\equiv& \mathrm{diag}(0,-m_D,m_D)=(-iU)^\dag M_D
U,\hs m_D = \sqrt{A^2+B^2+C^2}, \eea where the matrix $U$ is
easily obtained by \bea U=\fr{1}{m_D\sqrt{2(A^2+C^2)}}\left(
\begin{array}{ccc}
C\sqrt{2(A^2+C^2)} & iBC-Am_D & BC-iAm_D \\
-B\sqrt{2(A^2+C^2)} & i(A^2+C^2) & (A^2+C^2) \\
A\sqrt{2(A^2+C^2)} & iAB+Cm_D & AB+iCm_D \\
\end{array}
\right).\eea Resulted by the anti-Hermitianity of $M_D$, it is
worth noting that $M_\nu$ in the case of vanishing $M_{L,R}$
(\ref{treedirac}) is indeed diagonalized by the following unitary
transformation: \bea V=\fr{1}{\sqrt{2}}\left(
\begin{array}{cc}
U & U \\
-iU & iU \\
\end{array}
\right).\label{fin1}\eea

A new basis $(\nu_1,\nu_2,...,\nu_6)^T_L \equiv V^\dag X^T_L$,
which is different from $(\nu_{jL},\nu^c_{iR})^T$ of (\ref{csl}),
is therefore performed. The neutrino mass matrix (\ref{matran}) in
this basis becomes \bea V^\dag M_\nu V &=& \left(
\begin{array}{cc}
M_{\mathrm{diag}} & \ep \\
\ep & -M_{\mathrm{diag}} \\
\end{array}
\right),\label{ben1}\\
\ep &\equiv& U^\dag M_L U,\hs \ep^\dag= \ep,\label{ben3} \eea
where the elements of $\ep$ are obtained by \bea \ep_{11}
&=&\ep_{22}=\ep_{33}=0, \label{pp1}\\
\ep_{12}&=&i\ep^*_{13}=\left\{[ABm_D+iC(A^2-B^2+C^2)]f
+[(C^2-A^2)m_D+2iABC]r\right.\crn &&\left.+[iA(A^2
-B^2+C^2)-BCm_D]t\right\}[m^2_D\sqrt{2(A^2 +
C^2)}]^{-1},\label{pp2} \\
\ep_{23}&=&\left\{(A^2+C^2)\left[(Cm_D-iAB)t-(Am_D+iBC)f\right]\right.\crn
&&\left.-\left[B(A^2-C^2)m_D+iAC(A^2+2B^2+C^2)\right]r\right\}
\left[m^2_D(A^2+C^2)\right]^{-1}.\label{pp3}\eea Let us remind the
reader that (\ref{pp1}) is exactly given at the one-loop level
$M_{L}$ (\ref{ap2}) without imposing any approximation on this
mass matrix. Interchanging the positions of component fields in
the basis $(\nu_1,\nu_2,...,\nu_6)^T_L$ by a permutation
transformation $P^\dag \equiv P_{23}P_{34}$, that is, $\nu_{p}
\rightarrow (P^\dag)_{pq} \nu_q\ (p,q=1,2,...,6)$ with \be P^\dag
=\left(
           \begin{array}{cccccc}
             1 & 0 & 0 & 0 & 0 & 0 \\
             0 & 0 & 0 & 1 & 0 & 0 \\
             0 & 1 & 0  & 0 & 0 & 0 \\
             0 & 0 & 1 & 0 & 0 & 0 \\
             0 & 0 & 0 & 0 & 1 & 0 \\
             0 & 0 & 0 & 0 & 0 & 1 \\
           \end{array}
         \right),\ee the mass matrix (\ref{ben1})
can be rewritten as follows \bea P^\dag (V^\dag M_\nu V)P &=&
\left(
  \begin{array}{cc|cccc}
    0 & 0 & 0 & 0 & \ep_{12} & \ep_{13} \\
    0 & 0 & \ep_{12}& \ep_{13} & 0 & 0 \\ \hline
    0 & \ep_{21} & -m_D & 0 & 0 & \ep_{23} \\
    0 &  \ep_{31} & 0 &  m_D & \ep_{32} & 0 \\
    \ep_{21} & 0 & 0 & \ep_{23} & m_D  & 0 \\
    \ep_{31} & 0 & \ep_{32} & 0 & 0 & -m_D \\
  \end{array}
\right).\label{ben2}\eea

It is worth noting that in (\ref{ben2}) all the off-diagonal
components $|\ep|$ are much smaller than the eigenvalues $|\pm
m_D|$ due to the condition (\ref{b12}). The degenerate eigenvalues
$0$, $-m_D$ and $+m_D$ (each twice) are now splitting into three
pairs with six different values, two light and four heavy. The two
neutrinos of first pair resulted by the $0$ splitting have very
small masses as a result of exactly what a seesaw does
\cite{seesaw}, that is, the off-diagonal block contributions to
these masses are suppressed by the large pseudo-Dirac masses of
the lower-right block. The suppression in this case is different
from the usual ones \cite{seesaw} because it needs only the
pseudo-Dirac particles \cite{pseudo-Dirac} with the masses $m_D$
of the electroweak scale instead of extremely heavy RH Majorana
fields, and that the Dirac masses in those mechanisms are now
played by loop-induced $f,r,t$ (\ref{majo1}) as a result of the
SLB $u/\om$. Therefore, the mass matrix (\ref{ben2}) is
effectively decomposed into $M_{\mathrm{S}}$ for the first pair of
light neutrinos $(\nu_\mathrm{S})$ and $M_{\mathrm{P}}$ for the
last two pairs of heavy pseudo-Dirac neutrinos $(\nu_\mathrm{P})$:
\bea (\nu_1,\nu_4,\nu_2,\nu_3,\nu_5,\nu_6)^T_L &\rightarrow&
(\nu_\mathrm{S},\nu_\mathrm{P})^T_L=
V^\dag_{\textmd{eff}}(\nu_1,\nu_4,\nu_2,\nu_3,\nu_5,\nu_6)^T_L,\crn
V^\dag_{\textmd{eff}}(P^\dag V^\dag M_\nu VP)V_{\textmd{eff}}&=&
\mathrm{diag}\left( M_\mathrm{S},M_\mathrm{P}\right),\eea where
$V_{\textmd{eff}}$, $M_\mathrm{S}$ and $M_\mathrm{P}$ get the
approximations: \bea V_{\textmd{eff}}&\simeq&\left(
\begin{array}{cc}
1 & \mathcal{E} \\
-\mathcal{E}^+ & 1 \\
\end{array}
\right),\ \mathcal{E}\equiv \left(
  \begin{array}{cccc}
0 & 0 & \ep_{12} & \ep_{13} \\
\ep_{12}& \ep_{13} & 0 & 0 \\
\end{array}\right)\left(\begin{array}{cccc}
-m_D & 0 & 0 & \ep_{23} \\
0 &  m_D & \ep_{32} & 0 \\
0 & \ep_{23} & m_D  & 0 \\
\ep_{32} & 0 & 0 & -m_D \\
\end{array}\right)^{-1}\crn M_{\mathrm{S}}&\simeq& -\mathcal{E}\left(
\begin{array}{cc}
0 & \ep_{21} \\
0 & \ep_{31} \\
\ep_{21}& 0 \\
\ep_{31}& 0 \\
\end{array}
\right),\hs M_{\mathrm{P}}\simeq \left(\begin{array}{cccc}
-m_D & 0 & 0 & \ep_{23} \\
0 &  m_D & \ep_{32} & 0 \\
0 & \ep_{23} & m_D  & 0 \\
\ep_{32} & 0 & 0 & -m_D \\
\end{array}\right).\label{pse}\eea
The mass matrices $M_\mathrm{S}$ and $M_\mathrm{P}$, respectively,
give exact eigenvalues as follows \bea m_{\mathrm{S}\pm}&=& \pm
\fr{2\mathrm{Im}(\ep_{13}\ep_{13}\ep_{32})}{m^2_D-\ep^2_{23}}\simeq
\pm 2\mathrm{Im}\left(\fr{\ep_{13}\ep_{13}\ep_{32}}{m^2_D}\right),\label{p1}\\
m_{\mathrm{P}\pm}&=& -m_D\pm |\ep_{23}|,\hs m_{\mathrm{P'}\pm} =
m_D \pm |\ep_{23}|.\label{p3}\eea In this case, the mixing
matrices are collected into
$(\nu_{\mathrm{S}\pm},\nu_{\mathrm{P}\pm},\nu_{\mathrm{P}'\pm})^T_L
=V^\dag_{\pm}(\nu_\mathrm{S},\nu_\mathrm{P})^T_L$, where the
$V_{\pm}$ is obtained by \be V_{\pm} =\fr{1}{\sqrt{2}}\left(
\begin{array}{cccccc}
1 & -1 & 0 & 0 & 0 & 0 \\
1 & 1 & 0 & 0 & 0 & 0 \\
0 & 0 & \ka & -\ka & 0 & 0 \\
0 & 0 & 0 & 0 & 1 & 1 \\
0 & 0 & 0 & 0 & \ka & -\ka \\
0 & 0 & 1 & 1 & 0 & 0 \\
\end{array}
\right),\hs \ka\equiv
\fr{\ep_{23}}{|\ep_{23}|}=\exp(i\arg\ep_{23}).
 \ee It is to be noted that the degeneration in the Dirac one $|\pm m_D|$ is now
splitting severally.

From (\ref{p3}) we see that the four large pseudo-Dirac masses for
the neutrinos are almost degenerate. In addition, the resulting
spectrum (\ref{p1}), (\ref{p3}) yields two largest squared-mass
splittings, respectively, proportional to $m^2_D$ and $4 m_D
|\ep_{23}|$. From (\ref{pp3}) and (\ref{majo2}), we can evaluate
$|\ep_{23}|\simeq 3.95\times 10^{-9}\ m_D \ll m_D$ (where $A\sim B
\sim C \sim m_D/\sqrt{3}$ is understood). Because the splitting $4
m_D |\ep_{23}|$ is still much smaller than $\Delta
m^2_{\mathrm{sol}}$, this therefore implies that the fine-tuning,
as mentioned, is not realistic. (In detail, in Table
\ref{fine-tuning}, we give the numerical values of these
fine-tunings, where the parameters are given as before
(\ref{majo2}).)
\begin{table}[h]
\caption{\label{fine-tuning}The values for $h^\nu$ and two largest
splittings in squared-mass.} \bc
\begin{tabular}{|l|c|c|c|}
\hline
 Fine-tuning & $h^\nu$ & $m^2_D\
  (\mathrm{eV}^2)$ & $4 m_D
|\ep_{23}|\ (\mathrm{eV}^2)$ \\
 \hline
 $m^2_D \sim \Delta m^2_\mathrm{atm}$
 &  $8.30\times 10^{-14}$ & $2.50\times 10^{-3}$ & $3.95\times 10^{-11}$\\ \hline
 $m^2_D \sim \Delta
m^2_\mathrm{LSND}$ & $1.66\times 10^{-12}$ & $1.00$ & $1.58\times
10^{-8}$ \\ \hline
\end{tabular}
\ec
\end{table}

Similarly, for the two small masses, we can also evaluate
$|m_{\mathrm{S}\pm}|\simeq 4.29\times 10^{-28}\ m_D$. This shows
that the masses $m_{\mathrm{S}\pm}$ are very much smaller than the
splitting $|\ep_{23}|$. This also implies that the two light
neutrinos in this case are hidden for any $m_D$ value of
pseudo-Dirac neutrinos. Let us see the sources of the problem why
these masses are so small: (i) Vanishing of all the elements of
left-upper block of (\ref{ben2}); (ii) In (\ref{p1}) the resulting
masses are proportional to $|\ep|^3/m^2_D$, but not to
$|\ep|^2/m_D$ as expected from (\ref{ben2}). It turns out that
this is due to the antisymmetric of $h^\nu_{ab}$ enforcing on the
tree-level Dirac-mass matrix and the degenerate of $M_R=-M_L$ of
the one-loop level left-handed (LH) and RH Majorana-mass matrices.
It can be easily checked that such degeneration in Majorana masses
remains up to higher-order radiative corrections as a result of
treating the LH and RH neutrinos in the same gauge triplets with
the model Higgs content. For example, by the aid of (\ref{loops})
the degeneration retains up to any higher-order loop.

\centerline{{\it Radiative Corrections to $M_D$}}

As mentioned, the mass matrix $M_D$ requires the one-loop
corrections as given in Fig. \ref{hinh56}, and the contributions
are easily obtained as follows \bea
-i(M^{\mathrm{rad}}_D)_{ab}P_L&=&\int\fr{d^4p}{(2\pi)^4}
\left(-i2h^\nu_{ac}P_L\right) \fr{i(p\!\!\!/+m_c)}{p^2-m^2_c}
\left(ih^l_{cd}\fr{v}{\sqrt{2}}P_R\right)
\fr{i(p\!\!\!/+m_d)}{p^2-m^2_d}\crn
&\times&(ih^{l*}_{bd}P_L)\fr{-1}{(p^2-m^2_{\phi_1})^2}
\left(i\la_3\fr{u^2 +\om^2}{2}+i\la_4\fr{u^2}{2}\right)\crn
&+&\int\fr{d^4p}{(2\pi)^4}\left(ih^{l*}_{ac}P_L\right)
\fr{i(-p\!\!\!/+m_c)}{p^2-m^2_c}
\left(ih^l_{dc}\fr{v}{\sqrt{2}}P_R\right)
\fr{i(-p\!\!\!/+m_d)}{p^2-m^2_d}\crn
&\times&(i2h^{\nu}_{bd}P_L)\fr{-1}{(p^2-m^2_{\phi_3})^2}
\left(i\la_3\fr{u^2+\om^2}{2}
+i\la_4\fr{\om^2}{2}\right).\label{eq2}\eea We rewrite \bea
(M^{\mathrm{rad}}_D)_{ab}&=&-\fr{i\sqrt{2}h^\nu_{ab}}{v}
\left\{\left[\la_3(u^2+\om^2)+\la_4 u^2\right]m^2_b
I(m^2_b,m^2_{\phi_1})\right.\crn
&&\left.+\left[\la_3(u^2+\om^2)+\la_4 \om^2\right]m^2_a
I(m^2_a,m^2_{\phi_3})\right\},\hs (a,b \mbox{ not
summed}),\label{eq3}\eea where $I(a,b)$ is given in (\ref{ap4}).
With the help of (\ref{ap3}), the approximation for (\ref{eq3}) is
obtained by\be
(M^{\mathrm{rad}}_D)_{ab}\simeq-\fr{h^\nu_{ab}}{8\sqrt{2}\pi^2
v}\left\{\left[\la_3(u^2+\om^2)+\la_4
u^2\right]+\left[\la_3(u^2+\om^2)+\la_4
\om^2\right]\fr{m^2_a}{m^2_{H_2}} \right\}\nn \ee \be
=-\sqrt{2}h^\nu_{ab}\left(\fr{\la_3\om^2}{16\pi^2
v}\right)\left[1+\left(1+\fr{\la_4}{\la_3}\right)
\left(\fr{u^2}{\om^2}+\fr{m^2_a}{m^2_{H_2}}\right)+
\mathcal{O}\left(\fr{u^4}{\om^4},\fr{m^4_{a,b}}{m^4_{H_2}}
\right)\right].\label{dirac2}\ee Because of the constraint
(\ref{vevcons}) the higher-order corrections
$\mathcal{O}(\cdot\cdot\cdot)$ can be neglected, thus
$M^{\mathrm{rad}}_D$ is rewritten as follows\be
(M^{\mathrm{rad}}_D)_{ab}=-\sqrt{2}h^\nu_{ab}\left(\fr{\la_3\om^2}{16\pi^2
v}\right)\left(1+\delta_a\right),\hs \delta_a \equiv
\left(1+\fr{\la_4}{\la_3}\right)
\left(\fr{u^2}{\om^2}+\fr{m^2_a}{m^2_{H_2}}\right),
\label{epsi}\ee where $\delta_a$ is of course an infinitesimal
coefficient, i.e., $|\delta|\ll 1$. Again, this implies also that
if the fine-tuning is done the resulting Dirac-mass matrix get
trivially. It is due to the fact that the contribution of the term
associated with $\de_a$ in (\ref{epsi}) is then very small and
neglected, the remaining term gives an antisymmetric resulting
Dirac-mass matrix, that is therefore unrealistic under the data.

With this result, it is worth noting that the scale \be
\left|\fr{\la_3\om^2}{16\pi^2 v}\right|\label{sc} \ee of the
radiative Dirac masses (\ref{epsi}) is in the orders of the scale
$v$ of the tree-level Dirac masses (\ref{dirac1}). Indeed, if one
puts $|(\la_3\om^2)/(16 \pi^2 v)| = v$ and takes $|\la_3|\sim
0.1-1$, then $\om \sim 3-10\ \mathrm{TeV}$ as expected in the
constraints \cite{dln,ochoa}. The resulting Dirac-mass matrix
which is combined of (\ref{dirac1}) and (\ref{epsi}) therefore
gets two typical examples of the bounds: (i) $(\la_3\om^2)/(16
\pi^2 v)+v \sim \mathcal{O}(v) $; (ii) $(\la_3\om^2)/(16 \pi^2
v)+v \sim \mathcal{O}(0)$. The first case (i) yields that the
status on the masses of neutrinos as given above is remained
unchanged and therefore is also trivial as mentioned. In the last
case (ii), the combination of (\ref{dirac1}) and (\ref{epsi})
gives \be (M_D)_{ab}=\sqrt{2}h^\nu_{ab}(v\de_a).\label{fmr}\ee It
is interesting that in this case the scale $v$ for the Dirac
masses (\ref{dirac1}) gets naturally a large reduction, and we
argue that this is not a fine-tuning. Because the large radiative
mass term in (\ref{epsi}) is canceled by the tree-level Dirac
masses, we mean this as a finite renormalization in the masses of
neutrinos. It is also noteworthy that, unlike the case of the
tree-level mass term (\ref{dirac1}), the mass matrix (\ref{fmr})
is now {\it nonantisymmetric} in $a$ and $b$. Among the three
eigenvalues of this matrix, we can check that one vanishes (since
$\det M_D=0$) and two others massive are now {\it nondegenerate}
(splitting). Let us recall that in the first case (i) the
degeneration of the two nonzero-eigenvalues are, however, retained
because the combination of (\ref{dirac1}) and (\ref{epsi}) is
proportional to $h^{\nu}_{ab}v$.

In contrast to ({\ref{cont1}), in this case there is no large
hierarchy between $M_{L,R}$ and $M_D$. To see this explicitly, let
us take the values of the parameters as given before
(\ref{majo2}), thus $\la_3 \simeq -1.06$ and the coefficients
$\de_a$ are evaluated by \be \de_{e}\simeq 6.03\times
10^{-7},\hs\de_{\mu}\simeq 6.23\times 10^{-7},\hs
\de_{\tau}\simeq6.28\times 10^{-6}.\label{deta}\ee Hence, we get
\be |M_{L,R}|/|M_D| \sim 10^{-2}-10^{-3}.\label{her}\ee With the
values given in (\ref{deta}), the quantities $h^\nu$ and $m_D$ can
be evaluated through the mass term (\ref{fmr}); the neutrino data
imply that $h^\nu$ and $m_D$ are {\it in the orders of $h^e$ and
$m_e$} - the Yukawa coupling and mass of electron, respectively.

Because of the condition (\ref{her}) and the vanishing of one
eigenvalue of $M_D$, we can repeat the procedure as given above to
diagonalize the full matrix $M_\nu$ with $M_D$ given by
(\ref{fmr}) and $M_{L,R}$ by (\ref{majo}): First we can easily
find a mixing matrix $V$ as in (\ref{fin1}); Second in the new
basis we obtain the seesaw form as in (\ref{ben2}); Finally the
resulting mixing matrix and masses for the neutrinos are derived.
It is worth checking that the two largest squared-mass splittings
as given before can be approximately applied on this case of
(\ref{her}), such as $(m_D|\de|)^2$ and $4 (m_D |\de|) |\ep| $,
and seeing that they fit {\it naturally} the data.

\bc {\it \label{lnv}Mass Contributions from Heavy Particles}\ec

There remain now two questions not yet answered: (i) The
degeneration of $M_R=-M_L$; (ii) The hierarchy of $M_{L,R}$ and
$M_D$ (\ref{her}) can be continuously reduced? As mentioned, we
will prove that the new physics gives us the solution.

The mass Lagrangian for the neutrinos given by the operator
(\ref{heavyparticles}) can be explicitly written as follows \bea
\mathcal{L}^{\mathrm{LNV}}_{\mathrm{mass}}&=&s^\nu_{ab}\mathcal{M}^{-1}
(\langle\chi^\dag\rangle\bar{\psi}^c_{aL})(\langle\chi^\dag\rangle\psi_{bL})+\mathrm{H.c.}\crn
&=&
s^\nu_{ab}\mathcal{M}^{-1}\left(\fr{u}{\sqrt{2}}\bar{\nu}^c_{aL}
+\fr{\om}{\sqrt{2}}\bar{\nu}_{aR}\right)\left(\fr{u}{\sqrt{2}}\nu_{bL}
+\fr{\om}{\sqrt{2}}\nu^c_{bR}\right)+\mathrm{H.c.}\crn
&=&-\fr{1}{2}\bar{X}^c_L M^{\mathrm{new}}_{\nu} X_L +
\mathrm{H.c.},\eea where the mass matrix for the neutrinos is
obtained by \be M^{\mathrm{new}}_\nu \equiv-\left(
\begin{array}{cc}
\fr{u^2}{\mathcal{M}}s^\nu & \fr{u\om}{\mathcal{M}}s^\nu \\
\fr{u\om}{\mathcal{M}}s^\nu & \fr{\om^2}{\mathcal{M}}s^\nu\\
\end{array}
\right),\label{heve}\ee in which, the coupling $s^\nu_{ab}$ is
symmetric in $a$ and $b$. For convenience in reading, let us
define the submatrices of (\ref{heve}) to be $M^{\mathrm{new}}_L$,
$M^{\mathrm{new}}_D$ and $M^{\mathrm{new}}_R$ similar to that of
(\ref{matran}). Because of the condition $u^2\ll u\om \ll \om^2$,
the corresponding  submatrices $M^{\mathrm{new}}_L$,
$M^{\mathrm{new}}_D$ and $M^{\mathrm{new}}_R$ of (\ref{heve}) get
the right  hierarchies and  the two questions as mentioned are
{\it solved simultaneously}.

Intriguing comparisons between $s^\nu$ and $h^\nu$ are given in
order \ben \item $h^\nu$ conserves the lepton number; $s^\nu$
violates this charge.\item $h^\nu$ is antisymmetric and enforcing
on the Dirac-mass matrix; $s^\nu$ is symmetric and breaks this
property.\item $h^\nu$ preserves the degeneration of $M_R=-M_L$;
$s^\nu$ breaks the $M_R=-M_L$.\item A pair of $(s^\nu,h^\nu)$ in
the lepton sector will complete the rule played by the quark
couplings $(s^{q},h^{q})$ (see below).\item $h^\nu$ defines the
interactions in the standard model scale $v$; $s^\nu$ gives those
in the GUT scale $\mathcal{M}$.\een

Let us now take the values $\mathcal{M}\simeq 10^{16}\
\mathrm{GeV}$, $\om \simeq 3000\ \mathrm{GeV}$, $u\simeq 2.46\
\mathrm{GeV}$ and $s^\nu\sim \mathcal{O}(1)$ (perhaps smaller),
the submatrices $M^{\mathrm{new}}_L\simeq-6.05\times 10^{-7}
s^\nu\ \mathrm{eV}$ and $M^{\mathrm{new}}_D\simeq-7.38\times
10^{-4} s^\nu\ \mathrm{eV}$ can give contributions (to the
diagonal components of $M_L$ and $M_D$, respectively) but very
small. It is noteworthy that the last one $M^{\mathrm{new}}_R
\simeq -0.9s^\nu\ \mathrm{eV}$ can {\it dominate} $M_R$.

To summarize, in this model the neutrino mass matrix is combined
by $M_\nu+M^{\mathrm{new}}_\nu$ where the first term is defined by
(\ref{matran}), and the last term by (\ref{heve}); the submatrices
of $M_\nu$ are given in (\ref{majo}) and (\ref{fmr}),
respectively. Dependence on the strength of the new physics
coupling $s^\nu$, the submatrices of the last term,
$M^{\mathrm{new}}_{L}$ and $M^{\mathrm{new}}_{D}$, are included or
removed.

\subsubsection{\label{lfnvp}Some Remarks from Experimental
Constraints}

Conventional neutrino oscillations are insensitive to the absolute
scale of neutrino masses. Although the latter will be tested
directly in high sensitivity tritium beta decay studies and
neutrinoless double beta decay $(0\nu \bet \bet)$ as well as by
its effects on the cosmic microwave background and the large scale
structure of the Universe \cite{cosmos}. With the present of
sterile neutrinos in this model, the experimental constraints on
their masses may be also important and give us bounds on several
parameters such as the coupling $h^\nu$ and $\de_a$.

If the Liquid Scintillator Neutrino Detector experiment is
confirmed, the sterile-neutrino masses will get some values in
range of eV. In this case the coupling $h^\nu$ is also in orders
of $h^e$. The X-ray measurements yield an upper limit of sterile
neutrino mass \cite{cosmos2} $ m_s < 6.3 \ \textrm{keV}$. For all
the other cosmological constraints, the sterile neutrino masses
are in the range \cite{cosmos3} $ 2 \ \textrm{keV} < m_s < 8 \
\textrm{keV}$. In such cases the coupling $h^\nu$ will get bounds
in orders of $h^{\mu,\tau}$.

It is well-known that the radiative mass generation can also
induce the large lepton flavor violating processes such as $\mu
\rightarrow e \gamma$ as the similar one-loop effect. The possible
one-loop diagrams for this process are depicted in Fig.
(\ref{muegamma}). Suppose that $m^2_Y, m^2_{H_2}\gg
m^2_W=g^2v^2/4$ \cite{dlns} we get the approximation
\cite{hechengli} \be \mathrm{Br}(\mu\rightarrow e
\gamma)\equiv\fr{\Ga(\mu\rightarrow e \gamma)}{\Ga(\mu\rightarrow
e \widetilde{\nu}_e \nu_\mu)}
\simeq\fr{3s^4_W}{8\pi^3\alpha}\left(h^{\nu*}_{\mu\tau}h^\nu_{e\tau}\right)^2
\ee
\begin{figure}
\includegraphics{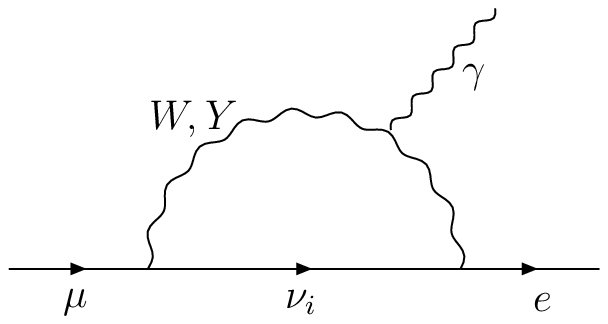}
\includegraphics{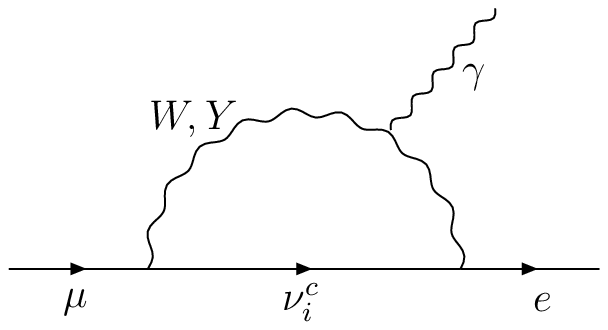}
\includegraphics{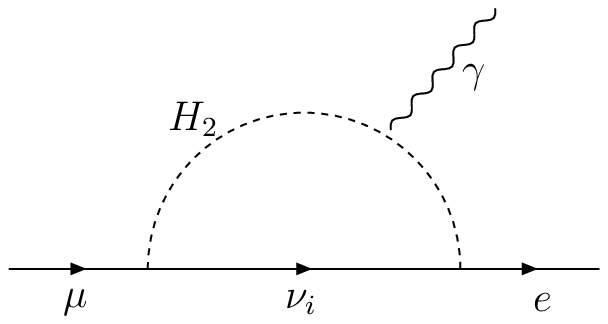}
\includegraphics{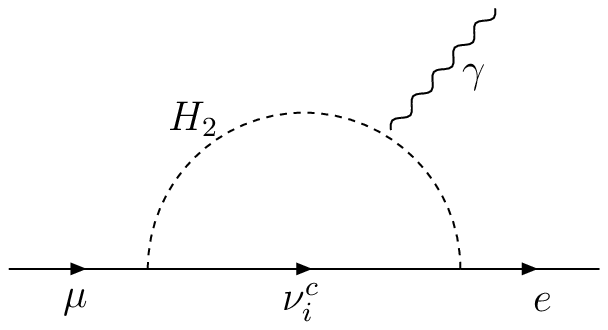}
\caption{\label{muegamma}One-loop contributions to the lepton
flavor violating decay $\mu \rightarrow e \ga$.}
\end{figure} Since $\mathrm{Br}(\mu\rightarrow e
\gamma)<1.2\times 10 ^{-11}$, $\alpha=1/128$ and $s^2_W=0.2312$
\cite{pdg}, the coupling $h^\nu$ is bounded by $h^\nu< 3.47 \times
10^{-3}$, where $h^\nu \equiv h^\nu_{e\tau}=h^\nu_{\mu \tau}$ set
is understood. Our above result, $h^\nu\sim h^e$, satisfies this
constraint. It can be shown that the value for $h^\nu$ also
satisfies constraints from such processes as $\mu\rightarrow 3e$
and $\mu e$ conversion (for more details, see \cite{kuka}).

\subsection{\label{quarkmasses}Quark Masses}

First we present the general quark mass spectrum. Some details on
the one-loop quark masses are given then.

\subsubsection{Quark Mass Spectra}

Note that in Ref. \cite{ponc1}, the authors have considered the
fermion mass spectrum under the $Z_2$ discrete symmetry which
discards the LNV interactions. Here the couplings of Eq.
(\ref{y2}) in such case are forbidden. Then it can be checked that
some quarks remain massless up to two-loop level. To solve the
mass problem of the quarks, the authors in Ref. \cite{ponc1} have
shown that one third scalar triplet has to be added to the
resulting model. In the following we show that it is not
necessary. The $Z_2$ is not introduced and thus the third one is
not required. The LNV Yukawa couplings are vital for the
economical 3-3-1 model.

The Yukawa couplings in (\ref{y1}) and (\ref{y2}) give the mass
Lagrangian for the up-quarks (quark sector with electric charge
$q_{\mathrm{up}}=2/3$) \bea
\mathcal{L}_{\mathrm{up}}^{\mathrm{mass}} &=&
\fr{h^U}{\sqrt{2}}\left( \bar{u}_{1L} u + \bar{U}_L \om \right) U_R
+ \fr{s^u_{a}}{\sqrt{2}}\left( \bar{u}_{1L} u + \bar{U}_L \om
\right) u_{aR} \crn && - \fr{v }{\sqrt{2}} \bar{u}_{\al L}\left(
h^u_{\al a} u_{aR} + s^U_{\al } U_R \right) + \mathrm{H. c.}
\label{upqmasst} \eea Consequently, we obtain the mass matrix for
the up-quarks  $(u_1, u_2, u_3, U)$ as follows \be
M_{\mathrm{up}} = \fr{1}{\sqrt{2}}\left(%
\begin{array}{cccc}
  -s^u_{1} u & -s^u_{2}u  & -s^u_{3}u & -h^U u \\
 h^u_{2 1} v  & h^u_{2 2} v & h^u_{2 3} v & s^U_{2} v \\
 h^u_{3 1} v  & h^u_{3 2} v & h^u_{3 3} v & s^U_{3} v  \\
  -s^u_{1} \om  & -s^u_{2} \om & -s^u_{3} \om & -h^U \om \\
\end{array}%
\right) \label{upqmasstu} \ee Because the first and the last rows
of the matrix (\ref{upqmasstu}) are proportional, the tree level
up-quark spectrum contains a massless one!

Similarly,  for the down-quarks ($q_{\mathrm{down}}= -1/3$), we
get the following mass Lagrangian \bea
\mathcal{L}_{\mathrm{down}}^{\mathrm{mass}} &=& \fr{h^D_{\al
\bet}}{\sqrt{2}}\left( \bar{d}_{\al L} u + \bar{D}_{\al L} \om
\right) D_{\bet R} + \fr{s^d_{\al a}}{\sqrt{2}}\left( \bar{d}_{\al
L} u + \bar{D}_{\al L} \om \right) d_{aR} \crn && + \fr{v
}{\sqrt{2}} \bar{d}_{1 L}\left(  h^d_{ a} d_{aR} + s^D_{\al}
D_{\al R} \right) + H. c. \label{upqmassdow} \eea Hence we get
mass matrix for the down-quarks $(d_1, d_2, d_3, D_2, D_3)$ \be
M_{\mathrm{down}} = - \fr{1}{\sqrt{2}}
\left(%
\begin{array}{ccccc}
  h^d_{1} v  &  h^d_{2} v  &  h^d_{3} v  &  s^D_{2} v  & s^D_{3} v \\
  s^d_{21} u &  s^d_{22} u &  s^d_{23} u &  h^D_{22} u &  h^D_{23} u \\
   s^d_{31} u &  s^d_{32} u &  s^d_{33} u  & h^D_{32} u  & h^D_{33} u  \\
   s^d_{21} \om & s^d_{22} \om & s^d_{23} \om & h^D_{22} \om & h^D_{23} \om  \\
  s^d_{31} \om & s^d_{32} \om & s^d_{33} \om & h^D_{32} \om & h^D_{33}\om \\
\end{array}%
\right)
 \label{upqmasstdow2} \ee
We see that the second and fourth rows of matrix in
(\ref{upqmasstdow2})   are proportional, while the third and the
last are the same. Hence, in this case there are two massless
eigenstates.

The masslessness of the tree level quarks in both the sectors
calls radiative corrections (the so-called mass problem of
quarks). These corrections start at the one-loop level. The
diagrams in the figure (\ref{figh4})
\begin{figure}\bc
\includegraphics[width=14cm,height=19cm]{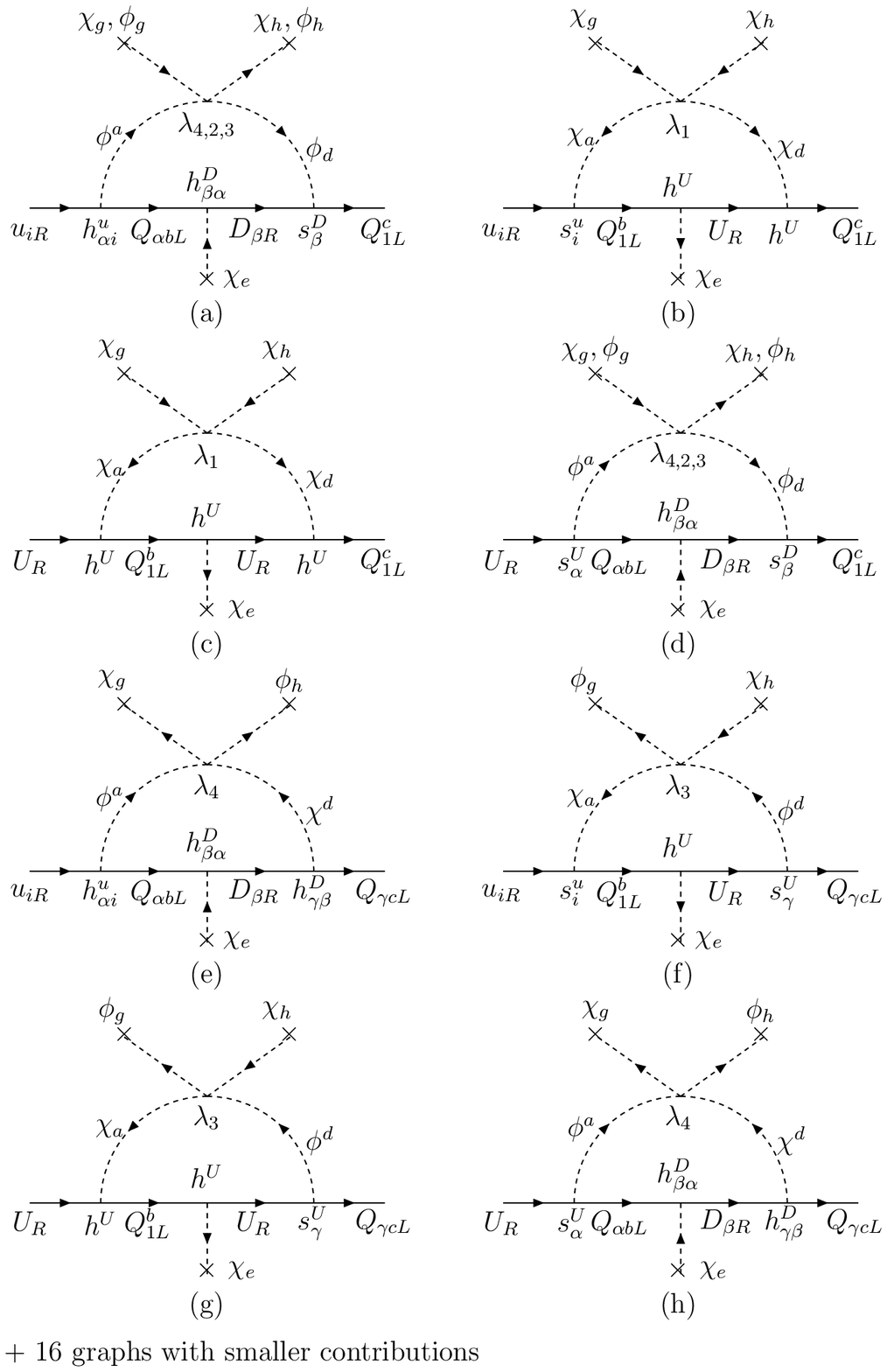}
\caption{\label{figh4}One-loop contributions to the up-quark mass
matrix (\ref{upqmasstu}).}\ec
\end{figure}contribute the up-quark
spectrum while the figure (\ref{figh5})
\begin{figure}\bc
\includegraphics[width=14cm,height=19cm]{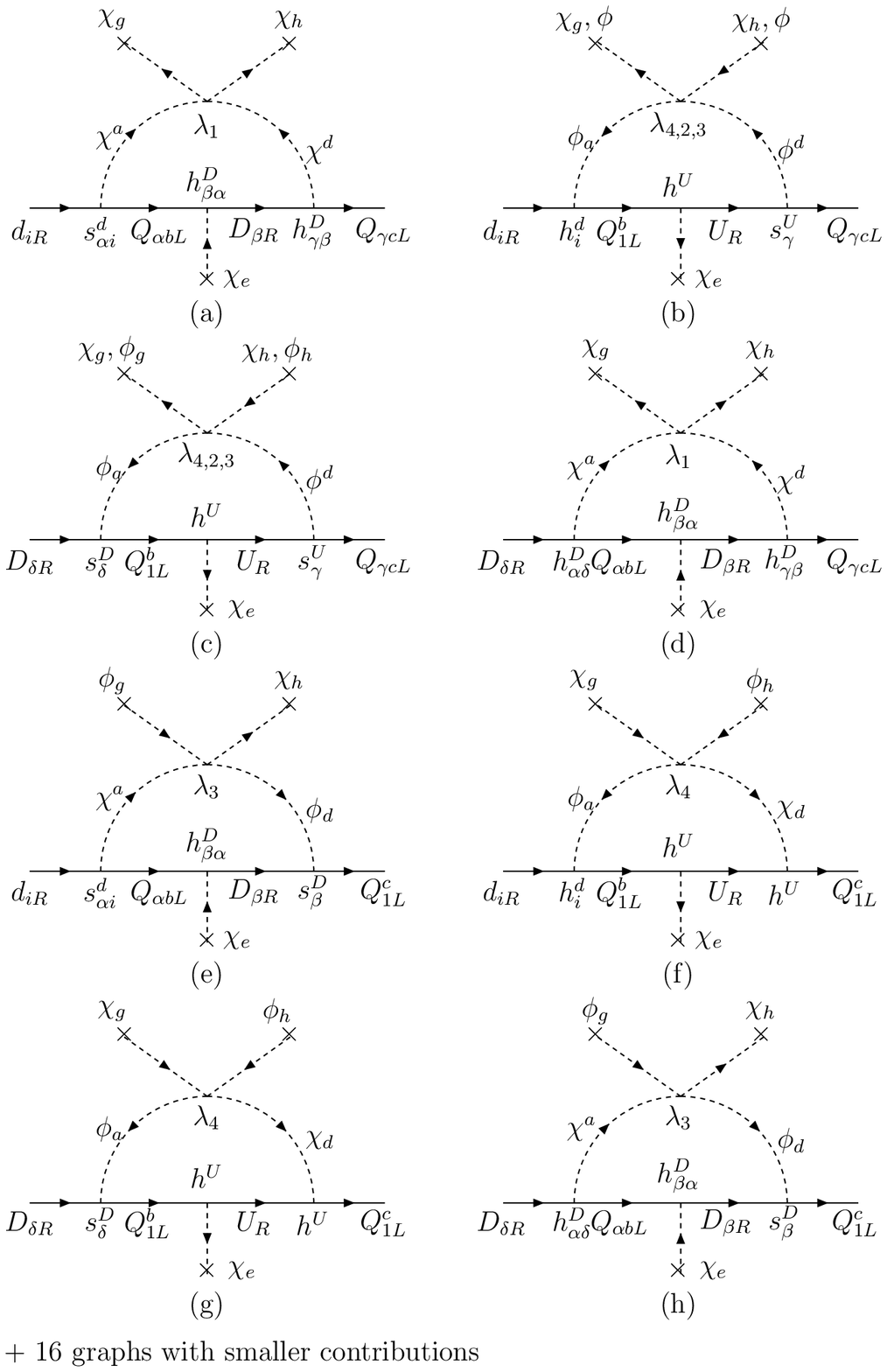}
\caption{\label{figh5}One-loop contributions to the down-quark
mass matrix (\ref{upqmasstdow2}).}\ec
\end{figure}
gives the down-quarks. Let us note the reader that the quarks also
get some one-loop contributions in the case of the $Z_2$ symmetry
enforcing \cite{ponc1}. The careful study of this radiative
mechanism shows that the one-loop quark spectrum is consistent.

\subsubsection{\label{oneloop}Typical Examples of the One-Loop Corrections}

To provide the quarks masses, in the following we can suppose that
the Yukawa couplings are flavor diagonal. Then the $u_2$ and $u_3$
states are mass eigenstates corresponding to the mass eigenvalues:
\be m_2= h^u_{22}\fr{v}{\sqrt{2}},\hs
m_{3}=h^u_{33}\fr{v}{\sqrt{2}}. \label{hh1} \ee The $u_1$ state
mixes with the exotic $U$ in terms of one sub-matrix of the mass
matrix (\ref{upqmasst})
\be M_{uU}=-\fr{1}{\sqrt{2}}\left(%
\begin{array}{cc}
  s^u_1 u & h^U u \\
  s^u_1 \om & h^U \om \\
\end{array}%
\right).\label{m11}\ee This matrix contains one massless quark
$\sim u_1$, $m_1=0$, and the remaining exotic quark $\sim U$ with
the mass of the scale $\om$.

Similarly, for the down-quarks, the $d_1$ state is a mass
eigenstate corresponding to the eigenvalue: \be m'_1 =
-h^d_{1}\fr{v}{\sqrt{2}}.\ee The pairs $(d_2,D_2)$ and $(d_3,D_3)$
are decouple, while the quarks of each pair mix via the mass
sub-matrices, respectively,\bea M_{d_2D_2}&=&
-\fr{1}{\sqrt{2}}\left(%
\begin{array}{cc}
  s^d_{22} u & h^D_{22} u \\
  s^d _{22}\om & h^D_{22} \om \\
\end{array}%
\right),\label{m22}\\ M_{d_3D_3}&=&-\fr{1}{\sqrt{2}}\left(%
\begin{array}{cc}
  s^d_{33} u & h^D_{33} u \\
  s^d _{33}\om & h^D_{33} \om \\
\end{array}%
\right).\label{m33}\eea These matrices contain the massless quarks
$\sim d_2$ and $d_3$ corresponding to $m'_2=0$ and $m'_3=0$, and
two exotic quarks $\sim D_2$ and $D_3$ with the masses of the
scale $\om$.

 With the help of the constraint (\ref{vevcons}), we identify
$m_1$, $m_2$ and $m_3$ respective to those of the $u_1=u$, $u_2 =
c$ and $u_3 = t$ quarks. The down quarks $d_1$, $d_2$ and $d_3$
are therefore corresponding to $d$, $s$ and $b$ quarks. Unlike the
usual 3-3-1 model with right-handed neutrinos, where the third
family of quarks should be discriminating \cite{longvan}, in the
model under consideration the {\it first} family has to be
different from the two others.

The mass matrices (\ref{m11}), (\ref{m22}) and (\ref{m33}) remain
the tree level properties for the quark spectra - one massless in
the up-quark sector and two in the down-quarks. From these
matrices, it is easily to verify that the conditions in
(\ref{vevcons}) and (\ref{dkhsyu}) are satisfied. First, we
consider radiative corrections to the up-quark masses.

\bc{\it Up Quarks}\ec

In the previous studies \cite{ponc1,violat}, the LNV interactions
have often been excluded, commonly by the adoption of an
appropriate discrete symmetry. Let us remind that there is no
reason within the 3-3-1 model to ignore such interactions. The
experimental limits on processes which do not conserve total
lepton numbers, such as neutrinoless double beta decay
\cite{0bedecay}, require them to be small.

If the Yukawa Lagrangian is restricted to
$\mathcal{L}_{\mathrm{LNC}}$ \cite{ponc1}, then the mass matrix
(\ref{m11}) becomes
\be M_{uU}=-\fr{1}{\sqrt{2}}\left(%
\begin{array}{cc}
  0 & h^U u \\
  0 & h^U \om \\
\end{array}%
\right).\label{m110}\ee In this case, only the element
$(M_{uU})_{12}$ gets an one-loop correction defined by the figure
(\ref{figh6}). Other elements remain unchanged under this one-loop
effect.\vs
\begin{figure}\bc
\includegraphics{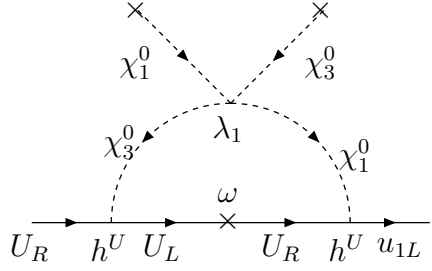}
\caption{\label{figh6}One-loop contribution under $Z_2$ to the
up-quark mass matrix (\ref{m110})}\ec
\end{figure}
The Feynman rules gives us \bea -i (M_{uU})_{12}P_R &=&\int
\fr{d^4 p}{(2\pi)^4}(ih^U
P_R)\fr{i(p\!\!\!/+M_{U})}{p^2-M^2_{U}}(- i M_U P_L)
\fr{i(p\!\!\!/+M_{U})}{p^2-M^2_{U}}(ih^U P_R)\crn &
&\times\fr{-1}{(p^2-M^2_{\chi_1})(p^2-M^2_{\chi_3})}(i 4
\la_1)\fr{ u\om}{2}.\nn\eea  Thus, we get \bea (M_{uU})_{12}
&=&-2i u\om \la_1 M_{U} (h^U)^2 \int \fr{d^4
p}{(2\pi)^4}\fr{p^2}{(p^2-M^2_U)^2(p^2-M^2_{\chi_3})(p^2-M^2_{\chi_1})}
\crn &\equiv&-2i u\om \la_1 M_{U} (h^U)^2
I(M^2_U,M^2_{\chi_3},M^2_{\chi_1}).\eea The integral $I(a,b,c)$
with $a,b\gg c$ is given in  the  \ref{ap1}.  Following  Ref.
\cite{dls},  we conclude that in an effective approximation,
$M^2_U,\ M^2_{\chi_3} \gg M^2_{\chi_1}$. Hence we have \bea
(M_{uU})_{12} &\simeq&-\fr{\la_1 t_\theta  M^3_{U}
}{4\pi^2}\left[\fr{M^2_U-M^2_{\chi_3}+M^2_{\chi_3}\ln\fr{
M^2_{\chi_3}}{M^2_U}}{(M^2_U-M^2_{\chi_3})^2}\right]\sim u, \crn
&\equiv& -\fr{1}{\sqrt{2}}R(M_U). \eea The resulting mass matrix
is given by
\be M_{uU}=-\fr{1}{\sqrt{2}}\left(%
\begin{array}{cc}
  0 & h^U u+R \\
  0 & h^U \om \\
\end{array}%
\right).\label{m111}\ee We see that one quark remains massless as
the case of the tree level spectrum. This result keeps up to
two-loop level, and can be applied to the down-quark sector as
well as in the cases of non-diagonal Yukawa couplings. Therefore,
under the $Z_2$, it is not able to provide consistent masses for
the quarks.

If the full Yukawa Lagrangian is used, the LNV couplings must be
enough small in comparison with the usual couplings   [see
(\ref{dkhsyu})]. Combining (\ref{vevcons}) and (\ref{dkhsyu}) we
have \be h^U\om \gg h^U u,\ s^u_1 \om \gg s^u_1 u.\ee In this
case, the element  $(M_{uU})_{11}$ of (\ref{m11}) gets the
radiative correction  depicted in Fig.(\ref{figh7}).
\begin{figure}\bc
\includegraphics{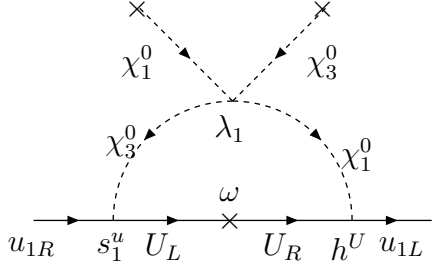}
\caption{\label{figh7}One-loop contribution to the up-quark mass
matrix (\ref{m11})}\ec
\end{figure} The resulting mass matrix is obtained by
\be M_{uU}=-\fr{1}{\sqrt{2}}\left(%
\begin{array}{cc}
  s^u_1( u +\fr{R}{h^U}) & h^U u \\
  s^u_1 \om & h^U \om \\
\end{array}%
\right).\label{m1111}\ee In contradiction with the first case, the
mass of $u$ quark is now non-zero and given by\be m_u \simeq
\fr{s^u_1}{\sqrt{2}h^U} R \label{mu2}.\ee

Let us note that the matrix (\ref{m1111}) gives an eigenvalue in
the scale of $\fr{1}{\sqrt{2}}h^U \om$ which can be identified
with that of the exotic quark $U$. In effective approximation
\cite{dls}, the mass for the Higgs $\chi_3$ is defined by
$M^2_{\chi_3}\simeq 2\la_1 \om^2$. Hereafter, for the parameters,
we use the following values $\la_1=2.0$, $t_\theta=0.08$ as
mentioned, and $\om =10\ \mathrm{TeV}$. The mass value for the $u$
quark is as function of $s^u_1$ and $h^U$. Some values of the pair
$(s^u_1,h^U)$ which give consistent masses for the $u$ quark is
listed in Table \ref{uvalues}.
\begin{table}[h]
\caption{\label{uvalues}Mass for the $u$ quark as function of
$(s^u_1,h^U)$.}\bc
\begin{tabular}{|c|c|c|c|c|c|}
  \hline
  $h^U$ & 2 & 1.5 & 1 & 0.5 & 0.1\\ \hline
  $s^u_1$& 0.0002 & 0.0003 & 0.0004 & 0.001 & 0.01 \\ \hline
  $m_u$ [MeV]& 2.207 & 2.565 & 2.246 & 2.375 & 2.025 \\
  \hline
\end{tabular}
\ec
\end{table}

Note that the mass values in the Table \ref{uvalues} for the $u$
quark are in good consistence with the data given in
Ref.~\cite{pdg}: $m_u \in 1.5\ \div \ 4 \ \mathrm{MeV}$.

\bc{\it Down Quarks}\ec

For the down quarks, the constraint, \be h^D_{\al\al} \om \gg
h^D_{\al\al} u,\ s^d_{\al\al} \om \gg s^d_{\al \al} u,\ee should
be applied. In this case, three elements $(M_{d_\al D_\al})_{11}$,
$(M_{d_\al D_\al})_{12}$ and $(M_{d_\al D_\al})_{21}$ will get
radiative corrections. The relevant diagrams are depicted in
figure (\ref{figh8}).
\begin{figure}
\includegraphics{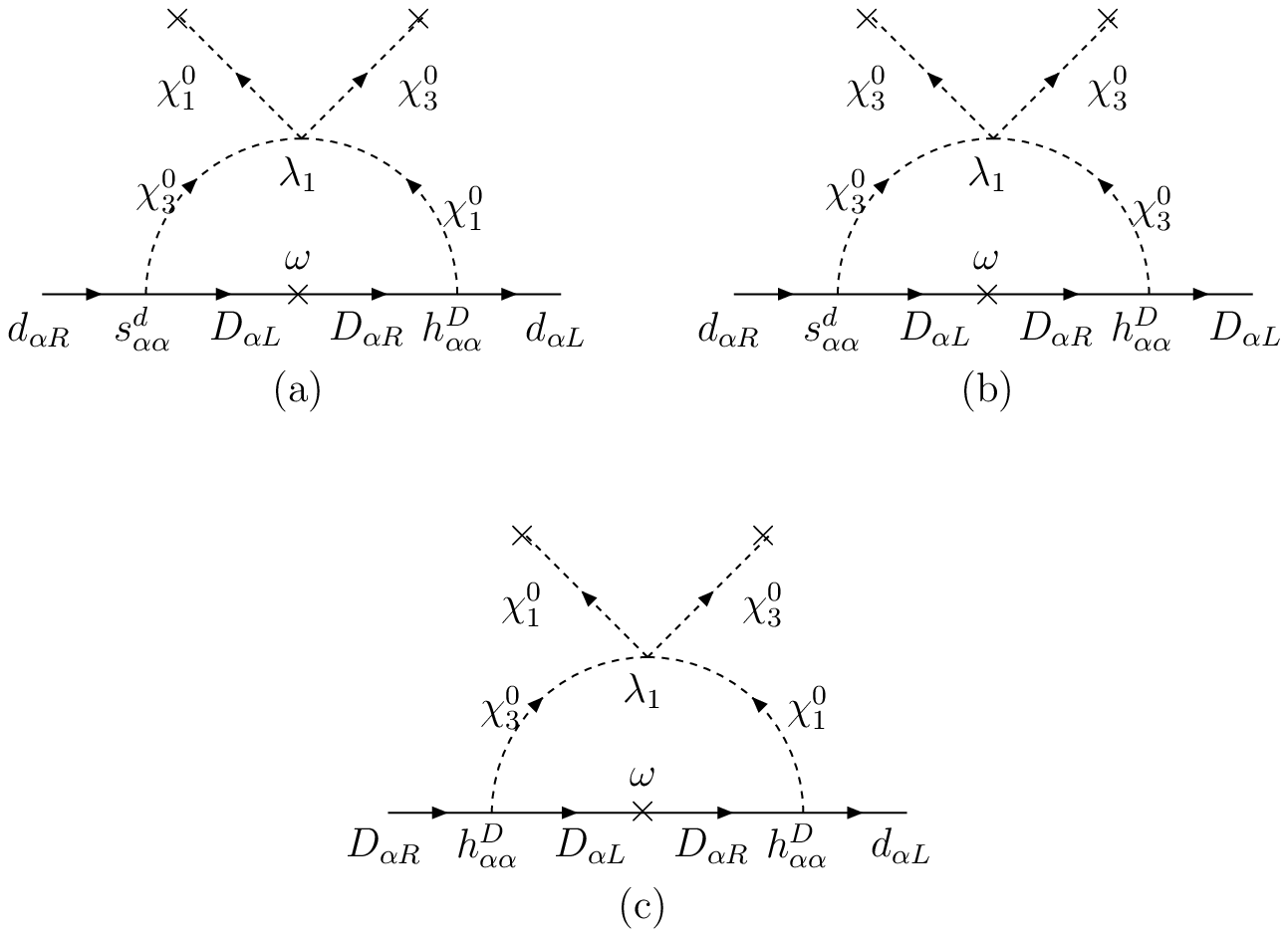}
\caption{\label{figh8}One-loop contributions to the down-quark
mass matrix (\ref{m22}) or (\ref{m33}).}
\end{figure}
It is worth noting that diagram \ref{figh8}(c) exists even in the
case of the $Z_2$ symmetry. The contributions are given by \bea
(M_{d_\al D_\al})_{11}&=&-\fr{s^d_{\al \al}}{\sqrt{2}h^D_{\al
\al}}R(M_{D_\al}),\\ (M_{d_\al D_\al})_{21}&=&-4i \la_1
\fr{s^d_{\al \al}}{h^D_{\al \al}}M^3_{D_{ \al}}
I(M^2_{D_{\al}},M^2_{\chi_3},M^2_{\chi_3})\crn &=&-\fr{\la_1
s^d_{\al\al}M^3_{D_{ \al}}}{4 \pi^2 h^D_{\al
\al}}\left[\fr{M^2_{D_{\al}}+M^2_{\chi_3}}{(M^2_{D_{\al}}
-M^2_{\chi_3})^2}-\fr{2M^2_{D_{\al}}M^2_{\chi_3}}
{(M^2_{D_{\al}}-M^2_{\chi_3})^3}\ln\fr{M^2_{D_{\al}}}
{M^2_{\chi_3}}\right]\crn
&\equiv&-\fr{1}{\sqrt{2}}R'(M_{D_{\al}}),
\\ (M_{d_\al D_\al})_{12}&=&-\fr{1}{\sqrt{2}}R(M_{D_\al}).\eea
We see that two last terms are much larger than the first one.
This is responsible for the masses of the quarks $d_2$ and $d_3$.
At the one-loop level, the mass matrix for the down-quarks is
given by
 \be M_{d_\al D_\al}=
-\fr{1}{\sqrt{2}}\left(%
\begin{array}{cc}
  s^d_{\al \al}( u+\fr{R}{h^D_{\al\al}}) & h^D_{\al\al} u+R \\
  s^d _{\al\al}\om + R' & h^D_{\al\al} \om \\
\end{array}%
\right).\label{domass}\ee

We remind the reader that a matrix (see also \cite{chengli})\be
\left(
  \begin{array}{cc}
    a & c \\
    b & D \\
  \end{array}
\right) \label{ptbac2}\ee with $D\gg b, c \gg a$ has two
eigenvalues \bea x_1 &\simeq&
\left[a^2-\fr{2bca}{D}+\fr{b^2c^2-(b^2+c^2)a^2}{D^2}\right]^{1/2},\crn
x_2 & \simeq & D. \label{ptbac22}\eea Therefore
 the mass matrix in (\ref{domass}) gives an eigenvalue in the scale of
$D \equiv \fr{1}{\sqrt{2}} h^D_{\al\al} \om $ which is of the
exotic quark $D'_\al$. Here we  have another
eigenvalue for the mass of $d'_\al$ \be m_{d'_\al}
=\fr{h^D_{\al\al}u+R}{\sqrt{2}h^D_{\al\al}\om}\left\{R'^2
-\fr{(s^d_{\al\al})^2}{(h^D_{\al\al})^2}\left[(s^d_{\al
\al}\om+R')^2+(h^D_{\al \al}u+R)^2\right]\right\}^{1/2}
\label{ptbac23}.\ee

Let us remember that $M^2_{\chi_3}\simeq 2\la_1 \om^2$, and the
parameters $\la_1=2.0$, $t_\theta=0.08$ and $\om =10\
\mathrm{TeV}$ as given above are used in this case. The
$m_{d_\al}$ is function of $s^d_{\al\al}$ and $h^D_{\al\al}$. We
take the value $h^D_{\al\al}=2.0$ for both the sectors, $\al=2$
and $\al=3$. If $s^d_{22}=0.015$ we get then the mass of the
so-called $s$ quark \be m_s = 99.3\ \mathrm{MeV}.\label{ms1}\ee
For the down quark of the third family, we put $s^d_{33}=0.7$.
Then, the mass of the $b$ quark is obtained by \be m_b =4.4\
\mathrm{GeV}.\label{mb1}\ee

We emphasize again  that Eqs. (\ref{ms1}) and (\ref{mb1}) are in
good consistence with the data given in Ref.~\cite{pdg}: $m_s \sim
95\ \pm 25 \ \mathrm{MeV}$ and  $m_b \sim 4.70 \pm 0.07 \
\mathrm{GeV}$.

\subsection{Summary}

The basic motivation of this section is to present the answer to one
of the most crucial questions: whether within the framework of the
model based on $\mathrm{SU}(3)_C\otimes \mathrm{SU}(3)_L \otimes
\mathrm{U}(1)_X$ gauge group contained minimal Higgs sector with
right-handed neutrinos, all fermions including quarks and neutrinos
can gain the consistent masses.

In this model, the masses of neutrinos are given by three different
sources widely ranging over the mass scales including the GUT's and
the small VEV $u$ of spontaneous lepton breaking. At the tree-level,
there are three Dirac neutrinos: one massless and two degenerate
with the masses in the order of the electron mass. At the one-loop
level, a possible framework for the finite renormalization of the
neutrino masses is obtained. The Dirac masses obtain a large
reduction, the Majorana mass types get degenerate in $M_R=-M_L$, all
these masses are in the bound of the data. It is emphasized that the
above degeneration is a consequence of the fact that the left-handed
and right-handed neutrinos in this model are in the same gauge
triplets. The new physics including the 3-3-1 model are strongly
signified. The degenerations and hierarchies among the mass types
are completely removed by heavy particles.

The resulting mass matrix for the neutrinos consists of two parts
$M_\nu+M^{\mathrm{new}}_\nu$: the first is mediated by the model
particles, and the last is due to the new physics. Upon the
contributions of $M^{\mathrm{new}}_\nu$, the different realistic
mass textures can be produced. For example, neglecting the last
term, the pseudo-Dirac patterns can be obtained. In another
scenario, that the bare coupling $h^{\nu}$ of Dirac masses get
higher values, for example, in orders of $h^{\mu,\tau}$, the VEV
$\om$ can be picked up to an enough large value $(\sim
\mathcal{O}(10^4-10^5)\ \mathrm{TeV})$ so that the type II seesaw
spectrum is obtained. Such features deserve further study. We have
also shown that the lepton flavor violating processes such as $\mu
\rightarrow e \gamma $, $\mu\rightarrow 3e$ and $\mu e$ conversion
get the  consistent values in the bounds of the current experiments.

In the first section we have shown that, in the considered model,
there are three quite different scales of vacuum expectation values:
$\omega \sim \mathcal{O}(1)\ \mathrm{TeV},\ v \approx 246\
\mathrm{GeV}$ and $ u \sim \mathcal{O}(1)\ \mathrm{GeV}$. In this
section we have added a new characteristic property, namely, there
are two types of Yukawa couplings with different strengths: the LNC
coupling $h$'s and the LNV ones $s$'s satisfying the condition: $ s
\ll h$. With the help of these key properties, the mass spectrum of
quarks is consistent without introducing the third scalar triplet.
With the given  set of parameters, the numerical evaluation shows
that in this model, masses of the exotic quarks also have different
scales, namely, the $U$ exotic quark ($q_U = 2/3$) gains mass $m_U
\approx 700 $ GeV, while the $D_\al$ exotic quarks ($q_{D_\al} =
-1/3$) have masses in the TeV scale: $m_{D_\al} \in 10 \div 80$ TeV.

Let us summarize our results: \ben \item {\it At the tree level}
\ben
\item  All charged leptons gain masses similar
to that in the standard model.
\item One neutrino is massless and other two are
degenerate in masses.
\item  Three quarks $u_1, d_2, d_3$ are massless.
\item All exotic quarks gain masses proportional to $\om$ - the
VEV of the first step of symmetry breaking. \een
\item {\it At the one-loop level}
\ben
\item  All above-mentioned fermions gain masses.
\item The light-quarks gain masses proportional to $u$ - the LNV parameter.
\item The exotic quark masses are separated:
$ m_U \approx 700\mathrm{GeV}, \ m_{D_\al}\in 10 \div
80\mathrm{TeV}$.
\item There exist two types of Yukawa couplings: the LNC and LNV with
quite different strengths.\een \een
 With the {\it positive} answer, the economical version
becomes one of the very attractive models beyond the standard model.

\section{\label{conclutons}Conclusion}

Finally, this is the time to mention some developments of the model
as reported on this work \cite{ponce,ponc1,dlns,dls,dln,dhhl,dls1}.
The idea to give VEVs at the top and bottom elements of $\chi$
triplet was given in \cite{ponce}. Some consequences such as the
atomic parity violation, $Z-Z'$ mixing angle and $Z'$ mass were
studied~\cite{ponc1}. However, in the above-mentioned works, the
$W-Y$ and $W_4-Z-Z'$ mixings were excluded. To solve the
difficulties such as the standard model coupling $ZZ h$ or quark
masses, the third scalar triplet was introduced. Thus, the scalar
sector was no longer minimal and the economical in this sense was
unrealistic!

In the beginning of the last year, there was a new step in
development of the model. In Ref~\cite{dlns}, the correct
identification of non-Hermitian bilepton gauge boson $X^0$ was
established. The $W-Y$ mixing as well as $W_4, Z, Z'$ one were also
entered into couplings among fermions and gauge bosons. The
lepton-number violating interactions exist in both charged and
neutral gauge boson sectors. However, the lepton-number violation
happens only in the neutrino and exotic quarks sectors, but not in
the charged lepton sector. The scalar sector was studied in
Ref.~\cite{dls} and all gauge-Higgs couplings were presented and all
similar  ones in the standard model were recovered. The Higgs sector
contains eight Goldstone bosons - the needed number for massive
gauge ones of the model. Interesting to note that, the $CP$-odd part
of Goldstone associated with the neutral non-Hermitian gauge boson
$G_{X^0}$ is decoupled, while its $CP$-even counterpart has the
mixing by the same way in the gauge boson sector.

In Ref.~\cite{dln}, the deviation $\de Q_{\mathrm{W}}$ of the weak
charge from its standard model prediction due to the mixing of the
$W$ boson with the charged bilepton $Y$ as well as of the $Z$ boson
with the neutral $Z'$ and the real part of the non-Hermitian neutral
bilepton $X^0$ is established.

The model is consistent with the effective theory and new
experiments because it can provide all fermions including the quarks
and neutrinos with the consistent masses \cite{dhhl,dls1}. The
exotic quarks and new bosons get masses in order of TeV. There are
two different scales of exotic quark masses: $ m_U \approx 700\
\mathrm{GeV}, \ m_{D_\al}\in 10 \div 80\ \mathrm{TeV}$.

It is worth  mentioning on advantage of the model: the new mixing
angle between the charged gauge bosons $\theta$ is connected with
one of the VEVs $u$ - the parameter of lepton-number violations.
There is no new  parameter, but it contains very simple Higgs
sector, hence the significant number of free parameters is reduced.
The Higgs self-couplings $\la_{1,2,4}$ are constrained by the scalar
masses, but the remainder $\la_3$ is fixed by the neutrino masses
\cite{dls1}. This means also that the generation of the neutrino
masses leads to a shift in mass of the Higgs boson from the standard
model prediction.

The model is rich in physics because it includes the right-handed
neutrinos, exotic quarks and new bosons, and also gives an possible
explanation of the generation question, electric charge quantization
and current neutrino mass problem. The suppersymmetric version has
being been considered \cite{supeco}. The new physics is at TeV scale
therefore the results can be verified in the next generation of
collides such as LHC and ILC.

\section*{Acknowledgments}

P.V.D. is grateful to Nishina Fellowship Foundation for financial
support. He would like to thank Prof. Y. Okada and Members of Theory
Group at KEK for warm hospitality during his visit where this work
was completed. This work was also supported by National Council for
Natural Sciences of Vietnam.

\appendix

\section{Mixing Matrices}

For convenience in calculating, in this appendix we give the
mixing matrices of the gauge and Higgs sectors.

\subsection{\label{matrantron}Neutral Gauge Bosons}
\be \left(%
\begin{array}{c}
  W_3 \\
  W_8 \\
  B \\
  W_4
\end{array}%
\right)= \left(%
\begin{array}{cccc}
  s_W & c_\va c_{\theta'}c_W  & s_\va c_{\theta'}c_W  & s_{\theta'}c_W \\
  -\fr{s_W}{\sqrt{3}} & \fr{c_\va(s^2_W-3c^2_Ws^2_{\theta'})
  -s_\va \la\kappa}{\sqrt{3}c_Wc_{\theta'}} &
  \fr{s_\va(s^2_W-3c^2_Ws^2_{\theta'})+
  c_\va \la \kappa}{\sqrt{3}c_Wc_{\theta'}} & \sqrt{3}s_{\theta'}c_W \\
  \fr{\kappa}{\sqrt{3}} & -\fr{t_W(c_\va\kappa
  +s_\va \la)}{\sqrt{3}c_{\theta'}} & -\fr{t_W(s_\va \kappa
 -c_\va \la)}{\sqrt{3}c_{\theta'}}  & 0 \\
  0 & -t_{\theta'}(c_\va\lambda
  -s_\va \kappa) & -t_{\theta'}(s_\va\la
  +c_\va \kappa) & \la
\end{array}%
\right)\left(%
\begin{array}{c}
  A \\
  Z^1 \\
  Z^2 \\
  W'_{4}
\end{array}%
\right), \ee where we have denoted  \bea s_{\theta'}\equiv
t_{2\theta}/(c_W\sqrt{1+4t^2_{2\theta}}),\hs
\kappa\equiv\sqrt{4c^2_W-1},\hs
\la\equiv\sqrt{1-4s^2_{\theta'}c^2_W}.\eea

\subsection{Neutral scalar bosons}
  \bea \left( \begin{array}{ccc} S_1 \\
S_2 \\ S_3
\end{array}\right) &=& \left( \begin{array}{ccc}  -s_\zeta
s_\theta & c_\zeta s_\theta
&  c_\theta \\
c_\zeta &
 s_\zeta & 0 \\
-s_\zeta c_\theta & c_\zeta c_\theta & -s_\theta
\end{array}\right) \left( \begin{array}{ccc}
H \\  H^0_1 \\ G_4 \end{array}\right), \label{potenn17} \eea

\subsection{Singly-charged scalar bosons}  \bea
 \left(%
\begin{array}{c}
  \phi^+_1 \\
  \chi^+_2 \\
  \phi^+_3
\end{array}%
\right)&=&\fr{1}{\sqrt{\om^2+c^2_\theta v^2}}\left(%
\begin{array}{ccc}
  \om s_\theta & c_\theta \sqrt{\om^2+c^2_\theta v^2} &
  \fr{v s_{2\theta}}{2} \\
  vc_\theta & 0 & -\om\\
  \om c_\theta & -s_\theta \sqrt{\om^2+c^2_\theta v^2} & v c^2_{\theta}
\end{array}%
\right)\left(%
\begin{array}{c}
  H^+_2 \\
  G^+_5 \\
  G^+_6
\end{array}%
\right).\label{potenn19}\eea

\section{\label{ap1}Feynman integrations} In this appendix, we
present evaluation of the integral \be I(a,b,c)\equiv
\int\fr{d^4p}{(2\pi)^4}\fr{p^2}{(p^2-a)^2(p^2-b)(p^2-c)},\label{tichphan}\ee
where $a,b,c>0$ and $I(a,b,c)=I(a,c,b)$.

\subsection{Case of $b\neq c$ and $b,c\neq a$}
We first introduce a well-known integral as follows\bea
\int\fr{d^4p}{(2\pi)^4}\fr{1}{(p^2-a)(p^2-b)(p^2-c)}
&=&\fr{-i}{16\pi^2}\left\{\fr{a\ln a}{(a-b)(a-c)}+\fr{b\ln
b}{(b-a)(b-c)}\right.\crn &&\left.+\fr{c\ln
c}{(c-b)(c-a)}\right\}.\label{interg}\eea Differentiating two
sides of this equation with respect to $a$ we have \bea
\int\fr{d^4p}{(2\pi)^4}\fr{1}{(p^2-a)^2(p^2-b)(p^2-c)}=\fr{-i}{16\pi^2}
\left\{\fr{\ln a +1}{(a-b)(a-c)}\right.\crn  \left.-\fr{a(2a -b
-c)\ln a }{(a-b)^2(a-c)^2}+\fr{b\ln b}{(b-a)^2(b-c)}+\fr{c\ln
c}{(c-a)^2(c-b)}\right\}.\label{intergd}\eea Combining
(\ref{interg}) and (\ref{intergd}) the integral (\ref{tichphan})
becomes \bea I(a,b,c)&=&
\int\fr{d^4p}{(2\pi)^4}\left[\fr{1}{(p^2-a)(p^2-b)(p^2-c)} +
\fr{a}{(p^2-a)^2(p^2-b)(p^2-c)}\right]\crn&=&
\fr{-i}{16\pi^2}\left\{\fr{a(2\ln a +1)}{(a-b)(a-c)} -\fr{a^2(2a
-b -c)\ln a }{(a-b)^2(a-c)^2}+\fr{b^2\ln
b}{(b-a)^2(b-c)}\right.\crn &&\left.+\fr{c^2\ln
c}{(c-a)^2(c-b)}\right\}.\eea

If $a,b\gg c$ or $c\simeq 0 $, we have an approximation as follows
\be I(a,b,c)\simeq
-\fr{i}{16\pi^2}\fr{1}{a-b}\left[1-\fr{b}{a-b}\ln\fr{a}{b}\right].
\label{ketqua}\ee

\subsection{Case of $b=c$ and $b\neq a$}
We put \be I(a,b)\equiv
I(a,b,b)=\int\fr{d^4p}{(2\pi)^4}\fr{p^2}{(p^2-a)^2(p^2-b)^2},\label{e1}\ee
where $I(a,b)=I(b,a)$.

Using the Feynman's parametrization, \be
\fr{1}{A^2B^2}=\fr{\Ga(4)}{\Ga(2)\Ga(2)}\int^1_0 dx
\fr{x(1-x)}{\left[xA+(1-x)B\right]^4},\ee we have \be
\fr{1}{(p^2-a)^2(p^2-b)^2}=6\int^1_0 dx
\fr{x(1-x)}{(p^2-M^2)^4},\ee where $M^2\equiv xa+(1-x)b$. The
equation (\ref{e1}) therefore become \be I(a,b)=6\int^1_0dx
x(1-x)\int\fr{d^4p}{(2\pi)^4}\fr{p^2}{(p^2-M^2)^4}.\label{e2}\ee

With the help of \be
\int\fr{d^4p}{(2\pi)^4}\fr{p^2}{(p^2-M^2)^4}=\fr{-i}{3(4\pi)^2}\fr{1}{M^2},\ee
Eq. (\ref{e2}) is given by \be
I(a,b)=\fr{-2i}{(4\pi)^2}\int^1_0dx\fr{x(1-x)}{xa+(1-x)b}.
\label{e3}\ee To obtain the integral we can put $t=xa+(1-x)b$, the
Eq. (\ref{e3}) is then rewitten\be
I(a,b)=\fr{2i}{(4\pi)^2(a-b)^3}\int^a_b dt
\left[t-(a+b)+\fr{ab}{t}\right].\ee Therefore we get \be
I(a,b)=-\fr{i}{16
\pi^2}\left[\fr{a+b}{(a-b)^2}-\fr{2ab}{(a-b)^3}\ln\fr{a}{b}\right].\label{ap4}\ee

If $b\gg a$ or $a\simeq 0$, we have the following approximation
\be I(a,b)\simeq-\fr{i}{16\pi^2b}.\label{ap3}\ee

Let us note that the above approximations $aI(a,b,c)$ (or
$bI(a,b,c)$) and $bI(a,b)$ are kept in the orders up to
$\mathcal{O}(c/a,c/b)$ and $\mathcal{O}(a/b)$, respectively.


\end{document}